\documentclass[a4paper,fleqn,usenatbib]{mnras}

\usepackage{Times}

\usepackage[T1]{fontenc}
\usepackage{ae,aecompl}
\usepackage{multirow}

\usepackage{graphicx}	
\usepackage{amsmath}	
\usepackage{amssymb}	

\title[Bo\"otes-HiZELS: a $z=0.4-4.7$ survey]{Bo\"otes-HiZELS: An optical to near-infrared survey of emission-line galaxies at $\bf z=0.4-4.7$ }

\author[J. Matthee et al.]{Jorryt Matthee$^{1}$\thanks{E-mail: matthee@strw.leidenuniv.nl}, David Sobral$^{2,1}$, Philip Best$^3$, Ian Smail$^4$, Fuyan Bian$^{5,6}$, \newauthor Behnam Darvish$^7$, Huub R\"ottgering$^1$, Xiaohui Fan$^8$\\
$^{1}$ Leiden Observatory, Leiden University, P.O.\ Box 9513, NL-2300 RA Leiden, The Netherlands\\
$^{2}$ Department of Physics, Lancaster University, Lancaster, LA1 4YB, UK\\
$^{3}$ Institute for Astronomy, University of Edinburgh, Royal Observatory, Blackford Hill, Edinburgh EH9 3HJ, UK\\
$^{4}$ Centre for Extragalactic Astronomy, Department of Physics, Durham University, South Road, Durham DH1 3LE UK \\
$^{5}$ Research School of Astronomy \& Astrophysics, Mt Stromlo Observatory, Australian National University, Weston Creek, ACT, 2611, Australia \\
$^{6}$ Stromlo fellow \\
$^{7}$ Cahill Center for Astrophysics, California Institute of Technology, 1216 East California Boulevard, Pasadena, CA 91125, USA \\
$^{8}$ Steward Observatory, University of Arizona, 933 North Cherry Avenue, Tucson, AZ, 85721, USA \\
}

\pubyear{2017}

\begin{document}
\label{firstpage}
\pagerange{\pageref{firstpage}--\pageref{lastpage}}
\maketitle

\begin{abstract}
We present a sample of $\sim 1000$ emission line galaxies at $z=0.4-4.7$ from the $\sim0.7$deg$^2$ High-$z$ Emission Line Survey (HiZELS) in the Bo\"otes field identified with a suite of six narrow-band filters at $\approx 0.4-2.1$ $\mu$m. These galaxies have been selected on their Ly$\alpha$ (73), {\sc [Oii]} (285), H$\beta$/{\sc [Oiii]} (387) or H$\alpha$ (362) emission-line, and have been classified with optical to near-infrared colours. A subsample of 98 sources have reliable redshifts from multiple narrow-band (e.g. [O{\sc ii}]-H$\alpha$) detections and/or spectroscopy. In this survey paper, we present the observations, selection and catalogs of emitters. We measure number densities of Ly$\alpha$, [O{\sc ii}], H$\beta$/{\sc [Oiii]} and H$\alpha$ and confirm strong luminosity evolution in star-forming galaxies from $z\sim0.4$ to $\sim 5$, in agreement with previous results. To demonstrate the usefulness of dual-line emitters, we use the sample of dual [O{\sc ii}]-H$\alpha$ emitters to measure the observed [O{\sc ii}]/H$\alpha$ ratio at $z=1.47$. The observed [O{\sc ii}]/H$\alpha$ ratio increases significantly from 0.40$\pm0.01$ at $z=0.1$ to 0.52$\pm0.05$ at $z=1.47$, which we attribute to either decreasing dust attenuation with redshift, or due to a bias in the (typically) fiber-measurements in the local Universe which only measure the central kpc regions. At the bright end, we find that both the H$\alpha$ and Ly$\alpha$ number densities at $z\approx2.2$ deviate significantly from a Schechter form, following a power-law. We show that this is driven entirely by an increasing X-ray/AGN fraction with line-luminosity, which reaches $\approx 100$ \% at line-luminosities $L\gtrsim3\times10^{44}$ erg s$^{-1}$. \end{abstract}

\begin{keywords}
galaxies: evolution -- galaxies: high-redshift -- galaxies: star formation -- galaxies: luminosity function, mass function -- galaxies: active \end{keywords}



\section{Introduction}
Understanding how and when galaxies grow their stellar mass and in some cases eventually stop forming stars are key goals of galaxy formation theory. However, since it is only possible to observe an individual galaxy at a single epoch, to assess their evolution it is crucial to homogeneously select equivalent samples of galaxies over a wide redshift range. Currently, different epochs in cosmic time are probed by different selections of galaxies. Moreover, the galaxy properties (such as star formation rates and estimates of dust attenuation) are measured with different tracers \citep[e.g.][]{Speagle2014}. Therefore, it is important to understand whether local calibrations can be extrapolated to high redshift. This requires large samples of galaxies with a well understood selection function and a large dynamic range in galaxy properties. 

Homogeneously selected samples of star-forming galaxies can be obtained with narrow-band (NB) surveys, that are very efficient in selecting emission-line galaxies. Using different NBs, galaxies can be selected on a particular emission-line across a range of redshifts with well defined line-luminosity and equivalent width limits. For example, in specific windows from the optical to the near-infrared, ground-based NB surveys can select H$\alpha_{\lambda 6563}$ emission-line galaxies up to $z\sim2.6$ \citep[e.g.][]{Bunker1995,Malkan1996,vanderWerf2000,Ly2007,Geach2008,Tadaki2011,Lee2012,Drake2013,Sobral2013,Sobral2015,StroeSobral2015}. The H$\alpha$ recombination-line is a reliable tracer of star-formation rate on $>10$ Myr time-scales \citep{Kennicutt1998}, and is less sensitive to attenuation due to dust than other shorter wavelength tracers \citep[e.g.][]{Garn2010,Ibar2013,Stott2013}. At redshifts $z>2.5$, the most commonly used rest-optical emission-lines are challenging to observe (but see e.g. \citealt{Khostovan2015}), while the rest-frame UV Lyman-$\alpha_{\lambda1216}$ (Ly$\alpha$) line, intrinsically the strongest emission-line emitted in H{\sc ii} regions, is efficiently observed up to $z\sim7$ \citep[e.g.][]{Rhoads2000,Dawson2007,Ouchi2008,Lee2014,Matthee2015,Santos2016}, but is extremely sensitive to resonant scattering and dust attenuation \citep[e.g.][]{Hayes2015}. 

Our High-$z$ Emission Line Survey (HiZELS, \citealt{Geach2008,Best2010,Sobral2013}) has been designed to observe multiple emission-lines in different NBs simultaneously. Hence, \cite{Sobral2012} used observations with NB921 at $\approx 920$ nm and NB$_{\rm H}$ at $\approx1620$ nm to jointly observe  [O{\sc ii}]$_{\lambda\lambda 3726,3729}$ and H$\alpha$ at $z=1.47$. At $z=2.2$, matched NB surveys have observed (combinations of) Ly$\alpha$+[O{\sc ii}]+[O{\sc iii}]$_{\lambda\lambda 4959,5007}$+H$\alpha$ at $z=2.2$ \citep{Lee2012,Nakajima2012,Oteo2015,Matthee2016,Sobral2016}. One of the advantages of this dual (or multiple) NB technique is that identification of the specific emission-line is secure, such that dual-emitters may be used to fine-tune colour selection criteria (in particular in fields where only limited multi-wavelength data are available). Another advantage is that various SFR estimators, for example [O{\sc ii}] or (to lesser extent) Ly$\alpha$, can be calibrated with joint H$\alpha$ observations.

Here we present the first results from Bo\"otes-HiZELS, which is a survey of a central 0.7 deg$^2$ region in the Bo\"otes field with a suite of six narrow-band filters, split into two sets: three red filters at $\approx$921, 1620 and 2120 nm from HiZELS that select rest-optical lines such as H$\alpha$\footnote{We note that narrow-band H$\alpha$ measurements measure the line-flux and EW of the combined H$\alpha$ and [N{\sc ii}] doublet depending on the precise redshift. Therefore, a correction needs to be applied to measured H$\alpha$ EWs and line-fluxes. For simplicity, we refer to H$\alpha$+[N{\sc ii}] emitters as H$\alpha$ emitters from now on.}, H$\beta$/[O{\sc iii}]\footnote{Typical photometric redshifts are not accurate enough to distinguish between H$\beta$ line-emitters and a line-emitter with one of the [O{\sc iii}] lines. Moreover, depending on the specific redshift, we either detect H$\beta$, or one or two of the [O{\sc iii}] lines in the narrow-band filter. The majority of H$\beta$/[O{\sc iii}] emitters are [O{\sc iii}]$_{\lambda 5007}$ emitters, as this line is typically the stronger line; see \cite{Sobral2015} and \cite{Khostovan2015} for details. Yet, to remind the reader of these caveats, we call these emitters H$\beta$/[O{\sc iii}] emitters throughout the paper.} and [O{\sc ii}], complemented by three blue filters at $\approx$ 392, 411 and 501 nm that select Ly$\alpha$ emitters. Using these narrow-band filters, we select samples of emission-line galaxies using their H$\alpha$ line at $z=0.4-2.2$, and from $z=0.8-4.7$ with H$\beta$/[O{\sc iii}], [O{\sc ii}] and Ly$\alpha$. These samples are used as targets for ongoing detailed spectroscopic follow-up studies.

This paper presents the selection and classification of line-emitters, their global properties such as number densities, the number of dual-NB emitters and X-ray detections. We compare our number densities to published luminosity functions for samples in the range $z\approx0.4-4.7$ and we study [O{\sc ii}]-H$\alpha$ emitters at $z=1.47$. These emitters can be used to measure whether the observed [O{\sc ii}]/H$\alpha$ ratio changes with redshift \citep[e.g.][]{Hayashi2012}, which is essential for studies employing [O{\sc ii}] as a SFR indicator at $z>1$ \citep[e.g.][]{Ly2012}. We also use the available deep X-ray coverage to study the X-ray fractions of line-emitters.

We present the observations and archival data used in this survey in \S$\ref{sec:2}$. The data reduction, characteristics and catalog production and selection of emitters are presented in \S$\ref{sec:3}$. \S$\ref{sec:4}$ presents our procedure for classifying emission-line galaxies. We present the number densities of classed line-emitters and compare these to published luminosity functions in \S$\ref{sec:6}$. In \S$\ref{sec:5}$ we investigate the properties of dual-NB line-emitters, the observed [O{\sc ii}]/H$\alpha$ ratio at $z=1.47$ and the X-ray fractions of HAEs and LAEs. Finally, \S$\ref{sec:8}$ presents our conclusions. 

We adopt a $\Lambda$CDM cosmology with $H_0$ = 70 km s$^{-1} $Mpc$^{-1}$, $\Omega_{\rm M} = 0.3$ and $\Omega_{\Lambda} = 0.7$. Magnitudes are in the AB system measured in 3$''$ apertures, unless noted otherwise.

\section{Observations \& data} \label{sec:2}
We observed a 0.7 deg$^2$ region in the Bo\"otes field with six narrow-band filters (NB392, stV, NB501, NB921, NB$_{\rm H}$ and NB$_{\rm K}$). The Bo\"otes field was chosen for the availability of deep multi-wavelength data (see e.g. \citealt{Lee2011,Bian2012,Bian2013,Beare2015}) over a relatively large area, avoiding the galactic plane and its observability from La Palma and Hawaii. In addition to publicly available broad-band (BB) imaging in the $U$, $B$, $R$, $I$, $J$ and $K$ bands (described in \S \ref{sec:2.3}), we also obtain the missing wavelength coverage with three broad-band filters $g$, $z$ and $H$ (see Table $\ref{tab:numbers}$ for an overview).

\subsection{Public/archival multi-wavelength data} \label{sec:2.3}
The Bo\"otes field has been imaged in the optical by the NOAO Deep Wide Field survey\footnote{http://noao.edu/noao/noaodeep/} in $Bw$, $R$ and $I$ \citep{Jannuzi1999} and by the LBT Bootes Field Survey in the $U$ and $Y$ bands \citep{Bian2013}. Near-infrared data in the $J$, $H$ and $Ks$ band are available from the Infrared Bootes Imaging Survey \citep{Gonzalez2010}, although we do not use the $H$ band data as our data are deeper. The general characteristics of the archival data used in this paper are listed in Table $\ref{tab:numbers}$. In addition, the field has been imaged in the X-ray by {\it Chandra} \citep{Murray2005}, in the UV by the {\it GALEX} Deep Imaging Survey \citep{Martin2005}, in the mid-infrared by {\it Spitzer}/IRAC \citep{Ashby2009} and in the far-infrared by {\it Herschel} as part of the {\it Herschel} Multi-tiered Extragalactic Survey \citep{Oliver2012}. However, the X-Ray, UV and mid- and far-infrared data are not explicitly used in the selection of line-emitters. In addition, \cite{Williams2016} presented deep low-frequency (150 MHz) radio observations in this field. Finally, spectroscopic follow-up of mostly X-ray selected sources (and hence AGN) has been performed by \cite{Kochanek2012}.

\begin{table*}
\caption{Description of the available (archival and new) multi-wavelength data in Bo\"otes-HiZELS, with narrow-band filters highlighted in bold. The abbreviations for the archival surveys are: LBFS -- LBT Bo\"otes Field Survey \citep{Bian2013}; NDWFS -- NOAO Deep Wide Field Survey \citep{Jannuzi1999}; IBIS -- Infrared Bootes Imaging Survey \citep{Gonzalez2010}. $\lambda_c$ is the central wavelength of the filter and $\Delta \lambda$ is the width between the full width half maxima of the filter transmission. The full width half maximum (FWHM) of the point spread function has been measured as described in \S$\ref{sec:2.1}$. We list the total exposure time per pixel and its variance. For NB921 and $z$, 60 \% of the coverage has the highest exposure time listed. Depths are measured by measuring the sky value in 100,000 3$''$ apertures on blank regions in the images as described in \S$\ref{sec:3.2.3}$. The coverage is after masking each individual filter for uncovered regions or regions with insufficient depth.}
\begin{tabular}{lrrrrrrp{3cm}rr}
\hline
Filter & Telescope & Survey & $\lambda_c$ & $\Delta \lambda$& FWHM & Exposure time & Dates&  Depth & Coverage  \\ 
 & & & [nm] & [nm] & [$''$] & [ks] & & [3$\sigma$, AB] & [deg$^2$] \\\hline
$U$ & LBT & LBFS & 359 & 54 & 1.2 & &  & 25.3 & 0.78 \\
\bf NB392 & INT & This survey & 392 & 5.2 & 1.8 & 12.4$\pm2.0$ & 2013 Jun 6-10; 2014  Feb 27, Mar 1-8,27 & 24.3 & 0.54 \\
$B_w$ & Mayall & NDWFS & 464 & 110 & 1.4 & & & 25.4 & 0.78\\
\bf stV & INT & This survey & 410 & 16 & 1.9 & 3.8$\pm0.8$ &2013 Jun 6-10; 2014 Mar 2, 5-7; 2016 Jun 11-13 & 24.1 & 0.63 \\
$g$ & INT & This survey & 485 & 129 & 1.6 & 6.0$\pm0.0$ &  2016 Jun 6-8  & 24.9 & 0.78\\
\bf NB501 & INT & This survey & 501 & 10 & 1.6 & 8.4$\pm0.1$ & 2015 Apr 11, 12, 16, 17; 2016 Jun 5, 6, 7, 10, 12 & 24.7 & 0.74\\
$R$ & Mayall& NDWFS & 602 & 160 & 1.1 & & & 25.0 & 0.78 \\
$I$ & Mayall & NDWFS & 754 & 170 & 1.1 & & & 24.4 & 0.78 \\
$z$ & Subaru & This survey & 878 & 113 & 0.8 & 1.0-2.1 & 2014 May 28, 29 & 24.3 & 0.76 \\
\bf NB921 & Subaru & This survey & 919 & 13 & 0.8 &  2.16-2.52 & 2014 May 28, 29 & 24.0 & 0.46 \\
$Y$ & LBT & LBFS & 984 & 42 & 0.8 & & & 23.1 & 0.78 \\
$J$ & NEWFIRM & IBIS & 1300 & 190 & 1.0 & & & 22.9 & 0.78 \\
$H$ & UKIRT & This survey & 1600 & 200 & 0.8 & 1.2$\pm0.1$ &2010 April 2-8 & 22.6 & 0.78 \\
\bf NB$_{\rm H}$ & UKIRT & This survey & 1620 & 21 & 0.8 & 15.4$\pm1.4$ & 2010 April 2-6 & 22.1 & 0.74  \\
$K$ & NEWFIRM & IBIS & 2260 & 280 & 1.2 & & & 22.1 & 0.78 \\
\bf NB$_{\rm K}$ & UKIRT & This survey & 2120 & 21 & 1.2 &  20.2$\pm0.0$ & 2010 April 8, 9, 14, July 18-23; 2011 Feb 16, 27, 28, Mar 9, 12, 16, 20, 22-26 & 22.3 & 0.73 \\
\hline\end{tabular}
\label{tab:numbers}
\end{table*}

\subsection{Optical observations} \label{sec:2.1}
Optical observations in two narrow-band filters (NB392, NB501), a medium band filter (stV) and the $g$ band were performed with the Wide Field Camera (WFC) on the 2.5m Isaac Newton Telescope, part of the Roque de los Muchachos Observatory on the island of La Palma, Spain. WFC has a mosaic of 4 CCDs with a combined field of view of 0.3 deg$^2$ and a 0.33$''$ pixel scale, see Table $\ref{tab:numbers}$. The survey was designed with four pointings, each with a C-NE-NW-SE-SW dither pattern (with 30$''$ offsets). Individual exposure times for narrow and medium-bands were either 0.2 or 1.0ks, depending on whether the telescope could successfully guide on a star (since the auto-guider CCD is behind the filter, this is challenging for narrow-band filters in extra-galactic fields) and the stability of the weather. The individual exposure times for the $g$ band were 0.6ks. 

Observations in the NB921 narrow-band filter and the $z$ filter were performed with Suprime-Cam (S-cam) on the 8.0m Subaru telescope of the National Astronomical Observatory of Japan. S-cam consists of a mosaic of 10 CCDs with a combined field of view of 0.255 deg$^2$ with a 0.2$''$ pixel scale. We imaged the field with the $z$ (NB921) filter with five (three) pointings. For NB921, we used individual 360s exposures dithered either 7 (2 pointings) or 6 (1 pointing) times. For $z$, we used individual 150s exposures of the same pointings as NB921 dithered 14 times, and 3 times 100s in the other two pointings. Observations were done sequentially to avoid contamination of the emission-line sample by variable sources and/or supernova \citep[i.e.][]{Matthee2014}.

\subsection{Near-infrared observations} \label{sec:2.2}
Near-infrared observations in $H$, NB$_{\rm H}$ and NB$_{\rm K}$ were performed with WFCAM on the UK Infrared Telescope (UKIRT) on Mauna Kea as part of the High-$z$ Emission Line Survey (HiZELS, e.g. \citealt{Sobral2013}). WFCAM has a ``paw-print'' configuration of four CCDs, with a total field of view of 0.21 deg$^2$ and a 0.4$''$ pixel scale. The field was imaged with 4 pointings in a dither sequence of 14 exposures with small offsets. Due to the high sky background in the near-infrared, the individual exposure times were 10s, 100s, and 60s for $H$, NB$_{\rm H}$ and NB$_{\rm K}$, respectively, to avoid saturation. In order to obtain the final depth, this dither sequence was typically repeated 9, 11 and 24 times for the respective filters.

\section{Data reduction \& Catalogue production}\label{sec:3}
\subsection{Data reduction} \label{sec:3.1}
\subsubsection{Optical} \label{sec:3.1.1}
We reduce data from the INT/WFC with a custom-made pipeline based on {\sc python} described in detail in \cite{Stroe2014} and \cite{Sobral2016} and we reduce Subaru/S-Cam data similarly with {\sc SDFred2} \citep{Ouchi2004}. In summary, we first bias subtract individual frames using a master bias from the median stack of bias frames for the corresponding night. We then create a master flat by median combining twilight flats and use it to flat-field the individual frames. Subsequently, we measure the FWHM of the PSF using unsaturated stars in individually reduced frames with {\sc SExtractor} \citep{Bertin1996}, and reject frames with PSF FWHM above the chosen target FWHM listed in Table $\ref{tab:numbers}$. This is particularly important for some exposures with the INT/WFC that have been observed in poor conditions. We then match the PSF of remaining frames before combining frames to the common mosaic by smoothing the images with a gaussian kernel.

\subsubsection{Near-infrared} \label{sec:3.1.2}
Near-infrared data from UKIRT/WFCAM have been reduced using PfHiZELS; see \cite{Sobral2009} and \cite{Sobral2013} for full details. The steps are similar to the steps in the optical data reduction, except for dark subtraction instead of bias subtraction and the master flat that is based on an iterative self-flat method using the science frames themselves, instead of relying on twilight flats; see \cite{Sobral2013}.

\subsubsection{Astrometric alignment} \label{sec:3.1.3}
The reduced frames are astrometrically registered to the 2MASS point source catalog \citep{Skrutskie2006} with {\sc Scamp} \citep{Scamp}. Frames are then co-added, resampled to a pixel-scale of 0.33$''$ and mapped to the {\sc mosaic} pointing with {\sc Swarp} \citep{Bertin2010}. We apply the same method for public data described in \S$\ref{sec:2.3}$.

While extracting initial catalogues, we encountered significant astrometric distortions of up to 1.2$''$ in the edges of the cameras of the public $B_w$, $R$ and $I$ data. These distortions significantly affect dual-mode photometry (described below in \S$\ref{sec:3.2.1}$). In order to obtain a more accurate astrometric solution for these data, we used the {\sc Scamp} software to remap the images to the SDSS DR7 astrometry \citep{Abazajian2009}. The astrometric differences between 2MASS and SDSS are minimal and no significant distortions affecting our photometry have been noticed after this correction.

\subsubsection{Photometric calibration} \label{sec:3.1.4}
We set the photometric zero-point ($ZP$) of the images to an arbitrary common $ZP=30$ by matching the {\sc Mag-auto} photometry in the combined images to the following available data: $g$ and $z$ are calibrated to SDSS and $H$ and $K$ to 2MASS. We then use the broad-bands available to calibrate the narrow-bands in the two following steps: we first calibrated NB921 to $z$, NB$_{\rm H}$ and NB$_{\rm K}$ are calibrated to $H$ and $K$, NB501 to $g$ from SDSS, and NB392 and stV to $Bw$. After this first step, we have to correct for the fact that most narrow-band central wavelengths are not in the center wavelength of the broad-band filters, which leads to a bias in line-flux measurements due to gradients in the continuum. This can be resolved by using colour information in adjacent broad-bands. For NB921, NB$_{\rm H}$ and NB$_{\rm K}$ we follow the corrections described in \cite{Sobral2013} and for NB392 we use the corrections from \cite{Matthee2016}. We derive the following correction for stV: stV$_{\rm cor} =$ stV$-0.23(U-B_w)+0.24$. For sources undetected in $U$ and $B_w$ we apply the median correction of $+0.04$. We do not apply a correction for NB501 as it is close to the center of the $g$ band. 

\begin{figure*}
	\includegraphics[width=18cm]{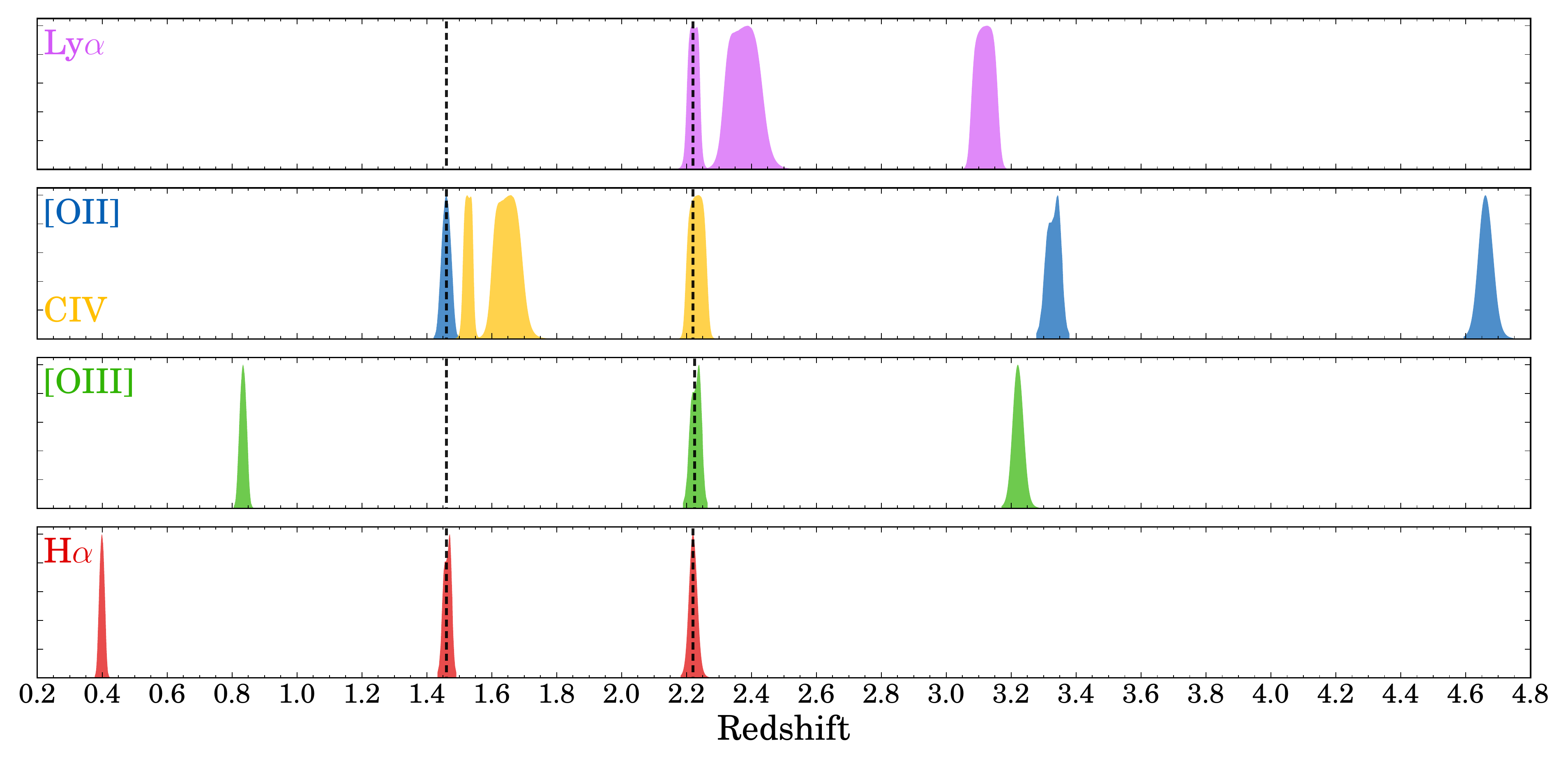}
    \caption{Redshift slices for different emission-lines probed by our Bo\"otes-HiZELS survey. As highlighted by the dashed lines, there is joint coverage of [O{\sc ii}] and H$\alpha$ at $z=1.47$ and of Ly$\alpha$, [O{\sc iii}] and H$\alpha$ at $z=2.22$. By coincidence, there is also matched volume coverage of C{\sc iv} at this redshift. }
    \label{fig:Redshifts}
\end{figure*}

\subsection{Catalogue production} \label{sec:3.2}
\subsubsection{Photometry} \label{sec:3.2.1}
Photometry of the optical-NIR filters listed in Table $\ref{tab:numbers}$ is performed with {\sc SExtractor} in dual-image mode. We create six catalogues, each with one of the six narrow-bands as detection image. Photometry is measured within circular apertures with a diameter of 3$''$. For each narrow-band we measure the narrow-band and the corresponding broadband magnitudes from images with their PSF matched to the narrow-band imaging. We also measure the magnitudes in all broad-bands with their PSF matched to the $g$ band PSF (1.6$''$ FWHM). The measurements with the PSF from the NB are used to select line-emitters and compute emission-line properties such as line-flux and equivalent width. The other measurements are used for colour-colour selections. 

We have produced a mask for each narrow-band individually, where we mask regions around bright, saturated stars, CCD bleeding, cross-talk in near-infrared detectors and regions with low S/N or incomplete coverage \citep[e.g.][]{Sobral2009,Santos2016}.

\subsubsection{Depths} \label{sec:3.2.3}
We estimate the depth of images by measuring the standard deviation of the total counts in 100,000 apertures with a diameter of 3$''$ placed at random (but avoiding sources) locations in our images. 3$\sigma$ depths range from $\sim25$ AB magnitude in blue broad-band filters to $\sim22$ AB magnitude in the near-infrared filters, see Table $\ref{tab:numbers}$.

\subsection{Selecting Line-emitters} \label{sec:3.3}
\begin{figure*}
\begin{tabular}{cc}
	\includegraphics[width=8.5cm]{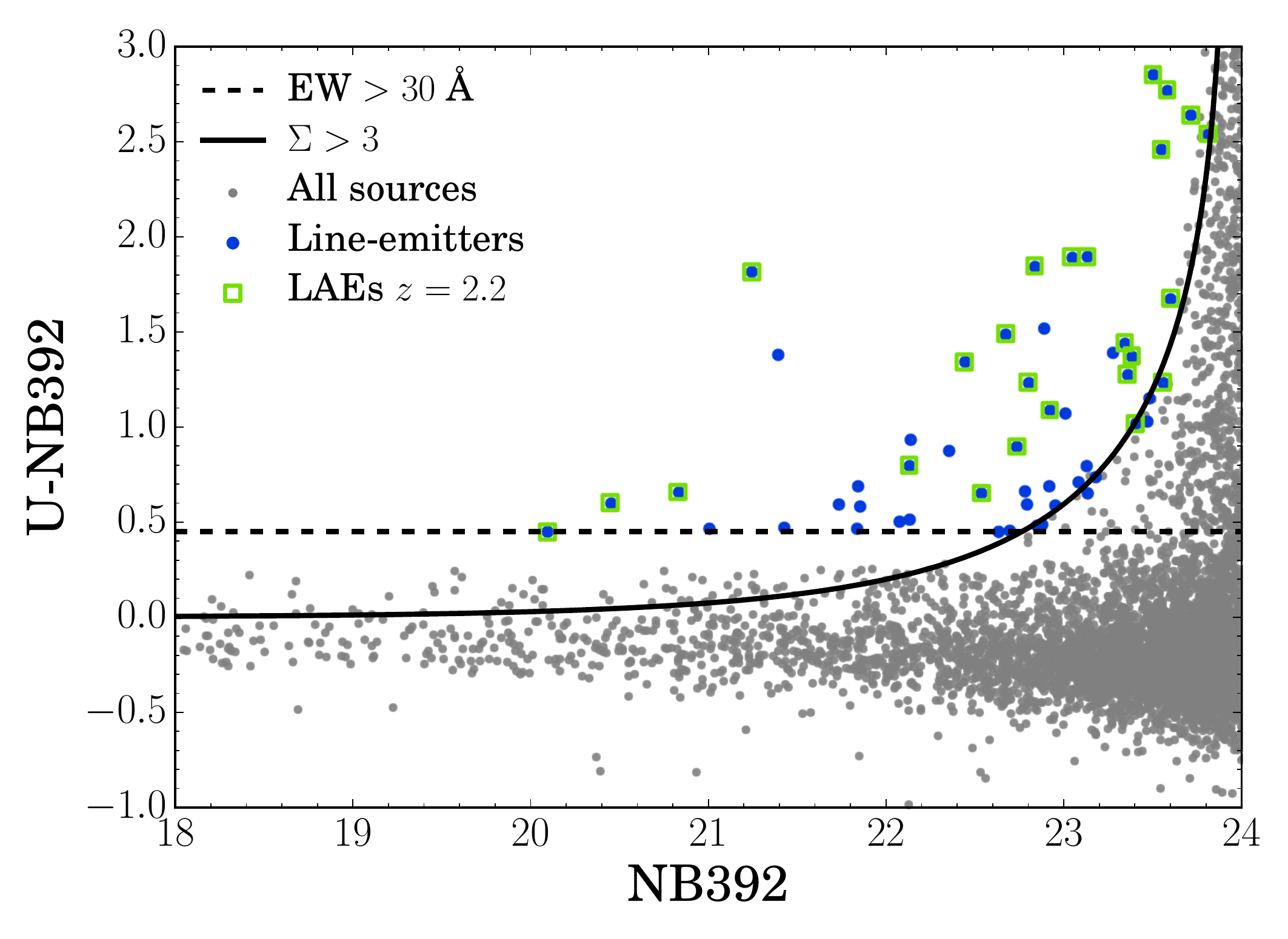}
	\includegraphics[width=8.5cm]{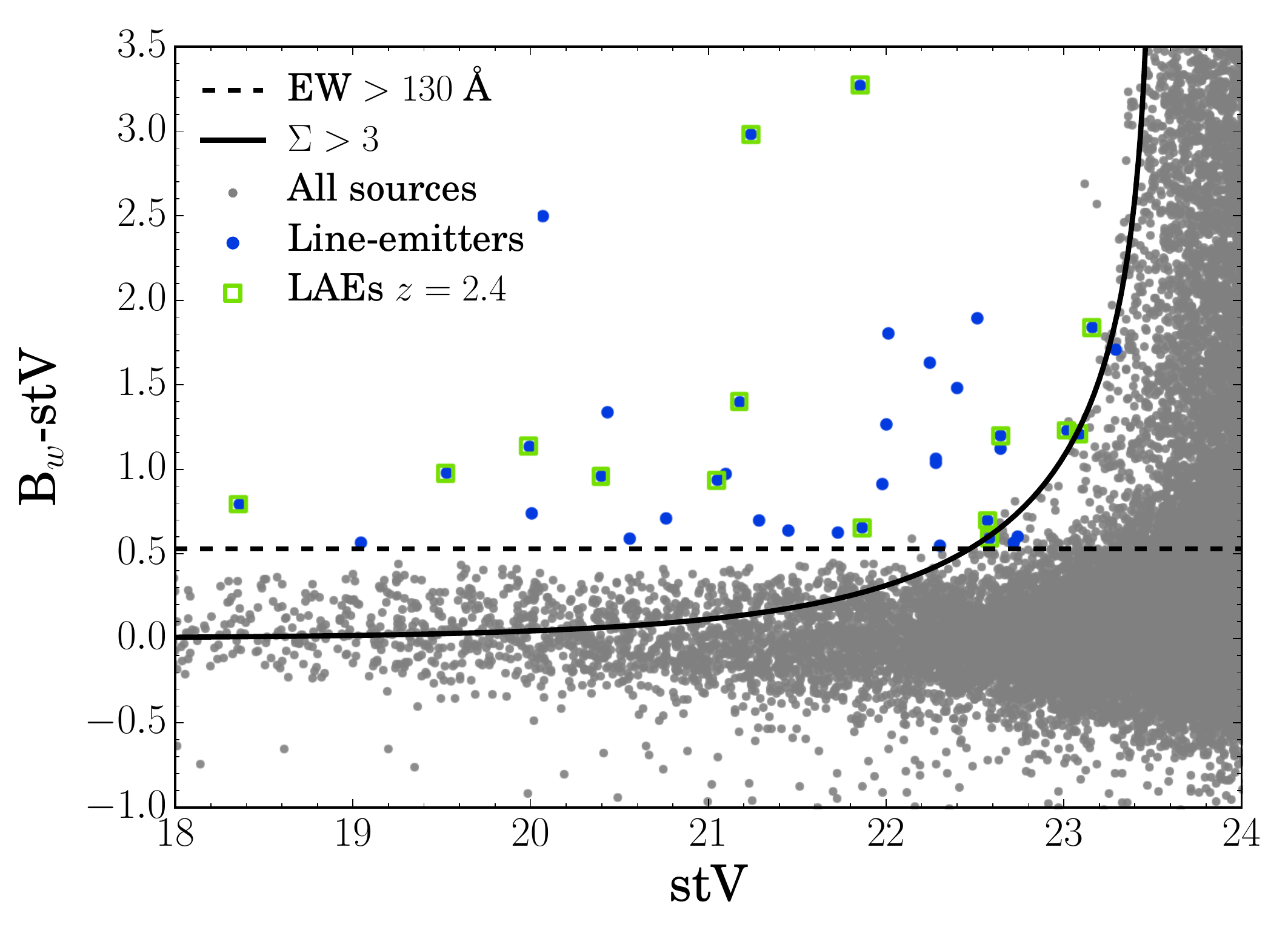}\\
	\includegraphics[width=8.5cm]{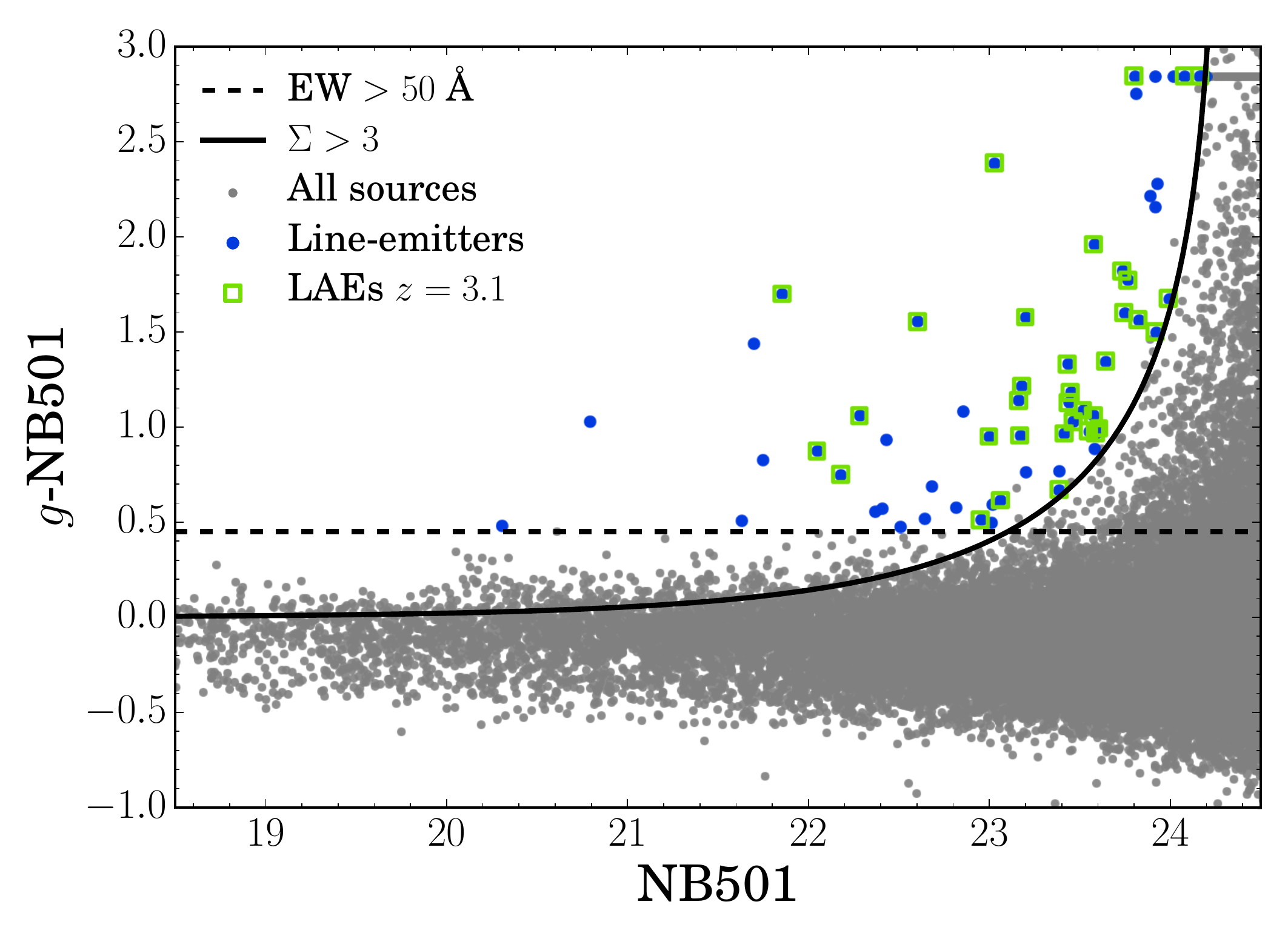}
	\includegraphics[width=8.5cm]{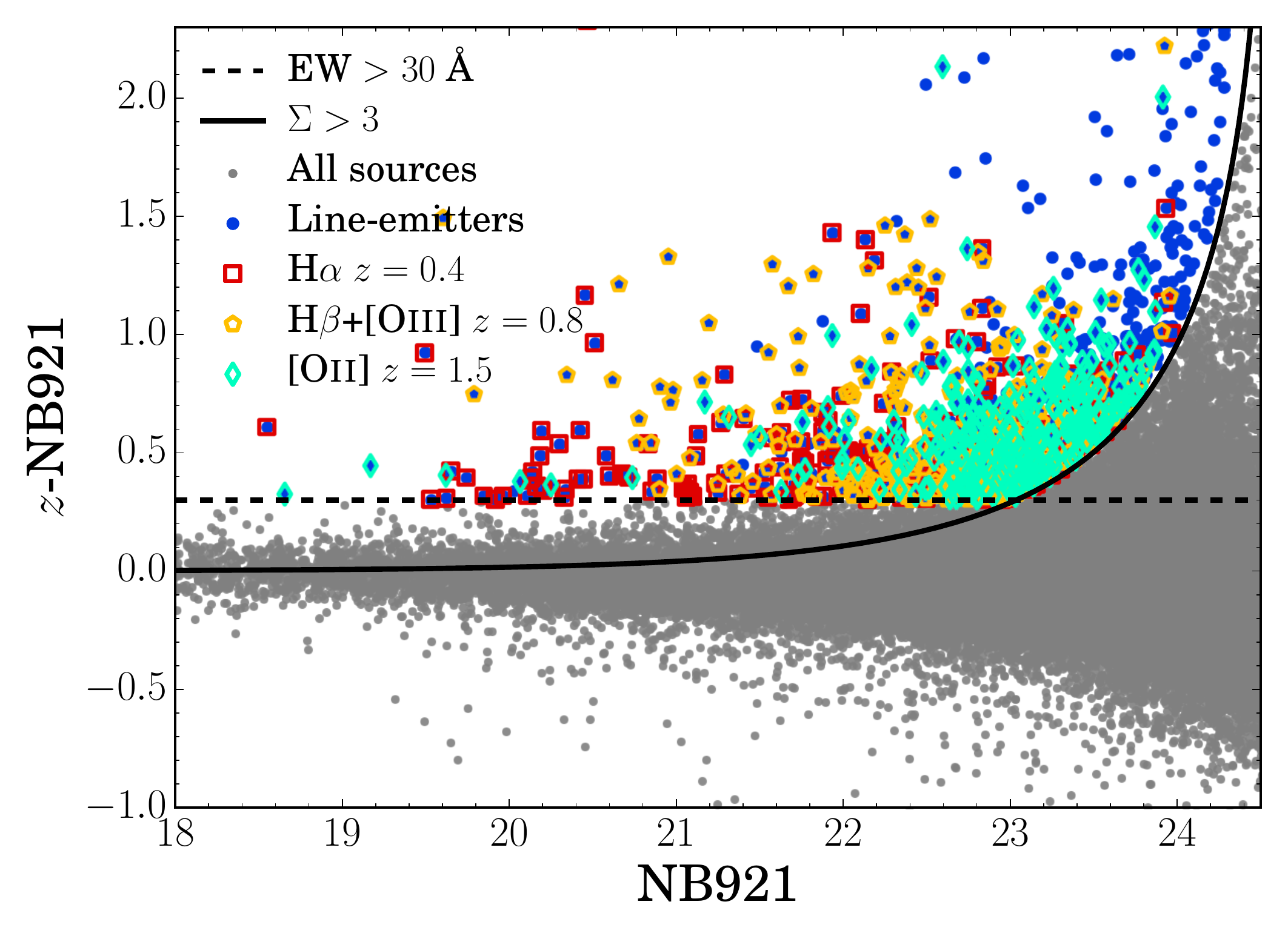}\\
	\includegraphics[width=8.5cm]{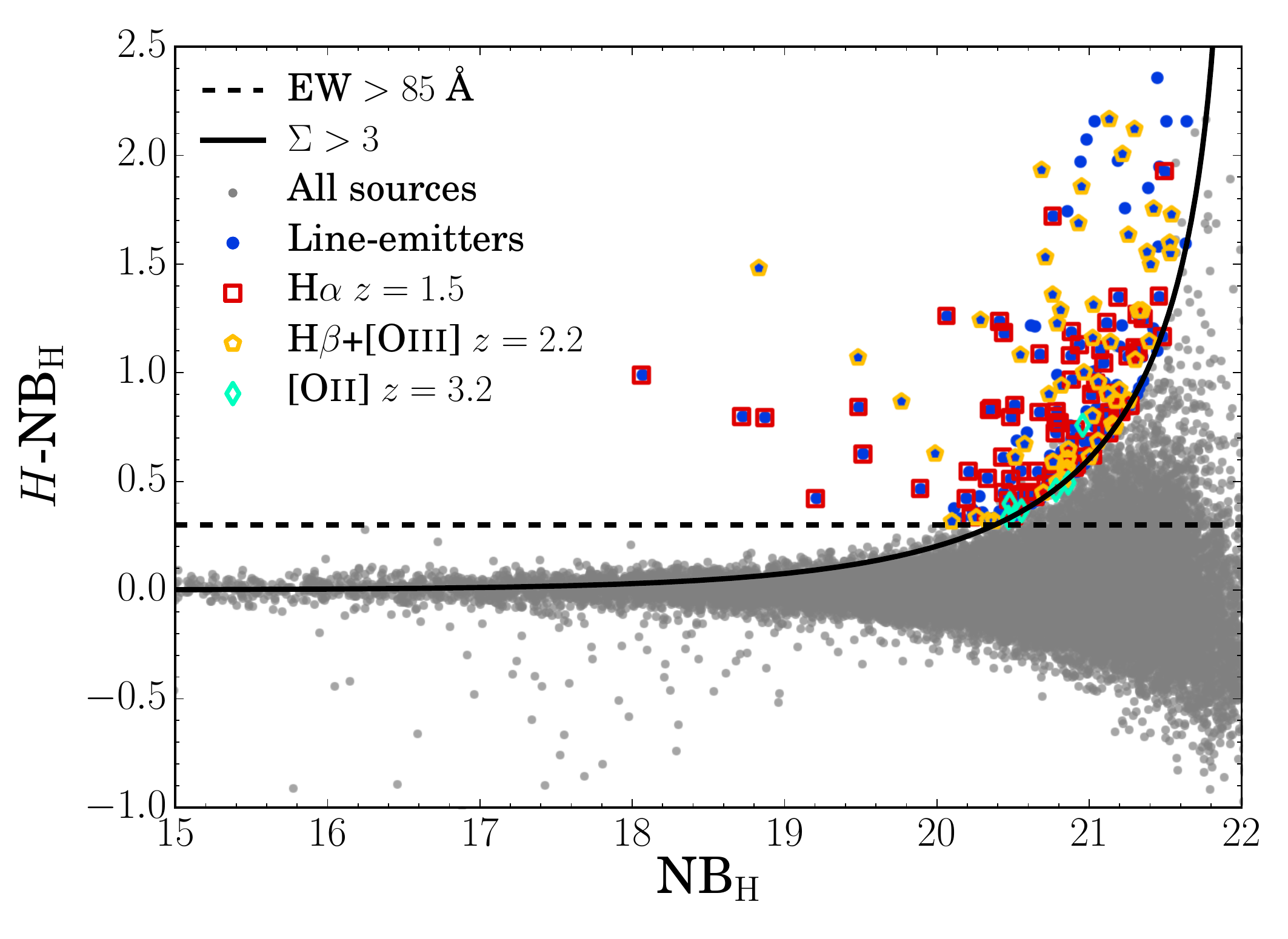}
	\includegraphics[width=8.5cm]{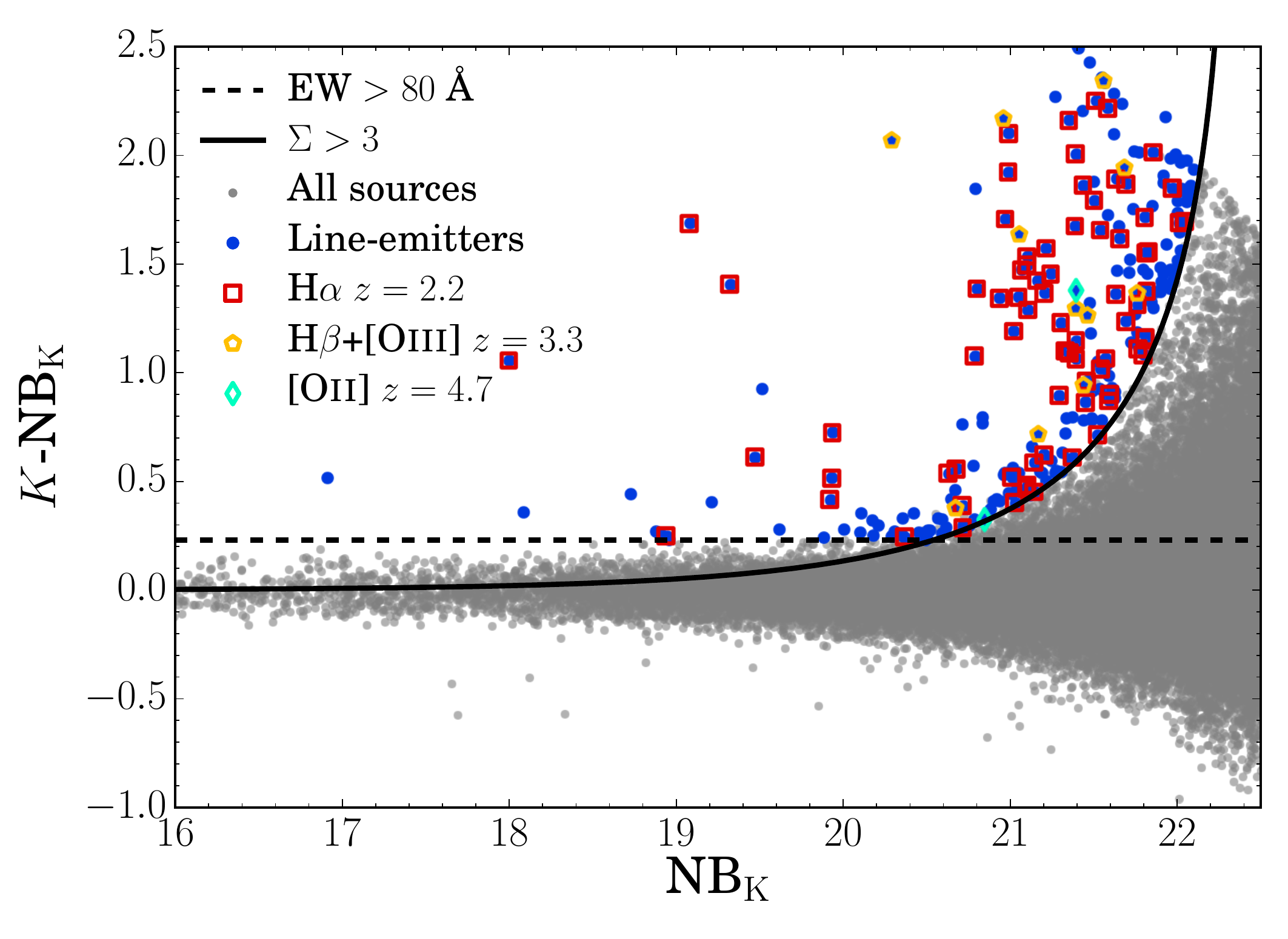}\\

	\end{tabular}
    \caption{Narrow-band excess diagrams in NB392, stV, NB501,NB921, NB$_{\rm H}$ and NB$_{\rm K}$. In grey we show all detected sources, while blue points are sources selected as line-emitters. The horizontal dashed line shows the imposed EW selection cut, while the solid line shows the excess significance criteria for the typical depth of the survey. In the three blue filters we mark Ly$\alpha$ selected sources with a green square. In the three red filters we mark H$\alpha$ emitters with a red square, H$\beta$/[O{\sc iii}] emitters with a yellow pentagon and [O{\sc ii}] emitters with a blue diamond.We note that we compute the excess significance locally, such that some sources may lie above the selection line, but are not selected as line-emitters because they are in shallower regions. It can be seen that LAEs detected in NB392 and NB501 are typically identified if a line-emitter has a high excess and faint magnitude (because most are likely faint star-forming galaxies with high EW), while this is not the case for LAEs identified in stV (which are typically bright AGN). In NB921 it can clearly be seen that most unidentified line-emitters are among the faintest magnitudes, and that there is a clear trend that higher redshift line-emitters are fainter and have higher observed EWs. In NB$_{\rm H}$ and NB$_{\rm K}$ H$\alpha$ emitters and H$\beta$/[O{\sc iii}] emitters have similar narrow-band magnitudes, but H$\beta$/[O{\sc iii}] emitters tend to have higher excess because they are at higher redshift.}
    \label{fig:excess}
\end{figure*}

Line-emitters are selected based on two criteria: the narrow-band excess (the equivalent width, EW) must be high enough and the excess must be significant. For the narrow-band filters NB392, stV, NB501, NB921, NB$_{\rm H}$ and NB$_{\rm K}$ we use the corresponding broad-band filters $U$, $B_w$, $g$, $z$, $H$ and $K$, see Fig. $\ref{fig:excess}$. In order to convert the photometric narrow-band excess to observed EW, we convert magnitudes ($m_i$) to flux densities in each filter ($f_i$) with the standard AB magnitude convention:

\begin{equation}
f_i = \frac{c}{\lambda^2_{i, \rm center}} 10^{-0.4(m_i +48.6)},
\end{equation}
where $c$ is the speed of light and $\lambda_{i, \rm center}$ is the central wavelength in each filter. Next, we use the following equations to convert the narrow-band and their corresponding broad-bands to EW:
\begin{equation}
\label{eq2}
EW_{\rm obs} = \Delta\lambda_{\rm NB} \frac{f_{\rm NB}-f_{\rm BB}}{f_{\rm BB}-f_{\rm NB} \frac{\Delta\lambda_{\rm NB}}{\Delta\lambda_{\rm BB}}}.
\end{equation}

Here, $f_{\rm NB}$ and $f_{\rm BB}$ are the flux-densities in the narrow-band and broad-band and $\Delta\lambda_{\rm NB}$ and $\Delta\lambda_{\rm BB}$ the filter-widths. In Eq. $\ref{eq2}$, the numerator is the difference in narrow-band and broad-band flux and the denominator is the continuum level, corrected for the contribution from the flux in the narrow-band. For sources without broad-band detection we set the EW to a lower limit. This lower limit ranges from 550 {\AA} for NB392 and NB501 to 2500 {\AA} for stV depending on the width of the NB filter and the depth of the BB data used to measure the continuum. The lower limit is around $\approx 1200$ {\AA} for near-infrared NBs. 

The excess significance ($\Sigma$) quantifies whether a certain narrow-band excess is due to errors in the narrow-band and broad-band photometry or not. Hence, we follow the methodology presented in \cite{Bunker1995} and the equation from \cite{Sobral2013} to compute $\Sigma$:

\begin{equation}
\Sigma = \frac{1-10^{-0.4(BB-NB)}}{10^{-0.4(ZP-NB)}\sqrt{(\sigma^2_{\rm box,BB}+\sigma^2_{\rm box,NB})}},
\end{equation}
where $BB$ is the broadband magnitude used for the continuum estimate, $NB$ is the narrow-band magnitude and $ZP$ is the zero-point of the images. $\sigma_{\rm box}$ is the root mean squared (rms) of background aperture values in the data of the respective filters (see \S$\ref{sec:3.2.3}$). 

The line-flux is computed using:
\begin{equation}
f_{\rm line}= \Delta\lambda_{\rm NB} \frac{f_{\rm NB}-f_{\rm BB}}{1-\frac{\Delta\lambda_{\rm NB}}{\Delta\lambda_{\rm BB}}}.
\end{equation}

We select line-emitters among narrow-band selected sources in all six narrow-band filters with the criterion that $\Sigma>3$. However, because each narrow-band has different filter characteristics, we do not apply a homogeneous excess (EW) selection threshold. For each filter, we apply the criterion that the observed EW is three times the standard scatter in observed EWs for sources detected at $>15 \sigma$. This means that we apply EW$>30, 130, 50, 30, 85, 80$ {\AA} for NB392, stV, NB501, NB921, NB$_{\rm H}$ and NB$_{\rm K}$ respectively. 

Before obtaining our final list of line-emitters, in each filter, we visually inspect all the sources in the narrow-band images for remaining spurious sources such as artefacts from bright stars, cosmic rays or mis-identifications by SExtractor. This can happen when the noise properties vary strongly locally, which is the case in small regions of the coverage by the NB392, stV and NB921 filters.

\section{Classifying line-emitters}\label{sec:4} 
Fig. $\ref{fig:Redshifts}$ shows the redshift ranges where our narrow-band filters sample the brightest emission-lines seen in normal star-forming galaxies and AGN. By a combination of design and coincidence, the HiZELS narrow-band filters coincide with several different emission-lines at specific redshifts. At $z=1.47$, the NB921/NB$_{\rm H}$ combination is sensitive to the [O{\sc ii}] and H$\alpha$ lines \citep{Sobral2012}. At $z=2.23,$ the Ly$\alpha$, [O{\sc iii}] and H$\alpha$ lines fall in the NB392/NB$_{\rm H}$/NB$_{\rm K}$ combination\footnote{In this case, it is certain that the emission-line in NB$_{\rm H}$ is [O{\sc iii}] and not H$\beta$.}. At $z=2.23$, the NB501 filter is also sensitive to C{\sc iv} emission. The redshifts of line-emitters detected in several narrow-bands (dual-emitters) can be estimated accurately and we refer to them as $z_{\rm dual-NB}$ in the remainder of this paper. As can be seen in Table $\ref{tab:dualemitters}$, dual-emitters are found as faint as $I \approx 25$, three magnitudes fainter than typical available spectroscopic redshifts, which were mostly for X-ray selected sources. 

Line-emitters that have no existing spectrosopic redshift or are not detected as dual-emitters are classified using colour-colour selections tuned to identify Lyman- and Balmer-breaks at various redshift intervals. For the blue narrow-bands, we use colour selections to identify Ly$\alpha$-emitters. For the red filters we use colour selections to identify H$\alpha$ emitters, H$\beta$/[O{\sc iii}] emitters and [O{\sc ii}] emitters (see also similar selections in \citealt{Sobral2013,Khostovan2015}). 

The main strategy to devise colour-criteria has been as follows: after removing stars (due to atmospheric features in the blue or the near-infrared, stars may be picked up with a narrow-band excess) using their $uJK$ colours \citep[e.g.][]{UVISTACAT}, we start with colour selections from the literature, which we slightly modify using the spectroscopically confirmed line-emitters and the dual-emitters. Colour-criteria are stated explicitly below and listed in Table $\ref{tab:criteria}$. We summarise the number of emitters and classified line-emitters in Table $\ref{tab:line-identifications}$.

\begin{table}
\centering
\caption{Spectroscopically and dual-NB confirmed emission-lines observed in narrow-band filters. We note that spectroscopic redshifts are highly biased towards AGN, and that the spectroscopic redshift distribution does not resemble the real redshift distribution, particularly for fainter line-emitters. Dual-NB redshifts are only available at $z=1.47$ and $z=2.23$, see Fig. $\ref{fig:Redshifts}$.}
\begin{tabular}{llrrr}
\hline
Filter & Emission-line & Redshift & \# $z_{\rm spec}$ & \# $z_{\rm dual NB}$  \\ \hline 
NB392 & Mg{\sc ii}$_{\lambda2798}$ & 0.39--0.41 & 8 & - \\
&C{\sc iii]}$_{\lambda1909}$ & 1.04--1.07 & 1 & - \\
& C{\sc iv}$_{\lambda1549}$ & 1.51--1.55 & 2 & - \\
& Ly$\alpha_{\lambda1216}$ & 2.20--2.24 & 2 & 5 \\ \hline 

stV & Mg{\sc ii}$_{\lambda2798}$ & 0.42--0.50 & 1 & - \\
&C{\sc iii]}$_{\lambda1909}$ & 1.10--1.20 & 3 & - \\
& He{\sc ii}$_{\lambda1640}$ &1.44--1.56 & 1 & - \\ 
& C{\sc iv}$_{\lambda1549}$ & 1.59--1.71 & 4 & - \\ 
& Ly$\alpha_{\lambda1216}$ & 2.30--2.45 & 6 & - \\ \hline 

NB501 & {\sc [Oii]$_{\lambda3727}$} & 0.32--0.36 & 1 & -  \\ 
& Mg{\sc ii}$_{\lambda2798}$ & 0.76--0.81 & 1 & - \\
& C{\sc iv}$_{\lambda1549}$ & 2.19--2.27 & 1 & 3 \\
& Ly$\alpha_{\lambda1216}$ & 3.06--3.17 & 5 & - \\ \hline

NB921 & H$\alpha_{\lambda6563}$ & 0.39--0.41 & 5 & - \\
 & {\sc [Oiii]$_{\lambda\lambda 4959,5007}$} & 0.82--0.87 & 1 & - \\  
&  {\sc [Oii]$_{\lambda3727}$} & 1.44--1.48 & 0 & 20 \\ 
& Mg{\sc ii}$_{\lambda2798}$ & 2.25--2.31 & 1 & 3 \\\hline

NB$_{\rm H}$ & H$\alpha_{\lambda 6563}$ & 1.44--1.48 & 6 & 21 \\
 & {\sc [Oiii]$_{\lambda\lambda 4959,5007}$} & 2.19--2.29 & 3 &16  \\ \hline 

NB$_{\rm K}$ & H$\alpha_{\lambda6563}$ & 2.21--2.25 & 8 & 20 \\ 
\hline\end{tabular}
\label{tab:redshifts}
\end{table} 

\begin{table*}
\caption{ Colour selection-criteria used to classify line-emitters. These criteria are based on the expected positions of the Lyman- and Balmer-breaks, see for example the $BzK$ criterion from \citet{Daddi2004}. These criteria are then fine-tuned using spectroscopic and dual-NB redshifts. In comparison to traditional criteria that use the $B$ band, we adjust the $B_w$ magnitude for the contribution from flux in the $U$ band. For samples at $z>2.4$, we also include a colour-criterion that removes very red objects for which the Lyman-break criterion has selected a strong Balmer break; see also \citet{Xue2016}. The colour-colour selections are illustrated in the figures in Appendix $\ref{sec:colsel}$. }
\begin{tabular}{llr}
\hline 
Filter & Emission-line & Colour-criterion \\ \hline
NB392 & Ly$\alpha$ $z=2.2$ & $ (2B_w-U)-z < 0.2 + 0.7 (z-K)$ \\ \hline 
stV & Ly$\alpha$ $z=2.4$ & $U-B_w > 0.3 (B_w-g)+0.2 \,\, \, \& \,\,\,  g-I < 1$ \\ \hline
NB501 & Ly$\alpha$ $z=3.1$ & $U-g > 1  \,\,\,\,\&\,\,\,\,  g-I < 1.5$ \\ \hline

NB921 & H$\alpha$ $z=0.40$ & $(2B_w-U)-I > 0.4+ 0.4 (Z-H) \,\&\, B_w-R > 1.3 (R-I) $\\ 
 & {\sc [Oiii]}/H$\beta$ $z=0.80$ &  $(2B_w-U)-I > 0.4+ 0.4 (Z-H) \,\&\, B_w-R < 1.3 (R-I) $ \\
 & {\sc [Oii]} $z=1.47$ & $(2B_w-U)-I < 0.4+ 0.4 (Z-H)$ \\ \hline
 
NB$_{\rm H}$ & $z>1$ & $(2B_w-U)-z < 0.4+0.8(z-K)$ or $z-K>2$ \\
& H$\alpha$ $z=1.47$ & $z>1$ \& $J-K < 2.1 (I-J) - 1$ \\ 
 & {\sc [Oiii]}/H$\beta$ $z=2.23$ & $z>1$ \& $J-K > 2.1 (I-J) - 1$ \\
 & {\sc [Oii]} $z=3.3$ & Not H$\alpha$ or {\sc [Oiii]}/H$\beta$ in NB$_{\rm H}$ \& $U-g > 1 \,\&\, g-I < 1.5$ \\ \hline

NB$_{\rm K}$ & $z>1$ & $(2B_w-U)-z < (z-K)-0.05$ or $z-K>2$ \\
& H$\alpha$ $z=2.23$ & $z>1$ \& $U-R<2$ \\ 
 & {\sc [Oiii]}/H$\beta$ $z=3.2$ & $z>1$ \& $U-g > 1 \,\&\, g-I < 1.5$ \\
 & {\sc [Oii]} $z=4.7$ & Not H$\alpha$ or {\sc [Oiii]}/H$\beta$  in NB$_{\rm K}$ \& $g-I > 1.5$ \\ \hline

\end{tabular}
\label{tab:criteria}

\end{table*}

\subsection{Line-emitters in NB392}\label{sec:4.1}
The narrow-band NB392 has specifically been designed to conduct a Ly$\alpha$ survey with a matched volume coverage to H$\alpha$ emitters identified with the HiZELS NB$_{\rm K}$ filter (H$_2$S1) at $z=2.23$ in order to study the Ly$\alpha$ escape fraction and its dependencies on galaxy properties, as described in detail in \cite{Matthee2016} and \cite{Sobral2016}.

We select 57 line-emitters with an excess criterion of EW$_{\rm obs} > 30 $ {\AA} ($U$-NB392 $>0.45$). For LAEs at $z=2.23$ this corresponds to EW$_0 > 9 $ {\AA} (it is possible to go to such low EWs because the width of NB392 is very narrow). Although Ly$\alpha$ surveys at $z\approx2-3$ typically invoke a higher EW criterion of $\sim25-30$ {\AA} \citep[e.g.][]{Ouchi2008,Nakajima2012}, we found that such a selection results in missing the most luminous LAEs at $z=2.2$ in the COSMOS and UDS fields \citep{Sobral2016}. This is because these sources are typically AGN, which have bright Ly$\alpha$ emission on top of a bright UV continuum. 

Using the spectroscopy available from AGES \citep{Kochanek2012}, we find 8 Mg{\sc ii} emitters at $z=0.4$ and 5 line-emitters at $z>1$ (including two LAEs at $z=2.2$), see Table $\ref{tab:redshifts}$. By matching the sample of line-emitters with the samples of line-emitters in NB$_{\rm H}$ and NB$_{\rm K}$ (see below), we add four other robust LAEs at $z=2.2$. Other line-emitters are classed using the criteria described in Table $\ref{tab:criteria}$. We spectroscopically identify four interlopers ($\approx15\pm7$ \% contamination, similar to the $10\pm4$ \% from \citealt{Sobral2016}). These comprise two C{\sc iv} emitters at $z=1.53$ and two AGN for which we measure Lyman-Werner and Lyman-Continuum radiation in the NB392 filter at $z=3.16$ and $z=3.57$. We also identify two dual-emitters that are missed by the colour-colour selection (see Fig. $\ref{fig:colsel_LAEz2}$). This results in a final sample of 25 LAEs at $z=2.2$.

\subsection{Line-emitters in stV}\label{sec:4.2} 
The stV medium-band filter is used to identify LAEs at $z\approx2.4$. Because the width of the filter is relatively broad, it is sensitive to line-emitters over a larger redshift space (and thus covers a larger volume), at the cost of being only sensitive to lines with high EW. We apply a selection criterion of EW$_{\rm obs} > 130$ {\AA} (which corresponds to $B_w - $stV$>0.54$). We note that due to the width of the filter, it is possible that multiple lines contribute to the observed EW and line-flux, such as the combination of Ly$\alpha$+N{\sc v}.\footnote{For example, for Type I AGN, N{\sc v}/Ly$\alpha$ is typically $\approx3$ \% \citep[e.g.][]{VandenBerk2001}, while for type II AGN (such as narrow-line Seyferts) N{\sc v}/Ly$\alpha$ can be as high as 50 \% (typically $\approx 20$ \%, e.g. \citealt{Alexandroff2013}).} It is therefore not straightforward to interpret measured EWs and line-fluxes and caution must be taken.

We find a total of 39 line-emitters, of which 15 have spectroscopic redshifts, see e.g. Table $\ref{tab:redshifts}$. As expected, these are dominated by LAEs at $z=2.3-2.45$, but also contains high-ionization lines as C{\sc iii}] and C{\sc iv} at $z\approx1.1-1.7$. After removal of one spectroscopic contaminant (a C{\sc iv} emitter at $z=1.613$) selected with the colour-criteria described in Table $\ref{tab:criteria}$ and illustrated in Fig. $\ref{fig:colsel_LAEz2}$), we obtain a sample of 16 LAEs. 

\subsection{Line-emitters in NB501} \label{sec:4.3}
The NB501 filter is used to select Ly$\alpha$ emitters at $z=3.1$. We apply EW$_{\rm obs} > 50$ {\AA} ($g-$NB501$ > 0.45$), corresponding to a Ly$\alpha$ rest-frame EW of $>12$ {\AA}.

We find a total of 65 line-emitters, of which only four have an archival spectroscopic redshift. This is because the majority of these line-emitters are faint with line-fluxes below $2\times10^{-16}$ erg s$^{-1}$ cm$^{-2}$. One spectroscopic confirmed line-emitter is a LAE, two are C{\sc iv} emitters at $z=2.24$ and $z=2.26$ (these are the dual-emitters B-HiZELS\_3 and B-HiZELS\_15, Table $\ref{tab:dualemitters}$, also detected as line-emitters in several other bands) and one is possibly [Ne{\sc v}] at $z=0.426$. 

LAEs are selected as relatively blue $U$ drop-outs \citep[e.g.][]{Hildebrandt2009} as described in Table $\ref{tab:criteria}$ and illustrated in Fig. $\ref{fig:colsel_LAEz3}$). One C{\sc iv} emitter at $z=2.24$ is mis-classed as a LAE and is removed from the sample. This leads to a sample of 31 LAEs. We note that our EW criterion of $EW_0 > 12 $ {\AA} is somewhat lower than the typical criterion used for selections of LAEs (EW$_{0} > 25 ${\AA}, e.g. \citealt{Ouchi2008,Yamada2012}). However, more than 90 \% of the identified LAEs have EW$_{0} > 25 ${\AA}. Contrary to the properties of LAEs at $z\sim2$, the additional LAEs with low EW are all faint in their UV continuum. This indicates an evolution in the properties of luminous LAEs from $z=2-3$, with an increasing Ly$\alpha$ EW$_0$ with redshift at fixed Ly$\alpha$ luminosity. 
Very recently, four additional LAEs from this sample have been confirmed at $z=3.1$ from our spectroscopic follow-up campaign (to be presented in Sobral et al. in prep), including the brightest LAE in our sample with a Ly$\alpha$ luminosity of $\approx10^{43.8}$ erg s$^{-1}$ ($\sim10\times L^{\star}$ at $z=3.1$, \citealt{Ouchi2008}) in a 3$''$ aperture and an EW$_0$ of $\sim150$ {\AA}. To the current surface brightness limit, it is extended over $\sim5''$ (40 kpc), and it may thus be classed a Ly$\alpha$ blob \citep[e.g.][]{Matsuda2004,Prescott2008,Dey2016}. This follow-up spectroscopy also identifies two interlopers: a red {\sc [Oii]} emitter at $z=0.35$ and a Mg{\sc ii} emitter at $z=0.81$, that we have removed from the sample.

\subsection{Line-emitters in NB921} \label{sec:4.4}
While the NB921 filter has been used to select LAEs at $z=6.6$ \citep[e.g.][]{Matthee2015}, it is also used to select H$\alpha$, H$\beta$/{\sc [Oiii]} and {\sc [Oii]} emitters at lower redshift \citep[e.g.][]{Ly2007,Drake2013,Sobral2013,Khostovan2015}. We select 1161 line-emitters with the excess criterion of EW$>30$ {\AA} (corresponding to $z-$NB921$>0.3$). 

Since our sample of line-emitters is selected from relatively deep narrow-band imaging (compared to the other narrow-bands in this survey), it is dominated by sources with fluxes fainter than $2\times10^{-16}$ erg s$^{-1}$ cm$^{-2}$ ($>94$ \% of line-emitters), down to fluxes of $2\times10^{-17}$ erg s$^{-1}$ cm$^{-2}$. Because of this, the number of spectroscopic redshifts is limited to only seven, of which five are H$\alpha$ emitters at $z=0.4$. However, the number of galaxies with a robust redshift due to emission-lines in multiple narrow-bands is significantly higher (23), see Table $\ref{tab:redshifts}$.

Among our sample of line-emitters, we use colour-criteria (Fig. $\ref{fig:colsel_NB921}$; based on \citealt{Sobral2013}) to identify H$\alpha$ emitters at $z=0.40$, H$\beta$/{\sc [Oiii]} emitters at $z=0.83$ and {\sc [Oii]} emitters at $z=1.47$, see Table $\ref{tab:criteria}$. We select 198 H$\alpha$ emitters, 304 H$\beta$/{\sc [Oiii]} emitters and 277 {\sc [Oii]} emitters (see Table $\ref{tab:line-identifications}$). Two Mg{\sc ii} emitters at $z=2.26$ are mis-identified as {\sc [Oii]} emitter, while one dual-emitter at $z=1.47$ is mis-identified as H$\beta$/{\sc [Oiii]} emitters.

\begin{figure*}
\begin{tabular}{cc}
	\includegraphics[width=8.8cm]{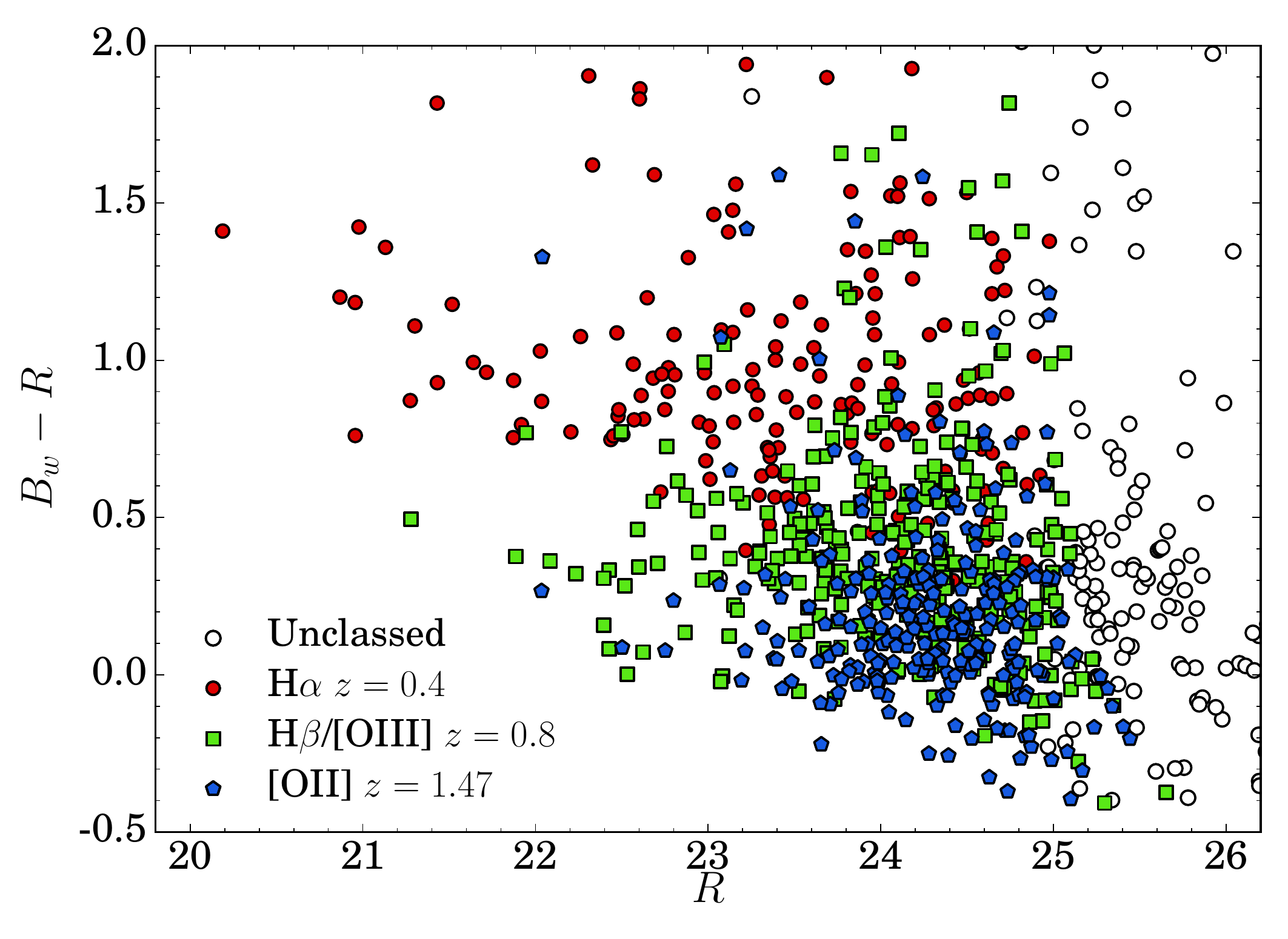} 
	\includegraphics[width=8.8cm]{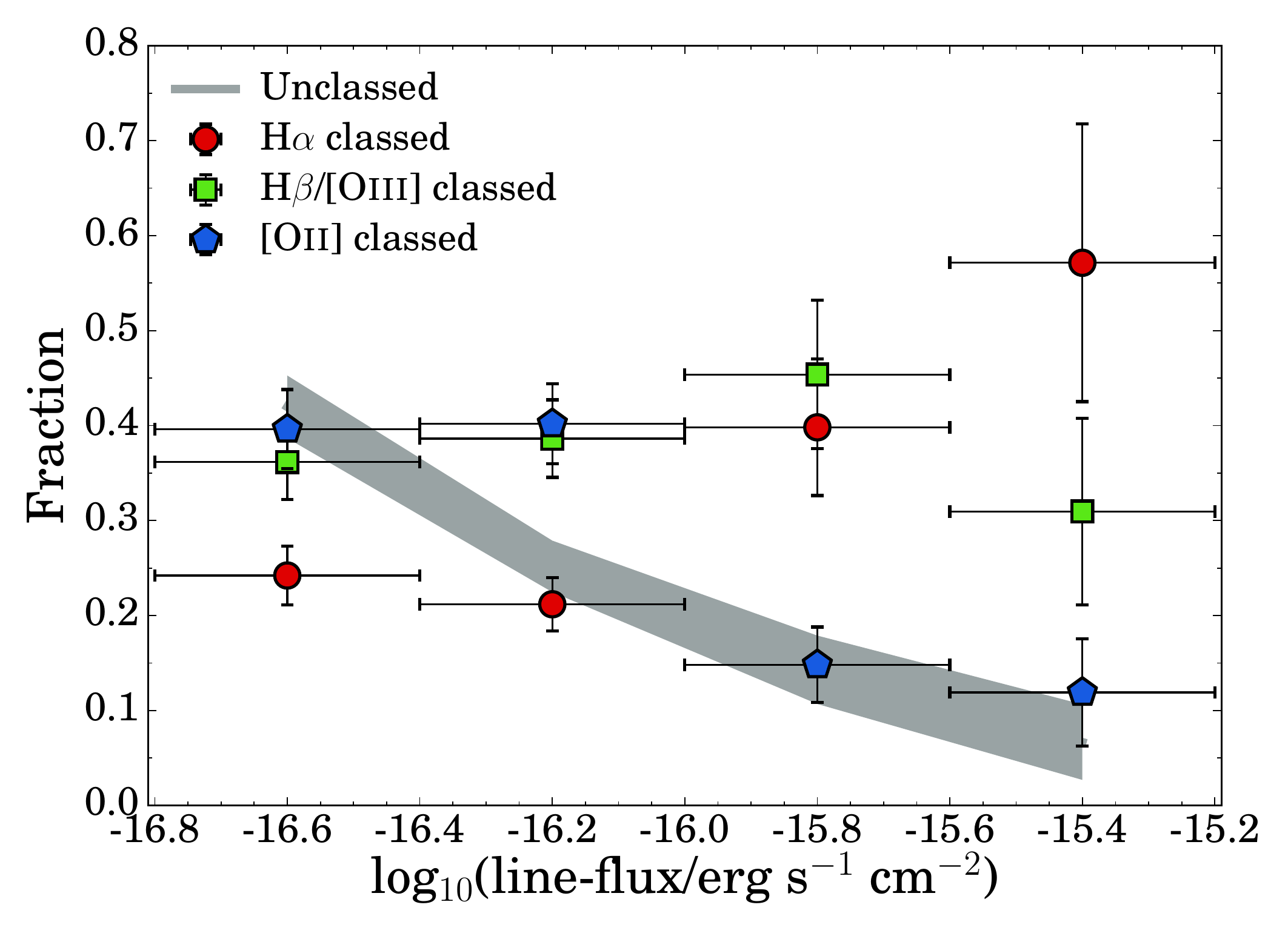} 
\end{tabular}
    \caption{Left: $B_w-R$ versus $R$ colour-magnitude diagram of line-emitters in NB921. This plot illustrates that the line-emitters that are not detected in enough broad-bands (such as $I$, $z$ or $H$) in order to be classified, are typically faint ($R>25.5$, note that these are low S/N detections) and lie most closely to the colour-magnitude parameter space probed filled with [O{\sc ii}] emitters. Right: The fraction of sources classed as H$\alpha$, H$\beta$/[O{\sc iii}] and [O{\sc ii}] among the classed NB921 line-emitters in bins of line-flux. The grey shaded area shows the fraction of sources that is unclassed and its poissonian uncertainty. All emitters at higher line-fluxes are classed. These fractions are used to correct for identification incompleteness (\S $\ref{sec:6.1.3}$).}
    \label{fig:NB921_identification}
\end{figure*}

Due to their faintness, 359 out of the 1161 line-emitters are not detected in a sufficient number of broadbands required for classification and can thus not be classed. We expect that most of these sources are faint H$\alpha$, H$\beta$/{\sc [Oiii]} or {\sc [Oii]} emitters. Based on the fraction of emitters in different classifications as a function of line-flux, we expect an increasing fraction of {\sc [Oii]} emitters at low line-fluxes (see also \cite{Sobral2012}). As illustrated in the left panel of Fig. $\ref{fig:NB921_identification}$, the majority of sources indeed has $B_w - R$ colours similar to {\sc [Oii]} emitters (but could not be classed due to their faintness in other broadband filters). We discuss this `identification-incompleteness' further in \S$\ref{sec:6.1.3}$.

\begin{table}
\centering
\caption{Line-identifications of the total $\sim 2000$ emitters (as described in \S$\ref{sec:4}$) in the Bo\"otes-HiZELS narrow-band filters.}
\begin{tabular}{llr}
\hline
Filter & Sub-sample & \# of sources  \\ \hline 
NB392 & $\Sigma > 3$, EW$>30$ {\AA} & 57 \\
& Ly$\alpha$ at $z=2.23$ & 25 \\ \hline
stV & $\Sigma > 3$, EW$>130$ {\AA} &  39 \\ 
& Ly$\alpha$ at $z=2.4$ & 16 \\ \hline
NB501 & $\Sigma > 3$, EW$>50$ {\AA} & 65 \\
& Ly$\alpha$ at $z=3.1$ & 32 \\ \hline

NB921 & $\Sigma > 3$, EW$>30$ {\AA} & 1161 \\
& H$\alpha$ at $z=0.40$ & 198 \\ 
& [O{\sc iii}]/H$\beta$ at $z=0.8$ & 304 \\ 
& [O{\sc ii}] at $z=1.47$ & 277 \\ \hline

NB$_{\rm H}$ & $\Sigma > 3$, EW$>85$ {\AA} & 301 \\
& H$\alpha$ at $z=1.47$ & 87 \\ 
& [O{\sc iii}]/H$\beta$ at $z=2.23$ & 72 \\ 
& [O{\sc ii}] at $z=3.3$ & 6 \\ \hline

NB$_{\rm K}$ & $\Sigma > 3$, EW$>80$ {\AA} & 255 \\
& H$\alpha$ at $z=2.23$ & 77 \\ 
& [O{\sc iii}]/H$\beta$ at $z=3.2$ & 11 \\ 
& [O{\sc ii}] at $z=4.7$ & 2 \\ 
\hline\end{tabular}
\label{tab:line-identifications}
\end{table}

\subsection{Line-emitters in NB$_{\rm H}$} \label{sec:4.5}
We select line-emitters detected in the NB$_{\rm H}$ filter with a narrow-band excess of EW$>85$ {\AA} (corresponding to $H-$NB$_H>0.3$). Since the near-infrared detectors of UKIRT/WFCAM contain significant amounts of crosstalk, we perform careful visual inspections of our sample of line-emitters, resulting in a sample of 301 line-emitters.

While the sample of line-emitters includes nine sources with spectroscopic redshifts (six H$\alpha$ at $z=1.4$ and three H$\beta$/{\sc [Oiii]} at $z=2.2$), the NB$_{\rm H}$ line-emitters are particularly suitable for identifying line-emitters with the dual-NB technique. These robustly identified line-emitters are used to adapt the colour selection criteria from \cite{Sobral2013} for the data available in this field. Using the criteria listed in Table $\ref{tab:criteria}$, we select 87 H$\alpha$ emitters at $z=1.47$, 72 H$\beta$/{\sc [Oiii]} emitters at $z=2.2$ and 6 [O{\sc ii}] emitters at $z=3.3$, see Table $\ref{tab:line-identifications}$. 99 sources are classed as low-redshift interlopers and 39 sources are too faint to be detected in the required broad-bands.   
Prior to the final identification, four spectroscopically confirmed H$\beta$/{\sc [Oiii]} emitters were classed as H$\alpha$ emitters, while five H$\alpha$ emitters were classed as H$\beta$/{\sc [Oiii]} emitter. These mis-identified emitters are typically very luminous and likely AGN, such that their colours are anomalous (see Fig. $\ref{fig:colsel_NBH}$). There are no such identified contaminants among the fainter dual-emitters.

\subsection{Line-emitters in NB$_{\rm K}$} \label{sec:4.6}
We select line-emitters detected in the NB$_{\rm K}$ filter with EW$>80$ {\AA} (corresponding to $K-$NB$_K>0.23$). In total, after visual inspections, we find 255 line-emitters, of which 18 have a spectroscopic redshift (including eight H$\alpha$ at $z=2.23$) and 20 are dual-emitters (all H$\alpha$ at $z=2.23$). Based on these robust redshifts and the colour selection criteria listed in Table $\ref{tab:criteria}$, we select 77 H$\alpha$ emitters at $z=2.2$, 11 H$\beta$/{\sc [Oiii]} emitters at $z=3.2$ and two [O{\sc ii}] emitters at $z=4.7$ (see Table $\ref{tab:line-identifications}$). 110 line-emitters are at $z<1.5$ and 55 line-emitters are too faint to be classified. 
We have not identified any spectroscopically confirmed contaminants before the final classification. However, the colour-colour criteria missed two X-ray detected dual-emitters at $z=2.23$ as illustrated in Fig. $\ref{fig:colsel_NBK}$.

\begin{table}
\centering
\caption{Survey volumes and flux completenesses for the various line-emitters in this survey.}
\begin{tabular}{llrr}
\hline
Filter & Emission-line & Volume & 50 \% completeness  \\ 
& & [10$^5$ Mpc$^3$] & [erg s$^{-1}$ cm$^{-2}$] \\ \hline 
NB392 &Ly$\alpha$ $z=2.2$ & 2.8 & $1.3\times10^{-16}$ \\
stV & Ly$\alpha$ $z=2.4$ & 9.5 & $4.7\times10^{-16}$ \\
NB501 & Ly$\alpha$ $z=3.1$ & 7.2 & $1.1\times10^{-16}$ \\
NB921 & H$\alpha$ $z=0.4$ & 0.2 & $1.0\times10^{-16}$ \\
 & H$\beta$/[O{\sc iii}] $z=0.8$ & 1.2 & $1.0\times10^{-16}$ \\
& [O{\sc ii}] $z=1.47$ & 1.7 & $1.0\times10^{-16}$  \\
NB$_{\rm H}$ & H$\alpha$ $z=1.47$ & 2.5 & $1.3\times10^{-16}$ \\
 & H$\beta$/[O{\sc iii}] $z=2.2$ & 5.2 & $1.3\times10^{-16}$ \\
NB$_{\rm K}$ & H$\alpha$ $z=2.2$ & 2.7 & $0.5\times10^{-16}$ \\
\hline\end{tabular}
\label{tab:volumes}
\end{table}

\section{Number densities} \label{sec:6}
\subsection{Method} \label{sec:6.1}
We measure the number densities of LAEs at $z=2.2, 2.4$ and $z=3.1$, H$\alpha$ emitters at $z=0.4, 1.47$ and $z=2.23$, H$\beta$/[O{\sc iii}] emitters at $z=0.8,2.2,3.2$ and [O{\sc ii}] emitters at $z=1.47,3.3,4.7$ as a function of their line-luminosity, in narrow luminosity bins (0.2-0.3 dex in this analysis). The luminosity is calculated using the line-flux (\S$\ref{sec:3.3}$) and assuming the luminosity distance corresponding to the redshift of peak filter transmission for the relevant emission-line. We calculate the comoving volume for each line/filter combination using the redshifts of half peak transmission, see Table $\ref{tab:volumes}$. For H$\beta$/{\sc [Oiii]}, we compute the volume following \cite{Khostovan2015}, who uses only the volume probed by the [O{\sc iii}]$_{\lambda 5007}$ line. We refer to this work and \cite{Sobral2015} for a detailed discussion on the contribution of H$\beta$ and [O{\sc iii}]$_{\lambda 4959}$. Luminosity-binned number densities were calculated by dividing the number of sources in each bin by the comoving volume, and then correcting these (as described in the following subsections) for the effects of the filter profile, flux incompleteness and identification incompleteness. Uncertainties on these number densities were estimated using Poissonian errors. To be conservative, we add in quadrature 20 \% of the flux-completeness correction and 20 \% of the identification-incompleteness correction (in the case of the red narrow-bands) to the error of each bin. We only show bins with $>40$ \% flux-completeness.

\subsubsection{Filter profile correction}\label{sec:6.1.1}
As described in \cite{Khostovan2015} and \cite{Sobral2013,Sobral2015}, observed number densities have to be corrected for the fact that the filter transmission curves are not a perfect top-hat. Because of this, luminous sources may be observed as faint sources if they lie at a redshift corresponding to the wings of the filter. Furthermore, at fixed flux-limit, fainter sources can only be observed over a smaller volume than more luminous sources. Following the method described in these papers, we compute the number density corrections using a simulation. This simulation assumes that sources are distributed randomly in redshift space and computes their observed luminosities based on the relevant filter transmission. We then obtain a volume correction for each luminosity bin. These corrections typically increase the number densities of the most luminous bins by at most 0.3 dex, while the number densities of fainter bins stay constant, or are decreased by at most 0.05 dex.

\begin{figure*}
\begin{tabular}{cc}
	\includegraphics[width=8.6cm]{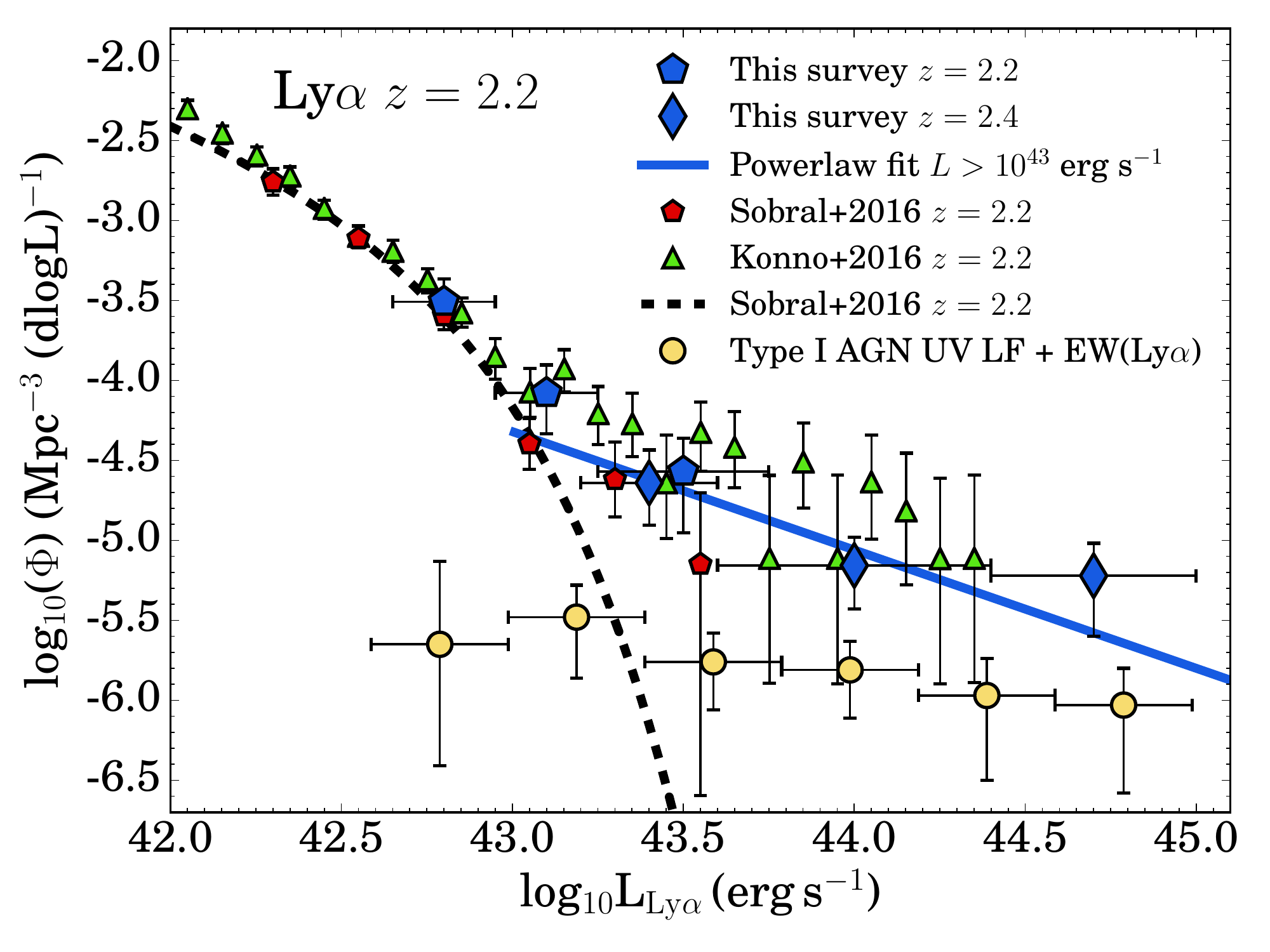} &
	\includegraphics[width=8.6cm]{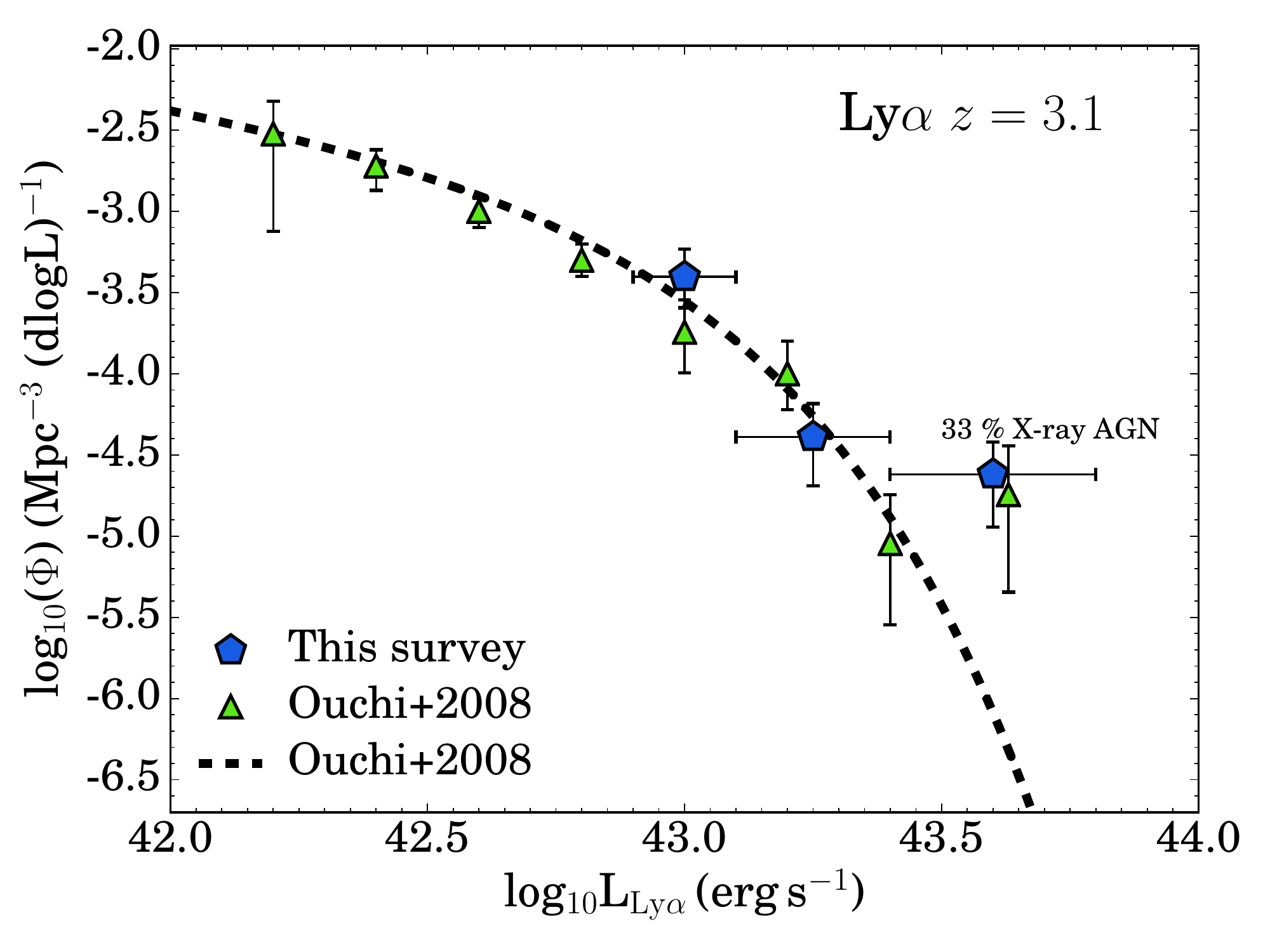} \\
	\end{tabular}
    \caption{Number densities and luminosity functions of LAEs at $z\approx2.2$ and $z=3.1$. The power-law like behaviour at the bright end is due to the contribution from AGN (for which we provide a fit in \S $\ref{sec:6.2.1}$). The yellow points show an estimate of the contribution from Type I AGN, based on the UV LF combined with a Ly$\alpha$ of 80 {\AA}. At $z\approx2.2$, we find good agreement between the luminosity function of LAEs in B-HiZELS and the luminosity function we measured in \citet{Sobral2016} and to the survey from \citet{Konno2016}. The number density of LAEs at $z=3.1$ is similar to that measured by \citet{Ouchi2008} in the UDS field.}
    \label{fig:LAE_LFs}
\end{figure*}

\subsubsection{Detection flux-completeness}\label{sec:6.1.2}
The flux-completeness of our selection is measured as a function of line-flux as follows: for the relevant line, we select galaxies that are not selected as a line-emitter, but do fulfil the colour criteria from \S$\ref{sec:4}$. We then artificially add line-flux (starting from $10^{-18}$ erg s$^{-1}$ cm$^{-2}$ in steps of 0.05 dex) and re-compute the line excess and excess significance for each step. After each step, we measure the fraction of sources that would be selected as line-emitter with the added line-flux. We tabulate the 50 \% completeness in Table $\ref{tab:volumes}$. Most narrow-band selections are 50 \% complete at $\sim1\times10^{-16}$ erg s$^{-1}$ cm$^{-2}$, with the exception of stV, which is only sensitive to brighter emission-lines.

\subsubsection{Identification incompleteness}\label{sec:6.1.3}
For the red narrow-bands, we also take into account that the broad-band data is not deep enough for a robust classification of all faintest line-emitters, which we call identification-incompleteness. We estimate corrections for this effect as follows: for each narrow-band filter, we measure the fraction of line-emitters that is classed as either H$\alpha$, H$\beta$/[O{\sc iii}] or [O{\sc ii}] emitter or as lower redshift source, as a function of line-flux and assume that this fraction can be extrapolated to the line-fluxes of the sources that are not classable.

We find that for line-emitters in NB921 the fraction of classed H$\alpha$, H$\beta$/[O{\sc iii}] and [O{\sc ii}] emitters is $25\pm5$, $35\pm5$ and $40\pm4$ \% respectively at fluxes $<6.3\times10^{-17}$ erg s$^{-1}$ cm$^{-2}$ and $42\pm10$, $38\pm10$ and $20\pm8$ \% respectively for fluxes between $6.3\times10^{-17}$ erg s$^{-1}$ cm$^{-2}$ and $4\times10^{-16}$ erg s$^{-1}$ cm$^{-2}$, see Fig. $\ref{fig:NB921_identification}$. This is expected, as sources with fainter fluxes are expected to be at higher redshift. For line-emitters in NB$_{\rm H}$, the corresponding fractions are $25\pm5$, $25\pm7$ and $4\pm2$ \% at fluxes below $4\times10^{-16}$ erg s$^{-1}$ cm$^{-2}$. This means that $\sim40$ \% of the unclassed sources is likely at $z<1$. Above this flux, all sources are classed. These fractions are in agreement with the estimate from \cite{Sobral2012}. Finally, for line-emitters in NB$_{\rm K}$, we estimate that below a flux of $1.5\times10^{-16}$ erg s$^{-1}$ cm$^{-2}$ a fraction of $40\pm8$ \% of the line-emitters is H$\alpha$, $7\pm3$ \% is H$\beta$/[O{\sc iii}] , while $55\pm6$ \% is at low redshift. All sources with a larger line-flux have been classed. We note that maximally 40 \% of the sources in a flux bin are unclassed. This maximum occurs in the faintest bin of the NB921 line-emitters (see Fig. $\ref{fig:NB921_identification}$). The typical fraction of unclassed sources at the discussed flux levels is 20 \%. We use the estimates described above to obtain the identification-incompleteness for each luminosity bin for the relevant emission-line. 

\begin{figure*}
\begin{tabular}{cc}
	\includegraphics[width=8.6cm]{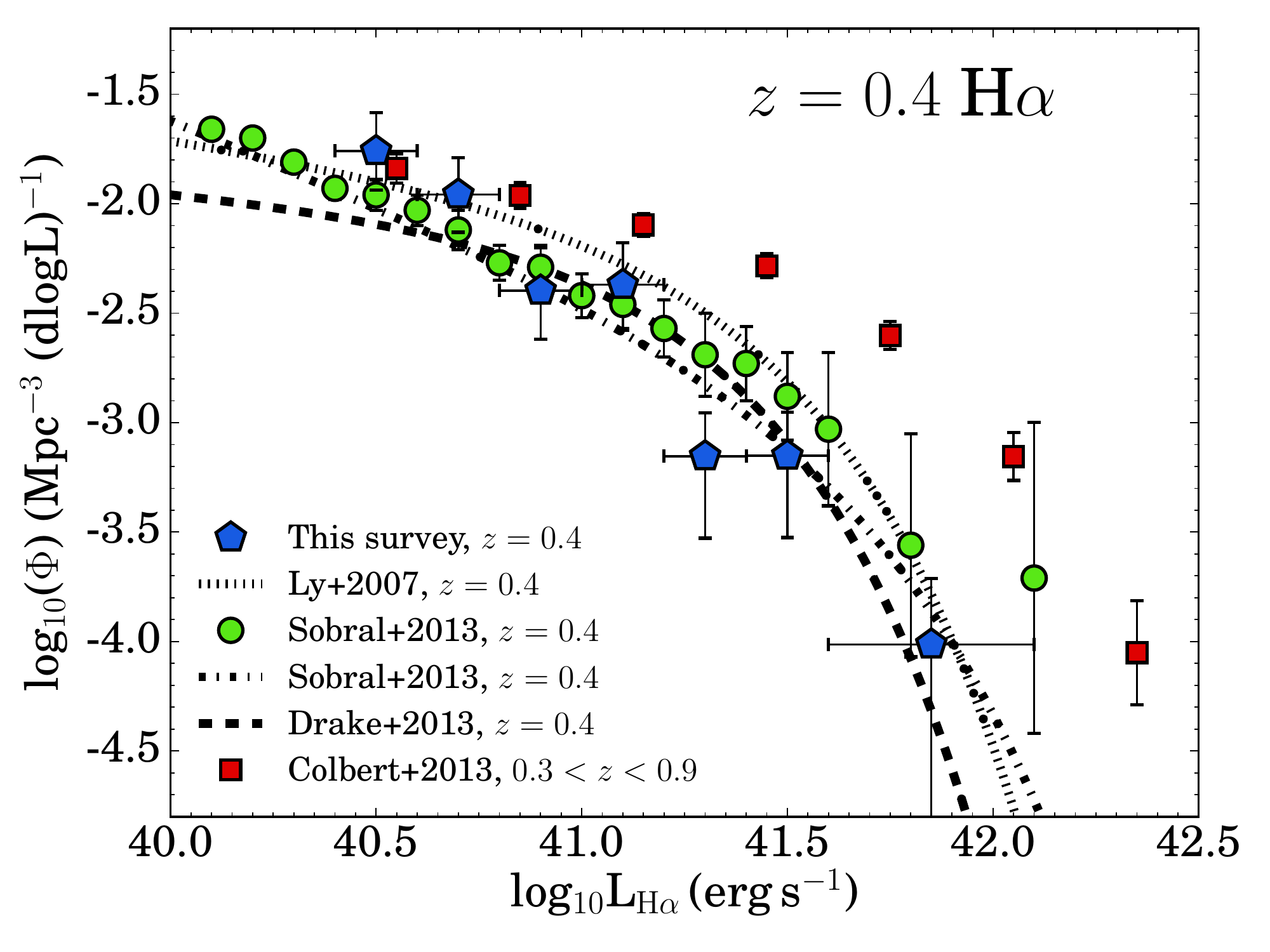} &
			\includegraphics[width=8.6cm]{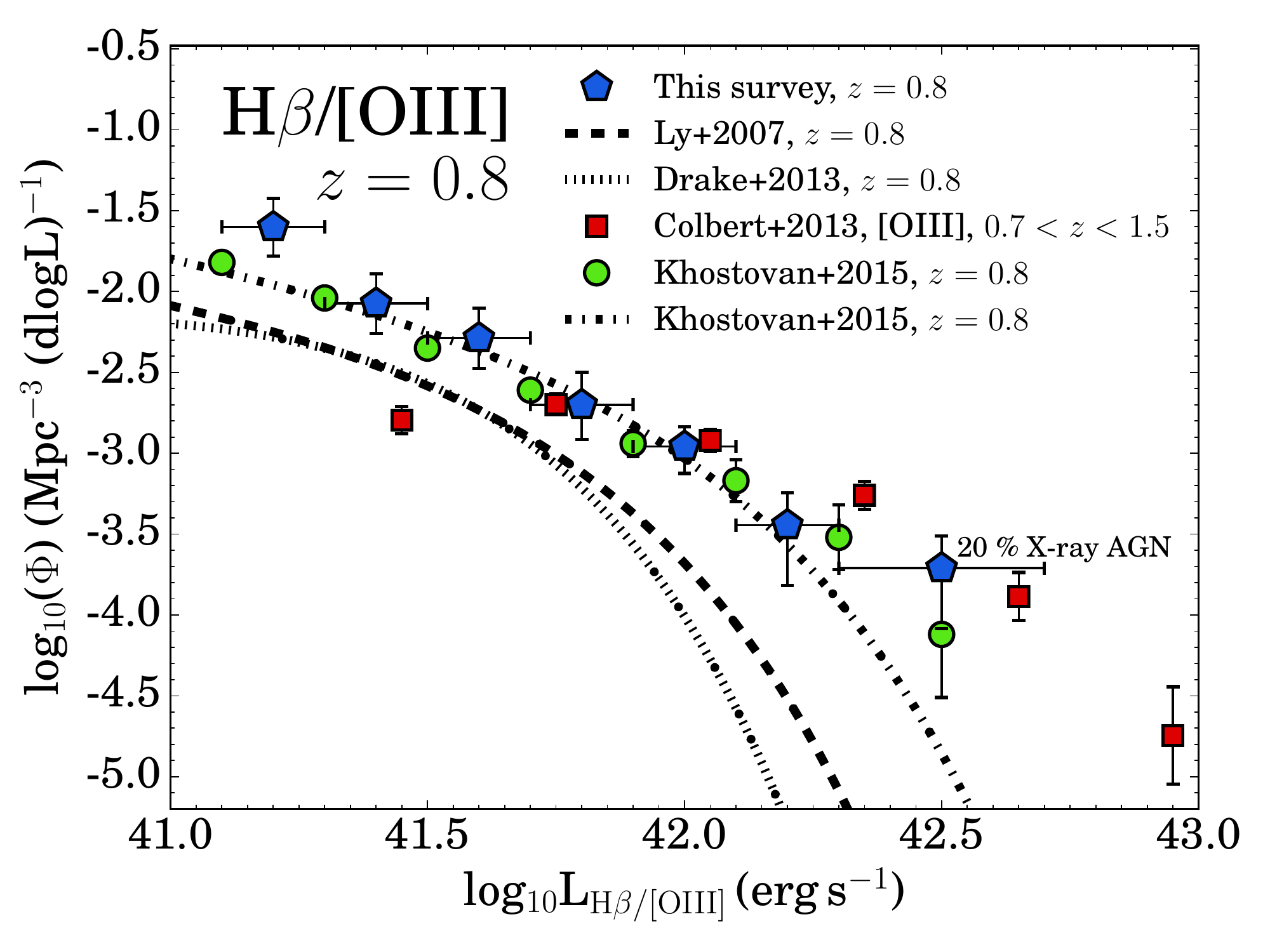} \\
		\includegraphics[width=8.6cm]{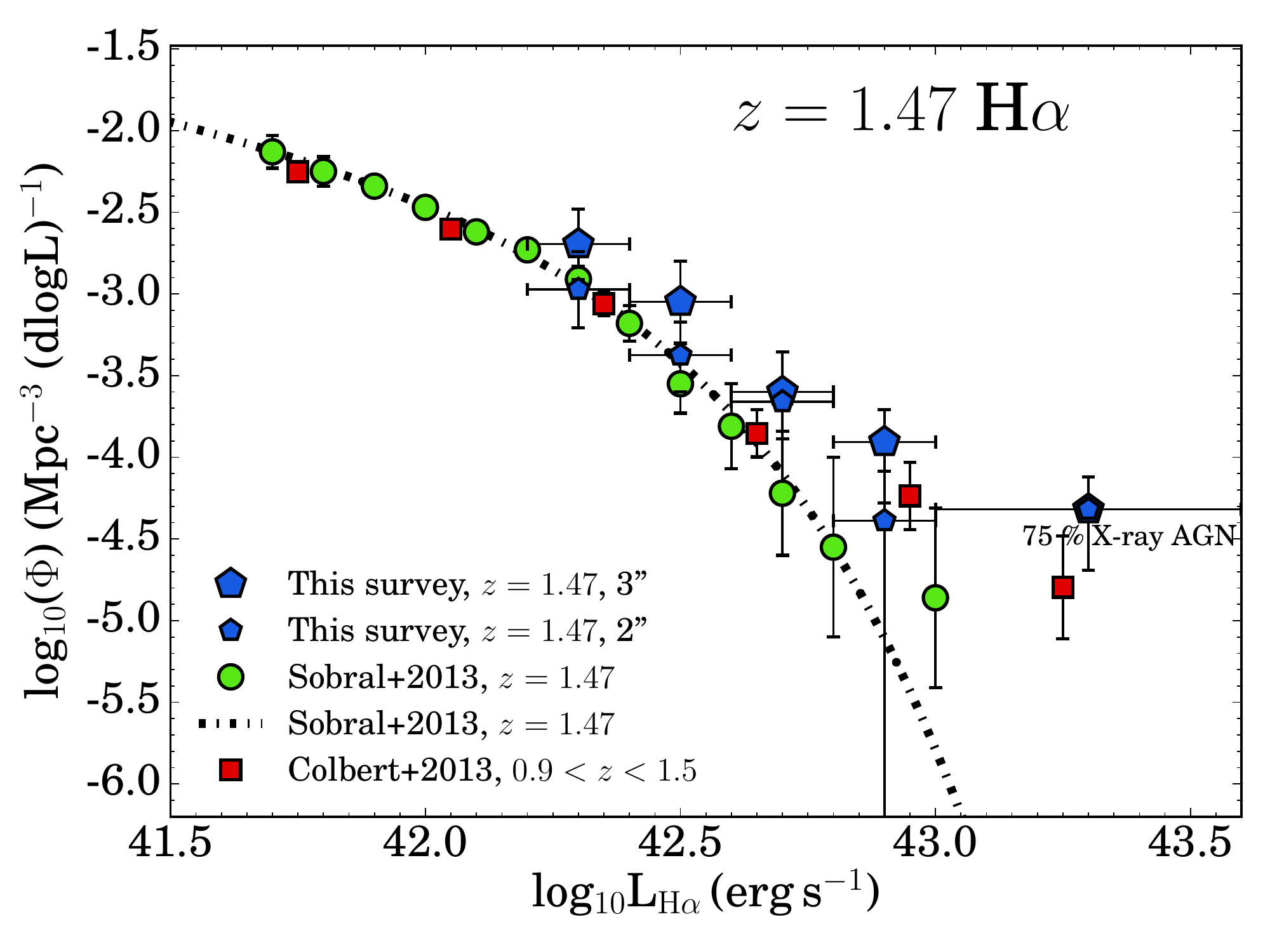} &
			\includegraphics[width=8.6cm]{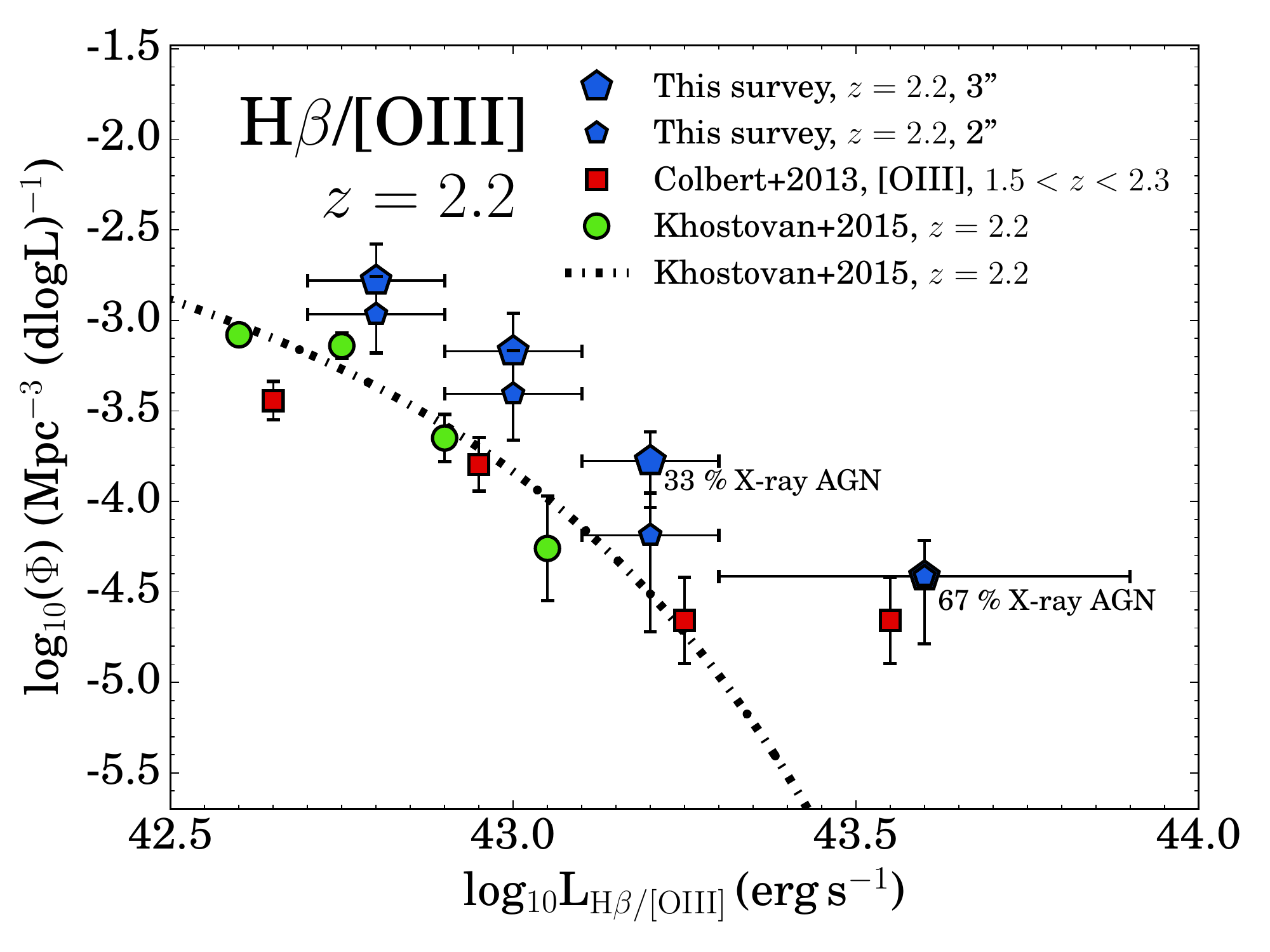} \\
	\includegraphics[width=8.6cm]{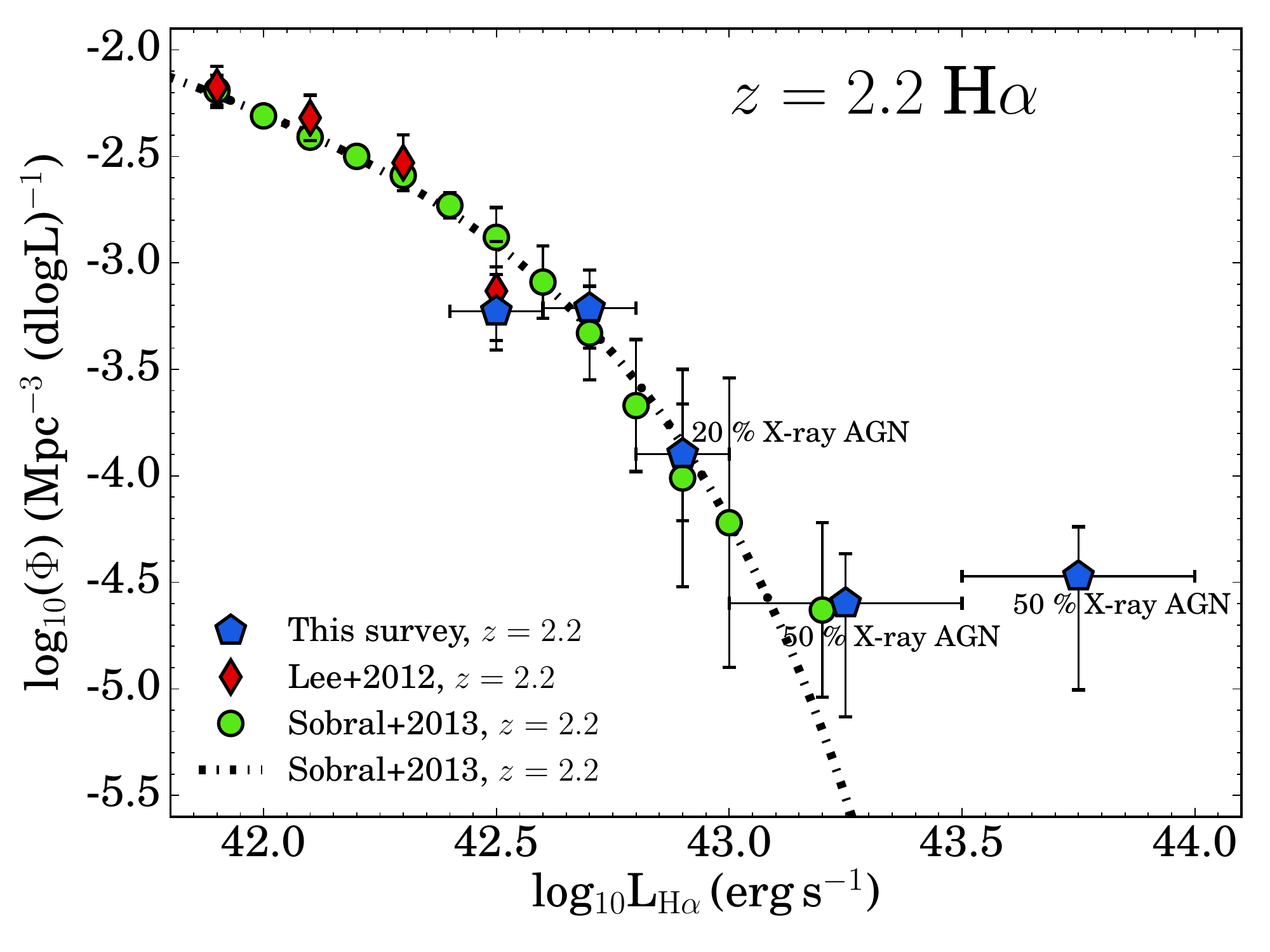} &
		\includegraphics[width=8.6cm]{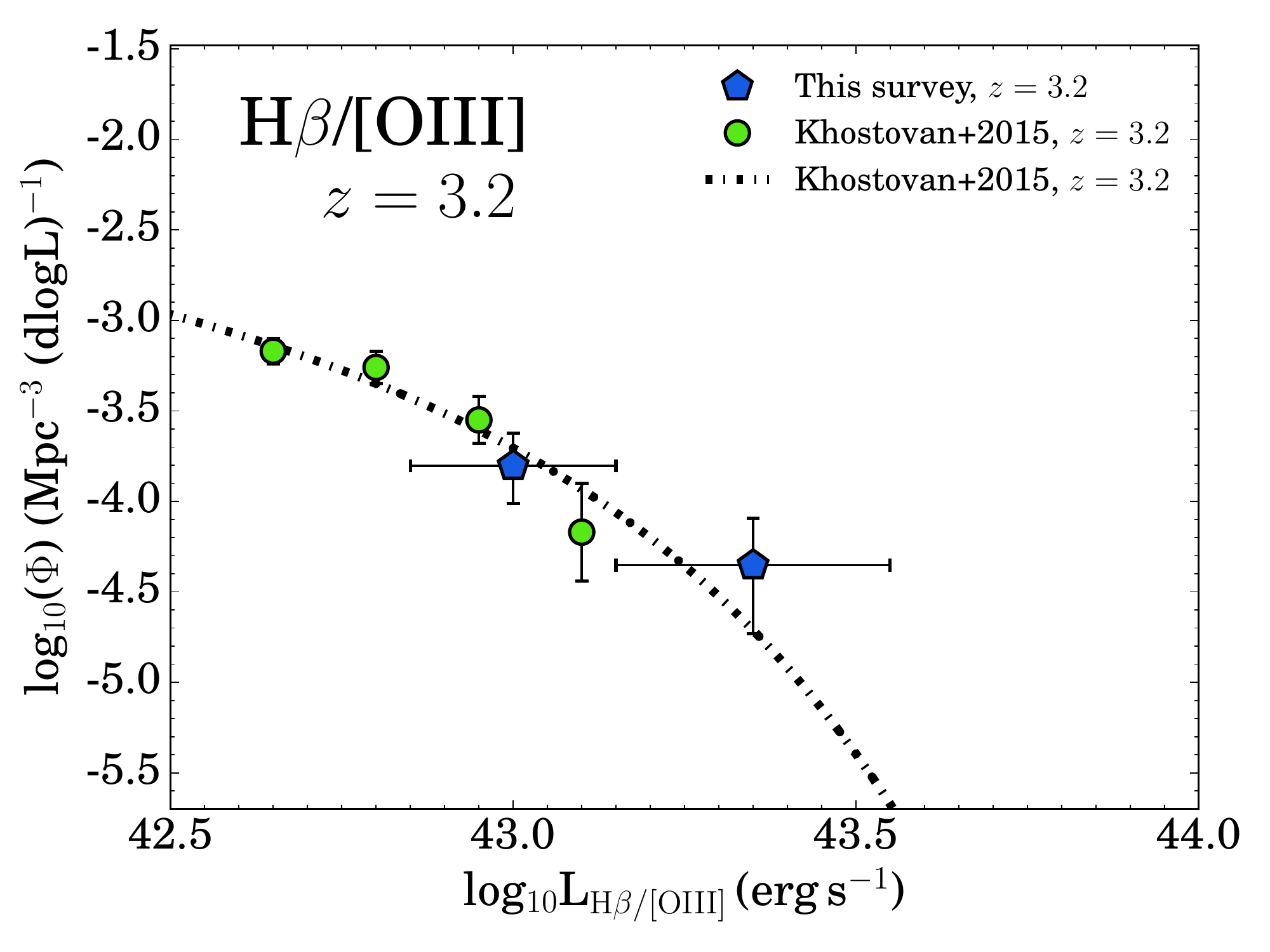} \\

	\end{tabular}
    \caption{Left column: Measured H$\alpha$ luminosity functions at $z=0.4, 1.47, 2.2$. H$\alpha$ luminosities are corrected for the [N{\sc ii}] contribution using a relation with H$\alpha$+[N{\sc ii}] EW (\citealt{Sobral2012}), but are not corrected for dust. At $z=0.4$ the LF is compared to the narrow-band survey from \citet{Ly2007} in the SDF field and \citet{Drake2013} in the UDS field and with the survey from \citet{Lee2012} at $z=2.2$. At all redshifts, the number densities are compared to the HiZELS survey results in the UDS+COSMOS fields from \citet{Sobral2013}. We also compare the LFs with those measured in a blind grism survey by \citet{Colbert2013}. Overall, there is reasonable agreement. The luminosity-offset at $z=1.47$ may be explained by aperture effects (see \S$\ref{sec:6.2.2}$). The brightest bin at $z=2.2$ is due to the presence of (spectroscopically confirmed) AGN. Right column: Measured H$\beta$/{\sc [Oiii]} luminosity functions at $z=0.8, 2.2, 3.2$, compared to  \citet{Ly2007} and \citet{Drake2013} at $z=0.8$ and to \citet{Khostovan2015} and \citet{Colbert2013} at higher redshift. There is a luminosity-offset at $z=2.2$ that is partially due to a larger aperture measuring a higher luminosity by $\approx0.14$ dex (this is the same filter as the H$\alpha$ luminosity function at $z=1.47$). The brightest bin at $z=2.2$ contains the same AGN as in the brightest bin of the H$\alpha$ luminosity function at $z=2.2$.}
    \label{fig:HAE_LFs}
\end{figure*}

\subsection{Comparison with previous surveys} \label{sec:6.2}
\subsubsection{Lyman-$\alpha$ emitters at $z=2.2-3.1$} \label{sec:6.2.1} 
We show the measured luminosity function (LF) of LAEs at $z=2.2-2.4$ and $z=3.1$ in Fig. $\ref{fig:LAE_LFs}$. The depth of our data allows us to constrain the LF to $\approx L^{\star}$. We find good agreement with earlier results from \cite{Sobral2016} and those from \cite{Konno2016} in different survey fields and slightly different colour-criteria (see also \citealt{An2016}). All surveys indicate that the Ly$\alpha$ LF at $z\approx2.2$ deviates from a Schechter function at bright luminosities.\footnote{ The \cite{Schechter1976} shape of the luminosity function is expressed as $\Phi(L)$ d$L$ = $\Phi^{\star} (\frac{L}{L^{\star}})^{\alpha} e^{-\frac{L}{L^{\star}}} d(\frac{L}{L^{\star}})$, where $\Phi^{\star}$ is the characteristic number density, $L^{\star}$ the characteristic luminosity and $\alpha$ the faint-end slope.} We note that all LAEs at $z=2.2$ with a luminosity above $10^{44}$ erg s$^{-1}$ are either spectroscopically confirmed or have a dual-NB redshift. As discussed in \cite{Sobral2016}, the power-law behaviour of the luminosity function is likely due to the contribution of AGN in addition to the normal Schechter function. Indeed, most ($\sim 80\pm40$ \%) luminous LAEs (L$_{\rm Ly\alpha} > 10^{43}$ erg s$^{-1}$) are AGN (either due to X-Ray detection or C{\sc iv} detection in NB501). We fully explore this in \S $\ref{sec:5.2}$.

Due to its larger probed cosmic volume, the stV filter is mostly sensitive to very luminous LAEs. Although all LAEs with a luminosity above $10^{44}$ erg s$^{-1}$ at $z=2.4$ are spectroscopically confirmed, we expect that the sample with luminosities $10^{43-44}$ erg s$^{-1}$ is contaminated, as the spectroscopic follow-up at these fluxes is not complete. Most importantly, we expect contaminants to be emission lines that are associated with AGN activity such as C{\sc iv}, C{\sc iii}] and He{\sc ii} at $z=1.15-1.65$ \citep[e.g.][]{Stroe2017}, which are challenging to identify with these colour-colour selections. We estimate the contamination at these flux levels by mimicking the selection of this survey in a similar medium-band in the COSMOS field (IA427, Santos et al. in prep). We select LAEs at $z=2.5$ with the same criteria (including broad-band depths) and estimate the number of interlopers using the most recent photometric redshifts \citep{Laigle2016} and a compilation of spectroscopic redshifts. We find that at luminosities $\sim10^{43-44}$ erg s$^{-1}$ there is a non-negligible contamination due to C{\sc iv},  C{\sc iii}] and He{\sc ii} emitters of 20$\pm10$ \%. At higher luminosities the contamination decreases to 4$\pm4$ \%. The plotted number densities are corrected for these contamination rates. 
We combine the $z=2.2-2.4$ data to fit a power-law function to the number density of LAEs at the bright end (L$_{\rm Ly\alpha} > 10^{43}$ erg s$^{-1}$), which results in: log$_{10}(\Phi) = 27.5^{+7.3}_{-7.4} - 0.74^{+0.17}_{-0.17}$log$_{10}$(L$_{\rm Ly\alpha}$), with a reduced $\chi^2$ of 1.1. This fit is slightly shallower than the power-law fitted by \cite{Sobral2016} based on a smaller volume, but consistent within 1$\sigma$.

We estimate the contribution of broad-line Type I AGN to the Ly$\alpha$ LF at $z\approx2.2$ based on the UV LF of Type I AGN at $2.0<z<2.5$ from \cite{Bongiorno2007} and the typical UV slope and Ly$\alpha$ EW of these AGN. Assuming f$_{\lambda} \propto \lambda^{-1.5}$ and Ly$\alpha$ EW$_0 =$ 80 {\AA} \citep[e.g.][]{VandenBerk2001,Hunt2004}, we convert the number densities as a function of $M_{1450}$ to number densities as a function of Ly$\alpha$ luminosity. As shown in Fig. $\ref{fig:LAE_LFs}$, it is clear that the number density of LAEs is higher than the number density of Type I AGN at fixed Ly$\alpha$ luminosity. This indicates that only a fraction of the luminous LAEs are likely Type I AGN. This estimate suggests that at Ly$\alpha$ luminosities $10^{43-44}$ erg s$^{-1}$ the fraction of Type I AGN is only $\sim 10$ \%, while the fraction is $\sim 20-30$ \% at higher Ly$\alpha$ luminosities. Because the AGN LF at the faintest UV magnitudes is relatively flat, these fractions do not depend strongly on the assumed values of the Ly$\alpha$ EW or UV slope. The low Type I AGN fraction indicates that the majority of luminous LAEs are narrow-line Type II AGN (see also \S$\ref{sec:5.2}$), or star-forming galaxies.

At $z=3.1$ we find that the Ly$\alpha$ LF agrees well with that from \cite{Ouchi2008}, who performed a deep Ly$\alpha$ survey over a similar area. Compared to similar NB501 data in the COSMOS field (with deeper ancillary data and relatively more spectroscopic follow-up; Matthee et al. in prep), we also find that the fraction of line-emitters (with similar line-flux and EW distributions) that are classed as LAE is similar: $51\pm9$ \% in Bo\"otes against $46\pm7$ \% in COSMOS. This also confirms evolution in L$^{\star}$ between $z=2.2-3.1$ from L$_{\rm Ly\alpha} \approx 4\times10^{42}$ erg s$^{-1}$ to  L$_{\rm Ly\alpha} \approx 9\times10^{42}$ erg s$^{-1}$. The number density of LAEs in the brightest bin (three out of the four sources in this bin are spectroscopically followed-up and confirmed) lies above the Schechter fit from \cite{Ouchi2008} (similar to the actual data-points from that survey), indicating the presence of AGN among these luminous sources, similar to $z\sim2$. Indeed, we find evidence for AGN activity for most LAEs in the most luminous bin, either due to an X-ray detection or due to the detection of high ionisation emission-lines as [Ne{\sc iv}]$_{\lambda 2424}$ or broad Si{\sc iv} and C{\sc iv} absorption features in the spectrum (Sobral et al. in prep).

\subsubsection{H$\alpha$ emitters at $z=0.4-2.2$} \label{sec:6.2.2} 
We show the number densities of H$\alpha$ emitters as a function of their observed H$\alpha$ luminosities. The H$\alpha$ luminosities are corrected for the contribution due to [N{\sc ii}] following a method based on observed H$\alpha$+[N{\sc ii}] EW \citep{Sobral2012}. H$\alpha$ luminosities are not corrected for attenuation due to dust and we compare our results with dust-uncorrected values from the literature. In case dust-uncorrected values are not provided, we convert dust-corrected values back using the prescriptions outlined in the relevant papers.

As illustrated in Fig. $\ref{fig:HAE_LFs}$, we find that the H$\alpha$ LF at $z=0.4, 1.47$ and $z=2.23$ are generally in good agreement with the number densities in the UDS+COSMOS parts of the HiZELS survey from \cite{Sobral2013}. Within the errors, the LF agrees well with the fitted relations from \cite{Drake2013} and \cite{Ly2007}. At $z=2.23$, the number densities complement the number densities at fainter luminosities from \cite{Lee2012}. This confirms strong evolution in L$^{\star}_{\rm H\alpha}$ from $z=0.4-2.23$. We note that at $z=0.4$ there could be some contamination at the faintest luminosities due to identification-incompleteness (see \S$\ref{sec:6.1.3}$). At $z=1.47$ the luminosities seem to be systematically higher by $\approx 0.1-0.15$ dex, increasing slightly with luminosity. This offset can partly be explained by different apertures used in the photometry. While \cite{Sobral2013} uses 2$''$ apertures for all measurements above $z>0.5$, we use 3$''$ measurements. Redoing the measurements with 2$''$ apertures (smaller symbols in Fig. $\ref{fig:HAE_LFs}$) results in good agreement at our faintest luminosities and reasonable agreement (within the error bars) at higher luminosities. At faint luminosities, the luminosity difference between the two apertures is typically 0.14 dex, while it is typically 0.1 dex at high luminosities. The highest luminosity bins at $z=1.47-2.23$ show number densities diverging from a Schechter function, contributing to the difference as well. The sources in these bins are all spectroscopically confirmed or have dual-NB redshifts and most are X-ray detected, as indicated in the corresponding panels. We discuss this in more detail in\S $\ref{sec:5.2}$.

Compared to the grism results at $0.3<z<0.9$ from \cite{Colbert2013}, the number densities at $z=0.4$ are offset (mostly in terms of luminosity). This can simply be explained by the evolution in the typical H$\alpha$ luminosity between $z=0.4$ and the median redshift of the \cite{Colbert2013} sample of $z\approx0.6$, as log $L^{\star}_{\rm H\alpha}$ increases with $0.45\times z$ over this redshift range \citep{Sobral2013}. The number densities at $z=1.47$ are in good agreement with the grism results at $0.9<z<1.5$, even though the median grism redshift is $z\approx1.2$. This indicates that there is little evolution in $L^{\star}_{\rm H\alpha}$ between $z=1.47$ and $z\approx1.2$. 

\begin{figure*}
\begin{tabular}{cc}
	\includegraphics[width=8.6cm]{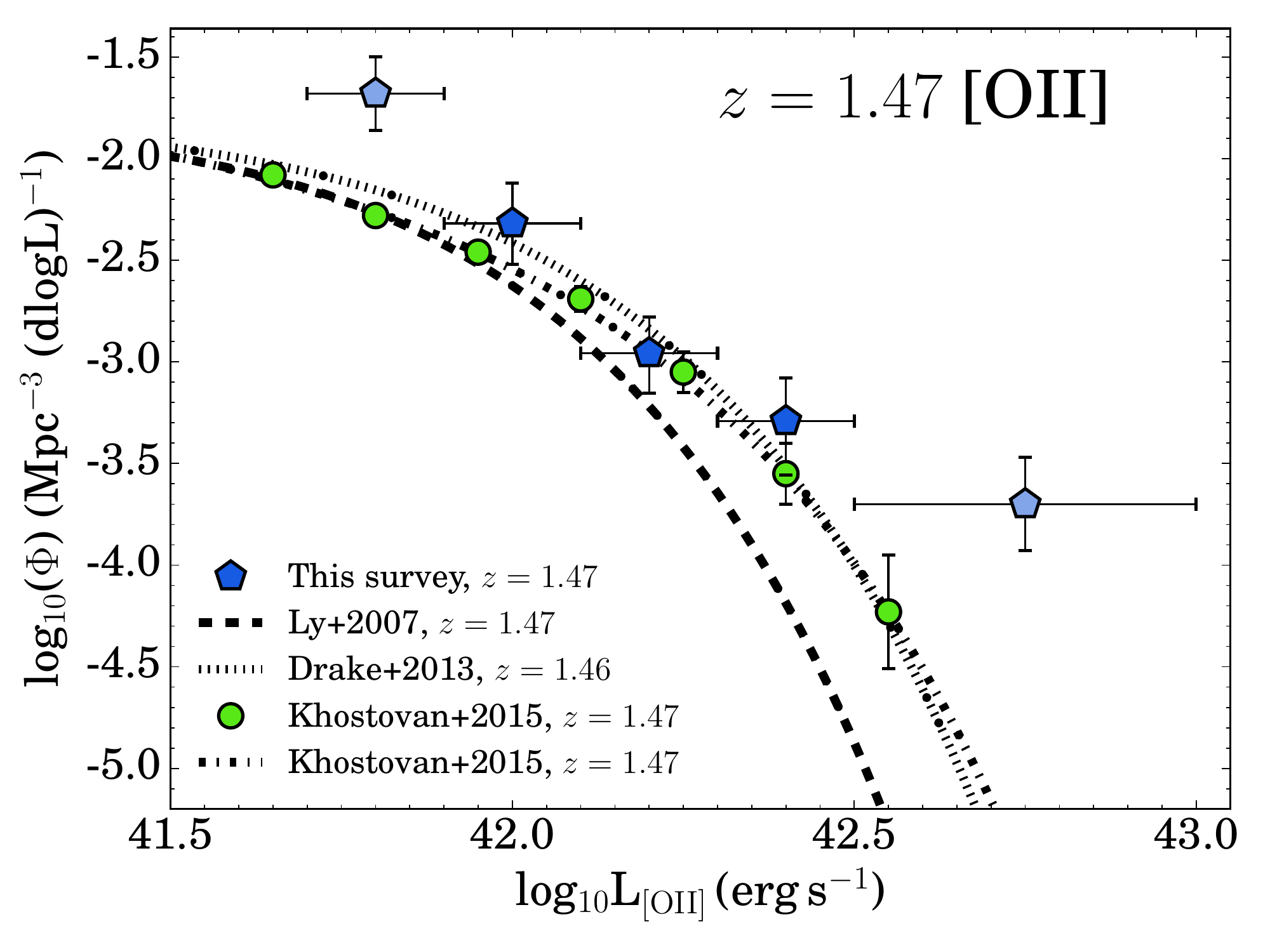} &
		\includegraphics[width=8.6cm]{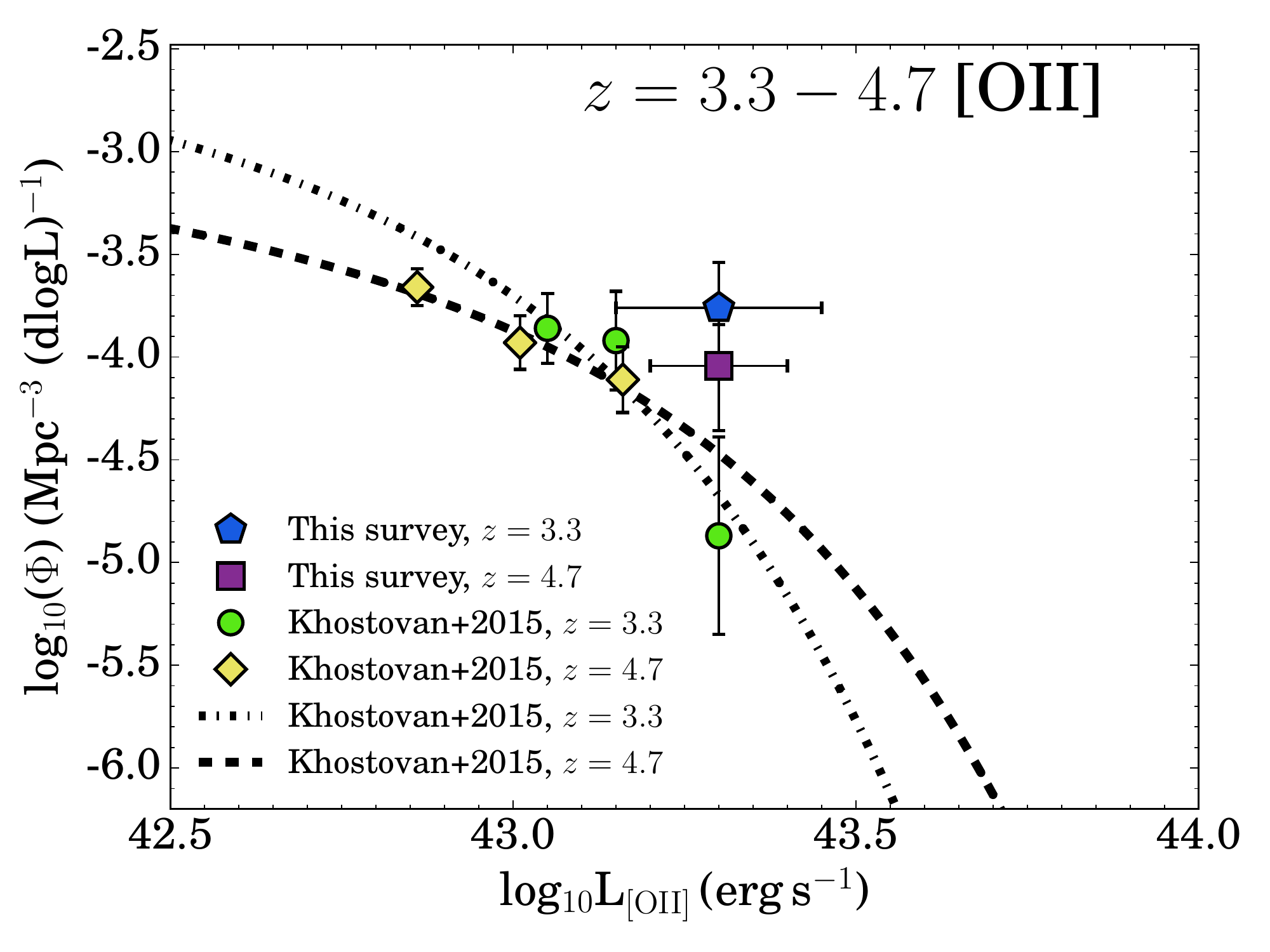}\\ 
\end{tabular}
    \caption{Measured [O{\sc ii}] luminosity functions at $z=1.47, 3.3, 4.7$. At $z=1.47$, the LF shows reasonable agreement with those from \citet{Ly2007}, \citet{Drake2013} and \citet{Khostovan2015}, except at the bright end. Since the sources in the brightest bin are not spectroscopically confirmed and identification-incompleteness in the faintest bin is large, we show these bins in a slightly lighter colour. We plot the number densities of [O{\sc ii}] emitters at $z=3.3$ and $z=4.7$ in a single panel. These number densities are slightly higher than \citet{Khostovan2015}, potentially indicating some contamination or cosmic variance.}
    \label{fig:O2_LF}
\end{figure*}

\subsubsection{H$\beta$/{\sc [Oiii]} emitters at $z=0.8-3.2$} \label{sec:6.2.3}
We also compare our idenfication of H$\beta$/{\sc [Oiii]} at $z=0.8, 2.2, 3.2$ with the analysis from \cite{Khostovan2015} in the COSMOS and UDS fields.

As illustrated in Fig. $\ref{fig:HAE_LFs}$, the number densities of H$\beta$/{\sc [Oiii]} emitters at $z=0.8$ are in good agreement with those from \cite{Khostovan2015}, and higher than those from \cite{Ly2007} and \cite{Drake2013}, which could be due to cosmic variance or systematics such as different apertures and estimates of volume and completeness. Using a large 10 deg$^2$ H$\alpha$, H$\beta$/{\sc [Oiii]} and  {\sc [Oii]} survey, \cite{Sobral2015} estimated empirically that the uncertainty in $L^{\star}$ and $\Phi^{\star}$ due to cosmic variance over the volume probed in these surveys at $z=0.8$ is $\approx 40-50$ \%. Such variance could easily explain the observed differences. At $z=2.2$, the number densities are systematically higher in luminosity compared to the literature results. Similarly as in \S $\ref{sec:6.2.2}$, we find that this is partly due to an aperture effect. As expected, the difference in luminosities measured with different apertures is slightly smaller at $z=2.2$ than at $z=1.47$, with a typical luminosity difference of 0.11 dex. As a consequence, even when matching apertures to \cite{Khostovan2015}, we still find an offset at fainter luminosities, which we attribute to cosmic variance. Although the number of H$\beta$/{\sc [Oiii]} emitters at $z=3.2$ is limited, their number densities agree well with those from \cite{Khostovan2015}. 

Unlike the H$\alpha$ number densities, the number densities of H$\beta$/{\sc [Oiii]} emitters are in good agreement with those from \cite{Colbert2013}, which could indicate that there is less evolution in the H$\beta$/{\sc [Oiii]} luminosity function than in the H$\alpha$ luminosity function between $0.7<z<1.5$. However, this could also be due to the contribution of H$\beta$ emitters (see \citealt{Sobral2015} and \citealt{Khostovan2015} for detailed discussions). Except for luminosities $>10^{43}$ erg s$^{-1}$, the number densities of H$\beta$/{\sc [Oiii]} emitters at $z=0.8$ are a factor 30 to 100 higher (at fixed {\sc [Oiii]} luminosity) than the number densities of Type II AGN at $z\sim0.71$ \citep{Bongiorno2010}.

\subsubsection{{\sc [Oii]} emitters at $z=1.47,3.3,4.7$} \label{sec:6.2.4}
Fig. $\ref{fig:O2_LF}$ compares the number density of [O{\sc ii}] emitters at $z=1.47, 3.3, 4.7$ to the other published results. At $z=1.47$, the number densities agree reasonably well with \cite{Drake2013} and \cite{Khostovan2015}, except for the faintest bin, although identification-incompleteness is significant in this bin. At the bright end, the Schechter fit from \cite{Ly2007} indicates a lower number density, potentially due to a lack of bright sources in a small survey volume. We note that the sources in the brightest bin are not spectroscopically confirmed and thus could be interlopers (for this reason, these points have a lighter colour in Fig. $\ref{fig:O2_LF}$). Although the number of [O{\sc ii}] emitters in our samples at $z=3.3$ and $z=4.7$ is limited to a handful, their number densities are slightly higher than \citet{Khostovan2015}, potentially indicating either contamination or cosmic variance. A further hint of contamination in this sample is that some have relatively low colour-excess (Fig. $\ref{fig:excess}$), which is unexpected for high-redshift sources, although this would also be consistent with a drop in typical EWs for [O{\sc ii}] emitters \citep{Khostovan2016}.

\section{Properties of line-emitters} \label{sec:5}
\subsection{Dual-emitters} 
In total, we detect 42 line-emitters that are line-emitters in multiple narrow-bands. We list the coordinates, redshifts and $I$ band magnitude of these line-emitters in Table $\ref{tab:dualemitters}$. The majority (20) of dual-emitters are H$\alpha$+[O{\sc ii}] emitters at $z=1.47$ detected in NB921 and NB$_{\rm H}$, followed by 17 H$\alpha$+[O{\sc iii}] emitters at $z=2.23$, of which two are also detected in Ly$\alpha$, three in C{\sc iv} and one in Mg{\sc ii}. Three galaxies are identified as LAE in NB392 at $z\sim$2.2--2.3 and are also detected in either NB$_{\rm H}$ ([O{\sc iii}]) or NB$_{\rm K}$ (H$\alpha$), of which one is also detected in NB501 (C{\sc iv}). 

\subsection{The most luminous H$\alpha$ emitters at $\bf z=2.23$} \label{sec:5.1}
B-HiZELS\_1 is the most luminous H$\alpha$ emitter known from HiZELS at $z=2.23$ \citep[i.e.][]{Brizels} and also detected as Ly$\alpha$ and [O{\sc iii}] emitter. Although its high luminosity (L$_{\rm H\alpha} = 7.9\times10^{43}$ erg s$^{-1}$) suggests that it is an AGN, it is X-ray undetected (L$_X <3\times10^{44}$ erg s$^{-1}$, or $<L^{\star}_{X}$, \citealt{LaFranca2005}). There is also a dual-emitter (B-HiZELS\_27), detected in {\sc [Oiii]} and H$\alpha$ at the same redshift only 6$''$ away (a projected distance of $\sim 50$ kpc). This source has an estimated H$\alpha$ EW$_0$ of $\gtrsim400$ {\AA} and {\sc [Oiii]} EW$_0 \gtrsim375$ {\AA}. This places it at the very high end of the H$\alpha$ EW distribution at $z=2.2$ \citep{Fumagalli2012,Sobral2014}, and the galaxy is thus likely a low mass extreme emission line galaxy \citep[e.g.][]{vanderWel2011}. These strong emission-lines could indicate that this galaxy may be undergoing high interaction-induced SFR combined with little extinction due to dust, or is a shocked gas cloud.

The second most luminous H$\alpha$ emitter at $z=2.2$ is B-HiZELS\_15 at $z=2.244$ (L$_{\rm H\alpha} = 3.6\times10^{43}$ erg s$^{-1}$). We also detect [O{\sc iii}], H$\alpha$ and C{\sc iv} emission-lines (there are no Ly$\alpha$ observations at its position). Although B-HiZELS\_15 also has a neighbouring galaxy (at $z=2.242$ and a projected separation of $\approx 50$ kpc) and the H$\alpha$ luminosity is only a factor two lower than that of B-HiZELS\_1, several other properties are different. B-HiZELS\_15 is X-ray detected, has a higher H$\alpha$ EW$_0$ (360 {\AA} versus 120 {\AA}) and is more than three magnitudes fainter in the optical and NIR continuum. This indicates a diversity in the properties of luminous H$\alpha$ emitters, similar to the results from \cite{Brizels}.

\begin{figure}
\includegraphics[width=8.6cm]{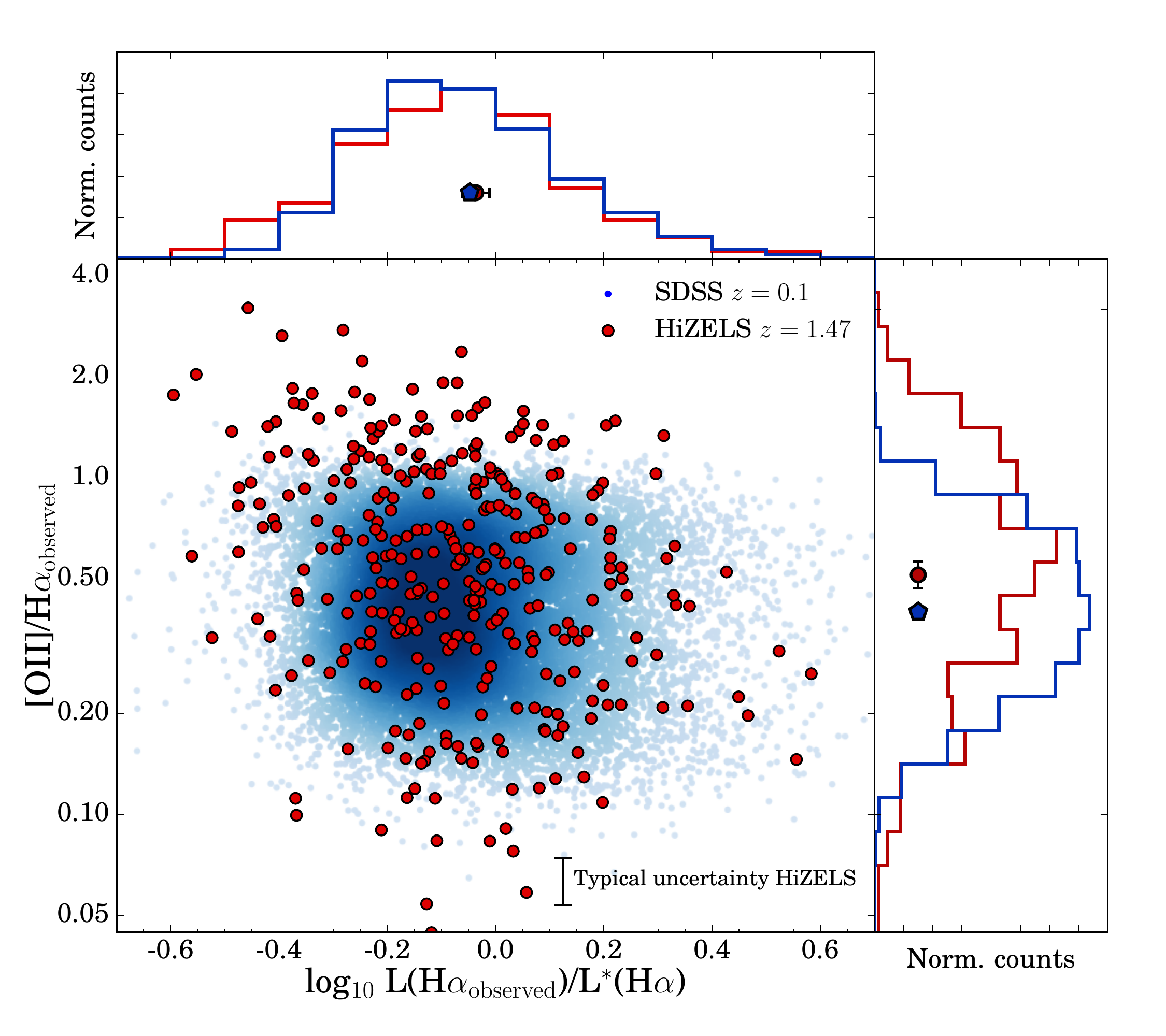} 
    \caption{Observed [O{\sc ii}]/H$\alpha$ ratio as a function of observed H$\alpha$ luminosity, normalized by the typical luminosity ($L^{\star}$) at either $z=1.47$ ($L^{\star} = 10^{42.16}$ erg s$^{-1}$, \citealt{Sobral2013}) or $z=0.1$ ($L^{\star} = 10^{41.4}$ erg s$^{-1}$, \citealt{Ly2007}). The blue points (coloured by density) show the ratios observed in SDSS at $z=0.1$, while the red points show the HiZELS measurements at $z=1.47$. The distribution of $L/L^{\star}$ is similar at both redshifts, while the median observed [O{\sc ii}]/H$\alpha$ increases from $0.40\pm0.01$ in the local Universe to $0.52\pm0.05$ at $z=1.5$. A KS test confirms that this increase is statistically significant. This increase could be due to evolution of the dust attenuation, or an effect from fiber-measurements in SDSS. }
    \label{fig:LO2LHa}
\end{figure}

\subsection{[OII]-H$\alpha$ view at $\bf z=1.47$} \label{sec:7}
One strength of the Bo\"otes-HiZELS survey is our sample of dual-emitters which can be used to study the relation between different star-formation rate indicators at $z=1.47$ and $z=2.23$, such as [O{\sc ii}], H$\alpha$ and continuum tracers such as the rest-frame UV, FIR and radio. Compared to H$\alpha$, the [O{\sc ii}] emission-line and UV continuum are more sensitive to dust attenuation and effects from metallicity and gas density \citep[e.g.][]{Kennicutt1998,Jansen2001,Ly2012}, which may all evolve with redshift. We exploit this sample to derive the  observed [O{\sc ii}]/H$\alpha$ ratio at $z=1.47$ and compare it to a reference sample from SDSS at $z=0.1$, to test claims based on smaller samples by e.g. \cite{Hayashi2012} and \cite{Sobral2012}.

We combine the sample of dual-emitters at $z=1.47$ in Bo\"otes with those from HiZELS in the UDS and COSMOS field (see \citealt{Sobral2013} for details), and remove any source that is detected in the X-rays. In total, this results in a sample of 340 dual-emitters at $z=1.47$. The majority of these are dominated by faint emitters observed in the deeper imaging in COSMOS. H$\alpha$ luminosities are corrected for the contribution from the adjacent [N{\sc ii}] doublet using the relation with EW described in \cite{Sobral2012} (see also \citealt{Sobral2015} for a spectroscopic validation). As a comparison sample at low redshift, we use a sample of emission-line measurements from a sample of star-forming galaxies at $z\approx0.1$ drawn from data from SDSS DR7 \citep{Abazajian2009} as described in \cite{Sobral2012}. In short, a sample of 16414 galaxies were selected at $0.07<z<0.1$ with observed H$\alpha$ luminosity $>10^{40.6}$ erg s$^{-1}$ and H$\alpha$ EW $>20$ {\AA}. For consistency with our sample at $z=1.47$, we do not remove AGN using the BPT diagnostic \citep{BPT}. Aperture corrections to emission-line measurements have been done following \cite{GarnBest2010} based on the ratio between the stellar mass in the fiber and the total stellar mass. We note that these corrections do not change line-ratios.

 \begin{figure*}
 \begin{tabular}{cc}
	\includegraphics[width=8.6cm]{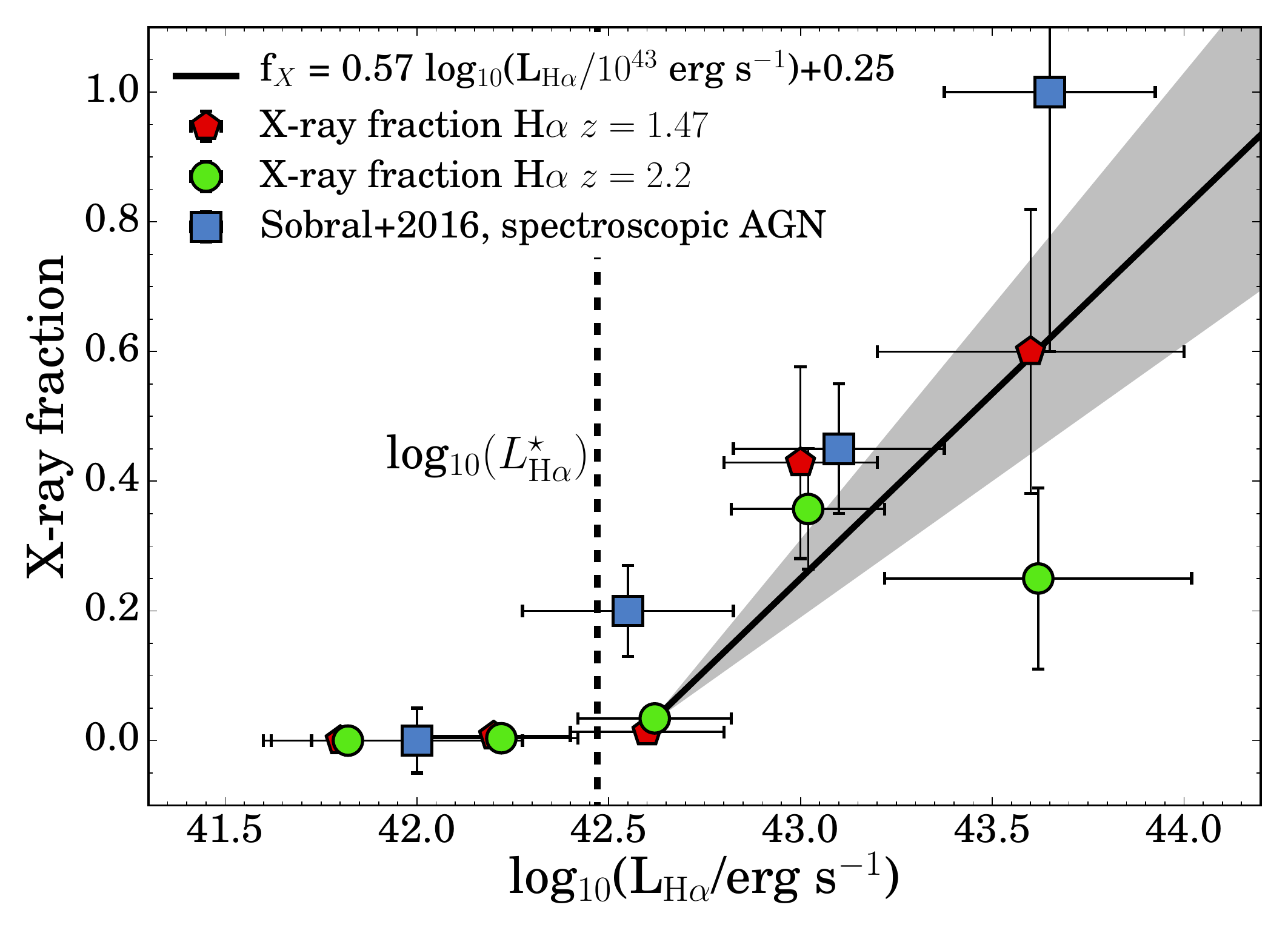}&
	\includegraphics[width=8.6cm]{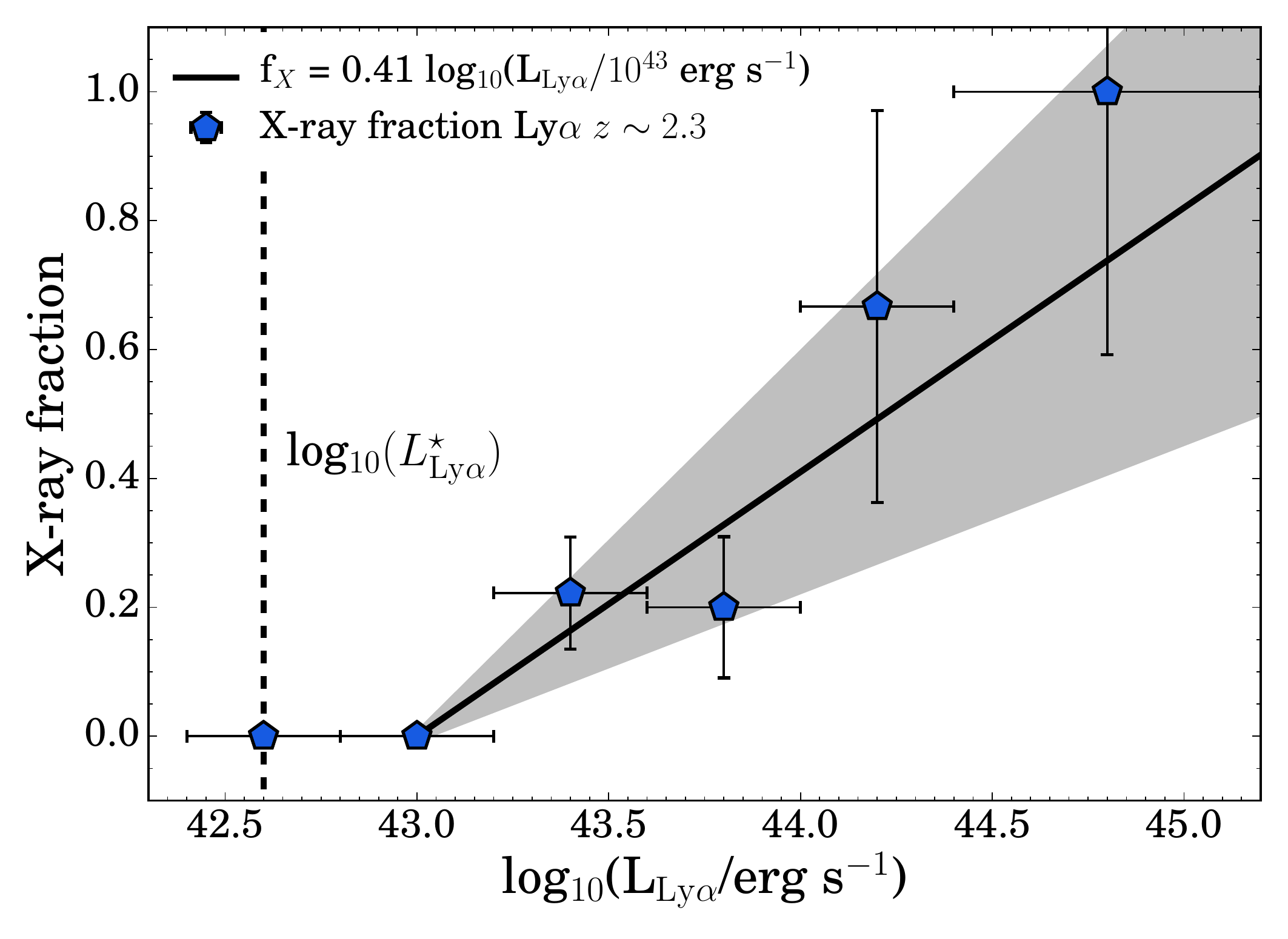}\\
\end{tabular}
    \caption{Left panel: X-ray fraction of HAEs combining Bo\"otes-HiZELS and the sample of emitters from COSMOS from HiZELS (\citealt{Sobral2013}) as a function of H$\alpha$ luminosity at $z=1.47-2.23$, compared to the AGN fraction measured with spectroscopy in \citet{Brizels}. The AGN fraction increases strongly with H$\alpha$ luminosity. At fixed H$\alpha$ luminosity, the observed X-ray fraction does not evolve strongly between $z=1.47-2.23$. Right panel: X-ray fraction of LAEs as a function of Ly$\alpha$ luminosity at $z\approx2.3$. The AGN fraction increases strongly with Ly$\alpha$ luminosity. The luminosities above which the X-ray fraction exceeds 20 \% correspond to the luminosities where the number densities start to diverge from Schechter, see e.g. Fig. $\ref{fig:LAE_LFs}$. 
}
    \label{fig:Xrayfrac}
\end{figure*}

In Fig. $\ref{fig:LO2LHa}$, we show the observed [O{\sc ii}]/H$\alpha$ ratio as a function of observed H$\alpha$ luminosity, normalized by the typical H$\alpha$ luminosity ($L^{\star}_{\rm H\alpha}$) at the specific redshift, both for the sample of dual-emitters and the local comparison sample. After correcting for the evolution in the typical H$\alpha$ luminosity of a factor of $\approx 6$, the distribution of H$\alpha$ luminosities is remarkably similar.
By computing the median ratio in 100,000 bootstrap resamples of the data, we measure [O{\sc ii}]/H$\alpha = 0.40\pm0.01$ with 95 \% confidence intervals at $z=0.1$ (slightly lower than the measurement of 0.45 in \citealt{Kennicutt1998}) and [O{\sc ii}]/H$\alpha = 0.55\pm0.07$, such that there is a slight increase of the median value with redshift (although the increase is within the observed scatter of $\approx 0.2-0.3$ dex; see also \citealt{Hayashi2012}). We note that our survey may miss the galaxies with lowest [O{\sc ii}]//H$\alpha$ ratio, in particular for the faintest H$\alpha$ emitters, which may result in a bias towards finding a higher [O{\sc ii}]/H$\alpha$ ratio at $z=1.47$. However, if we restrict the analysis to brighter sources ($>0.5\times L^{\star}_{\rm H\alpha}$), we find [O{\sc ii}]/H$\alpha = 0.52\pm0.05$ at $z=1.47$, while the SDSS results remain unchanged. This indicates that this selection effect is likely not driving the differences. A one dimensional Kolmogorov-Smirnov (KS) test of the observed [O{\sc ii}]/H$\alpha$ ratios confirms that the distributions are significantly different, with a KS-statistic of 0.20 ($\approx10^{-10}$ significance) and a P-value of $4\times10^{-11}$. This indicates that, even though the spread in values is relatively large ($\approx 0.4-0.5$ dex), the median observed [O{\sc ii}]/H$\alpha$ ratio increases slightly, but statistically significantly, between $z=0.1-1.47$. 

We test whether the observed difference can be caused by systematic errors. At $z=1.47$ there is a systematic uncertainty due to the relative filter transmissions at the different wavelengths, that leads to an increase in the scatter and a small bias towards higher [O{\sc ii}]/H$\alpha$ values. Based on the simulation that is discussed in detail in \cite{Sobral2012}, we estimate that this systematic increase is only of the order of $\approx 5$ \%, insufficient to explain the offset of the median ratio. If we remove AGN in the SDSS sample using the BPT criterion as defined in \cite{Kauffmann2003}, we find [O{\sc ii}]/H$\alpha = 0.42\pm0.01$. Finally, if we fully mimic the H$\alpha$ measurement (and its correction for the contribution of the [N{\sc ii}] in the narrow-band), we measure [O{\sc ii}]/H$\alpha = 0.45\pm0.01$. Thus, none of these effects can explain the observed difference, but they further highlight that the evolution is small, and thus only our large statistical sample can measure evolution.

A higher observed [O{\sc ii}]/H$\alpha$ ratio is expected when there is less attenuation due to dust, since [O{\sc ii}] is attenuated more than H$\alpha$ \citep[e.g.][]{Reddy2015}. For example, it could be that galaxies at $z=1.47$ are less dusty. Indeed, if we restrict the sample of local galaxies to those with A$_{\rm H\alpha} < 1.3$ (median A$_{\rm H\alpha} = 0.82$, compared to a median A$_{\rm H\alpha} = 0.91$ for the full sample), we find a similar observed [O{\sc ii}]/H$\alpha$ ratio of $0.49\pm0.01$. However, results from {\it Herschel} stacking of H$\alpha$ emitters at $z=1.47$ (\citealt{Thomson2017}; see also \citealt{Ibar2013}) indicate that their extinction properties are similar to local galaxies, with a similar relation between stellar mass and A$_{\rm H\alpha}$ \citep{GarnBest2010}. We note that because the samples are matched in L/L$^{\star}_{\rm H\alpha}$ there are likely no significant mass differences between the samples \citep[e.g.][]{Sobral2014}. A more detailed analysis of the extinction properties of this sample is beyond the scope of this paper.
Finally, another explanation is that the SDSS fiber-measurements are biased towards higher extinction (and thus lower [O{\sc ii}]/H$\alpha$ ratios), because they measure the line-ratios in the central 3-4 kpc of galaxies, that are observed to be dustier/more evolved \citep[e.g.][]{Sanchez2014}, while the 3$''$ measurements at $z=1.47$ measure flux out to radii of 8-13 kpc (depending on UDS/COSMOS or Bo\"otes). Therefore, the observed offset between $z=0.1$ and $z=1.47$ could also be an observational effect due to dust, age and/or metallicity gradients within galaxy. These observational issues can be overcome with large IFU or matched NB surveys in the local Universe, as for example the J-PAS project \citep{JPAS}.

\subsection{X-ray fraction \& the power-law component of the luminosity function} \label{sec:5.2}  
We investigate the AGN fractions of H$\alpha$ emitters at $z=1.47$ and $z=2.23$ and LAEs at $z=2.4$ by matching our samples to source catalogs from X-ray ({\it Chandra}, 0.5--7.0 keV, depth $7.8\times10^{-15}$ erg s$^{-1}$ cm$^{-2}$, 1$''$ resolution and matching radius, \citealt{Kenter2005}). In addition, we include the H$\alpha$ emitters at $z=1.47$ and $z=2.23$ in the COSMOS field from \cite{Sobral2013} and match these with the {\it Chandra} COSMOS point source catalog (0.5--7.0 keV, $5.7\times10^{-16}$ erg s$^{-1}$ cm$^{-2}$, \citealt{Elvis2009}); see also \cite{Calhau2017} for their detailed properties. These X-ray flux limits correspond to luminosity limits of 0.3--3$\times10^{44}$ erg s$^{-1}$. If such an X-ray luminosity would have its origin in star formation, it would require SFRs of $>10^{3.5-5}$ M$_{\odot}$ yr$^{-1}$ \citep[e.g.][]{Lehmer2016}, which is unlikely. This clearly indicates an AGN origin of all X-ray detections discussed in this section.

In total, we detect 21 HAEs in the X-ray, of which 10 are at $z=1.47$ and 11 are at $z=2.23$ and half of the X-ray detected HAEs are in Bo\"otes. We find that the detection rate depends strongly on the H$\alpha$ luminosity, see Fig. $\ref{fig:Xrayfrac}$. This has also been observed using spectroscopic follow-up by \cite{Brizels}, who found that the majority ($80\pm30$ \%) of luminous HAEs are broad-line AGN. Relatively independent of redshift, roughly half of the most luminous HAEs are X-ray detected. Note that the other half may easily be undetected due to the short duty cycle of X-ray AGN \citep[e.g.][]{Shankar2009,Fiore2012}, although it could also indicate that roughly half of the luminous HAEs are optically thick to X-rays. By combining the data-points above L$^{\star}$ at $z=1.47-2.23$, we find a best fit $f_{X} = 0.57^{+0.15}_{-0.15} $log$_{10}$(L$_{\rm H\alpha}/10^{43}$ erg s$^{-1}$) + 0.25$^{+0.06}_{-0.06}$, that we illustrate in Fig. $\ref{fig:Xrayfrac}$. 

We combine the sample of LAEs at $z=2.2-2.4$ identified with the NB392 and the stV filter to investigate the X-ray fraction of LAEs as a function of luminosity. Out of the 41 LAEs, eight are X-ray detected (L$_X \gtrsim 3\times10^{44}$ erg s$^{-1}$). The X-ray fraction of LAEs increases strongly with line luminosity, from $\approx 0$\% at L$^{\star}_{\rm Ly\alpha}$ to $\approx 100$ \% at $\gtrsim3\times10^{44}$erg s$^{-1}$. We note that the Ly$\alpha$ luminosities at which the X-ray fraction exceeds 20 \% correspond to the luminosities at which the number densities start to deviate from a Schechter function, as observed in \cite{Konno2016}, \cite{Sobral2016} and this work (Fig. $\ref{fig:LAE_LFs}$). For both H$\alpha$ and Ly$\alpha$, the X-ray fraction increases above L$^{\star}$. For Ly$\alpha$, we find a best fit relation of: $f_{X} = 0.41^{+0.18}_{-0.18} $log$_{10}$(L$_{\rm Ly\alpha}/10^{43}$ erg s$^{-1}$) above L$_{\rm Ly\alpha} > 10^{43}$ erg s$^{-1}$.

\section{ Conclusions} \label{sec:8}
We presented the first results from the Bo\"otes-HiZELS survey, which uses six narrow-bands to select emission-line galaxies from $z=0.4-4.7$ in a 0.7 deg$^2$ region in the Bo\"otes field. We described the observations, data-reduction, extraction of catalogs and selection of line-emitters, and how multi-wavelength data has been used to classify different populations of line-emitters. The main results are:

\begin{enumerate}
\item We identify 362 candidate H$\alpha$ emitters (HAEs) at $z=0.4, 1.47, 2.23$, 387 H$\beta$/{\sc [Oiii]} emitters at $z=0.8, 2.23, 3.3$, 285 {\sc [Oii]} emitters at $z=1.47, 3.3, 4.7$ and 73 Ly$\alpha$ emitters (LAEs) at $z=2.23, 2.3, 3.1$. 
\item Using a suite of matched narrow-band filters, we identify 42 galaxies with emission-lines in multiple narrow-bands, providing 22/18 new robust redshift identifications of {\sc [Oii]}/H$\alpha$ and {\sc [Oiii]}/H$\alpha$ emitters at $z=1.47/2.23$, without pre-selection on AGN activity or $I$ band magnitude, see \S$\ref{sec:5}$ and Table $\ref{tab:dualemitters}$. 56 additional line-emitters have a spectroscopic redshift.
\item In general, the number densities of line-emitters as a function of luminosity we derive agree remarkably well with luminosity functions observed in other survey fields (\S$\ref{sec:6}$, Figures $\ref{fig:LAE_LFs}$, $\ref{fig:HAE_LFs}$ and $\ref{fig:O2_LF}$), confirming strong evolution in L$^{\star}_{\rm H\alpha}$ from $z=0.4-2.2$ and evolution in L$^{\star}_{\rm Ly\alpha}$ from $z=2.2-3.1$. 
\item We confirm the result from \cite{Konno2016} and \cite{Sobral2016} that the luminosity function of LAEs at $z\approx2.2$ diverges from a Schechter function at the bright end, L$_{\rm Ly\alpha} \gtrsim 10^{43}$ erg s$^{-1}$. At these luminosities, the luminosity function follows log$_{10}(\Phi) = 27.5 - 0.74$log$_{10}$(L$_{\rm Ly\alpha}$).  Such a departure from a Schechter function is also clearly observed at the highest H$\alpha$ luminosities (L$_{\rm H\alpha} \gtrsim 10^{43.5}$ erg s$^{-1}$) at $z=2.2$ (\S$\ref{sec:6.2}$ and \S$\ref{sec:6.2.2}$). 
\item Combining our sample of dual-emitters with those from the COSMOS and UDS fields from HiZELS, we compare the observed [O{\sc ii}]/H$\alpha$ ratio of 340 star-forming galaxies at $z=1.47$ with those from a reference sample in the local Universe (\S$\ref{sec:7}$). We measure a median ratio of [O{\sc ii}]/H$\alpha = 0.40\pm0.01$ at $z=0.1$ and [O{\sc ii}]/H$\alpha = 0.55\pm0.07$ ([O{\sc ii}]/H$\alpha = 0.52\pm0.05$ if we restrict the sample to sources with slightly higher S/N, see Fig. $\ref{fig:LO2LHa}$). The $\approx 0.1$ dex offset can potentially be attributed to a lower dust attenuation at $z=1.47$, or biases in the fiber-measurements in the local Universe, which measure the ratio at the central 3-4 kpc of galaxies, while the measurements at $z=1.47$ are integrated over $\approx 10$kpc.
\item By exploiting {\it Chandra} X-Ray data, we show that the H$\alpha$ and Ly$\alpha$ luminosities at which the number densities start to diverge from pure Schechter form at similar luminosities to where the X-ray fractions start to increase, from $\sim 20$ \% to $\sim100$ \% (Fig. $\ref{fig:Xrayfrac}$). We also show that, under basic assumptions, the majority of luminous LAEs are not broad-line Type I AGN (Fig. $\ref{fig:LAE_LFs}$), and more likely narrow-line Type II AGN. 
\end{enumerate}

The sample of identified line-emitters can be used to study various properties of star-forming galaxies. In particular, the relatively large sample of H$\alpha$ emitters at $z=1.5-2.2$ can be used to test various SFR indicators (H$\alpha$, rest-UV, radio, FIR) in future papers. 

\section*{Acknowledgments}
We thank the referee for their constructive comments which helped improve the quality of this work. JM acknowledges the support of a Huygens PhD fellowship from Leiden University and JM and DS acknowledge financial support from a NWO/VENI grant awarded to David Sobral. IRS acknowledges support from STFC (ST/L00075X/1), the ERC Advanced Grant DUSTYGAL (321334) and a Royal Society/Wolfson Merit Award. HR acknowledges support from the ERC Advanced Investigator program NewClusters 321271. PNB is grateful to STFC for support via grant ST/M001229/1. B.D. acknowledges financial support from NASA through the Astrophysics Data Analysis Program (ADAP), grant number NNX12AE20G. We thank Ana Afonso, Jo\~ao Calhau, Leah Morabito, Iv\'an Oteo, S\'ergio Santos and Aayush Saxena for their assistance with observations. This work is based on observations obtained using the Wide Field Camera (WFCAM) on the 3.8m United Kingdom Infrared Telescope (UKIRT), as part of the High-redshift(Z) Emission Line Survey (HiZELS; U/CMP/3 and U/10B/07), using Suprime-Cam on the 8.2m Subaru Telescope as part of program S14A-086 and using the WFC on the 2.5m Isaac Newton Telescope, as part of programs 2013AN002, 2013BN008, 2014AC88, 2014AN002, 2014BN006, 2014BC118 and 2016AN001, using ISIS and AF2+WYFFOS on the 4.2m William Herschel Telescope, as part of programs 2016AN004 and 2016BN011 and using DEIMOS on the 10m Keck II Telescope as part of program C267D and on observations made with ESO Telescopes at the La Silla Paranal Observatory under ESO programme IDs 098.A-0819 and 179.A-2005. This work made use of images and/or data products provided by the NOAO Deep Wide-Field Survey \citep{Jannuzi1999} which is supported by the National Optical Astronomy Observatory (NOAO). NOAO is operated by AURA, Inc., under a cooperative agreement with the National Science Foundation.
We have benefited greatly from the public available programming language {\sc Python}, including the {\sc numpy, matplotlib, pyfits, scipy} and {\sc astropy} packages \citep{ASTROPY}, the astronomical imaging tools {\sc SExtractor, Swarp} and {\sc Scamp} and the {\sc Topcat} analysis program \citep{Topcat}.




\bibliographystyle{mnras}

\bibliography{bib_bmns.bib}


\appendix
\section{Photometric consistency check} \label{sec:3.2.2} 
In our catalogue production steps (\S $\ref{sec:3.2}$), we only include objects in our catalogues that have a physically plausible NB excess. Because the BB covers the same wavelength as the NB, flux in the NB must also be observed in the BB, otherwise the object is either unreal (such as a cosmic ray, artefacts, etc.) or variable (such as variable stars, supernovae or AGN). It is fairly straightforward to compute the faintest possible BB magnitude given a NB magnitude: 
\begin{equation}
BB_{max} = NB - 2.5 {\rm log}_{10}(\frac{\lambda_{c,BB}^2 \Delta\lambda_{NB}}{\lambda_{c,NB}^2 \Delta\lambda_{BB}} ) + 0.5
\end{equation}
in this equation, $\lambda_{c, X}$ is the central wavelength of filter $X$, and $\Delta\lambda_{X}$ the width of filter $X$. We conservatively add 0.5 magnitude to take into account uncertainties in the photometry and relative filter transmissions. We remove any source for which the excess is larger than $BB_{max} - NB$. For example, for NB921 and $z$ band, this equation results in $BB_{max} = NB + 3.4$. This means that if a source has a NB921 magnitude of 20, it must have a $z$ magnitude of 23.4 or brighter. It is possible that the implied $BB_{max}$ is below the background. In that case, we exclude sources for which the BB is not detected at 2$\sigma$. This consistency check removes most spurious objects such as cosmic rays and detector artefacts.

\section{Colour-colour selections} \label{sec:colsel}
\begin{figure*}
\begin{tabular}{cc}
	\includegraphics[width=8.6cm]{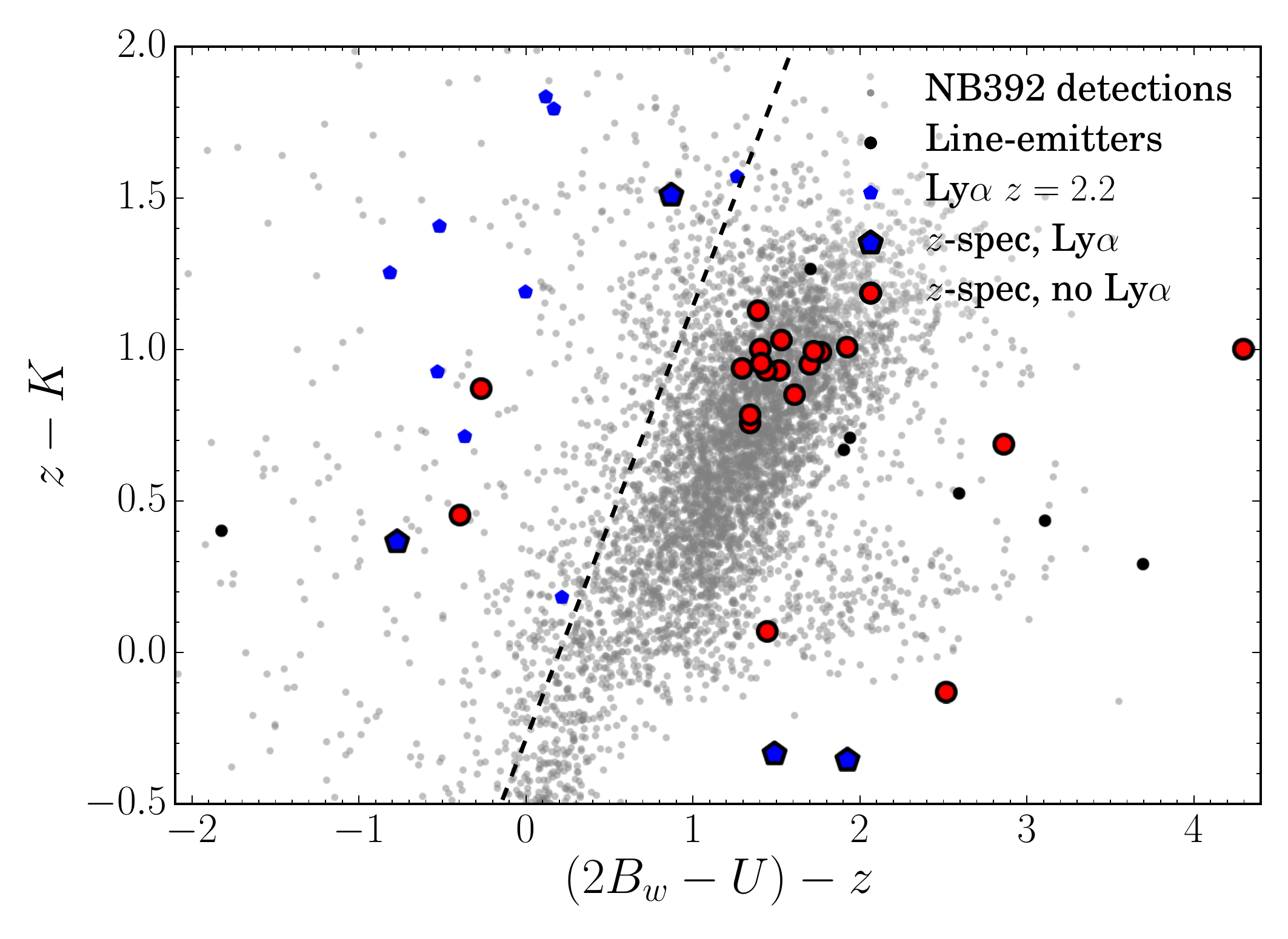} &
			\includegraphics[width=8.6cm]{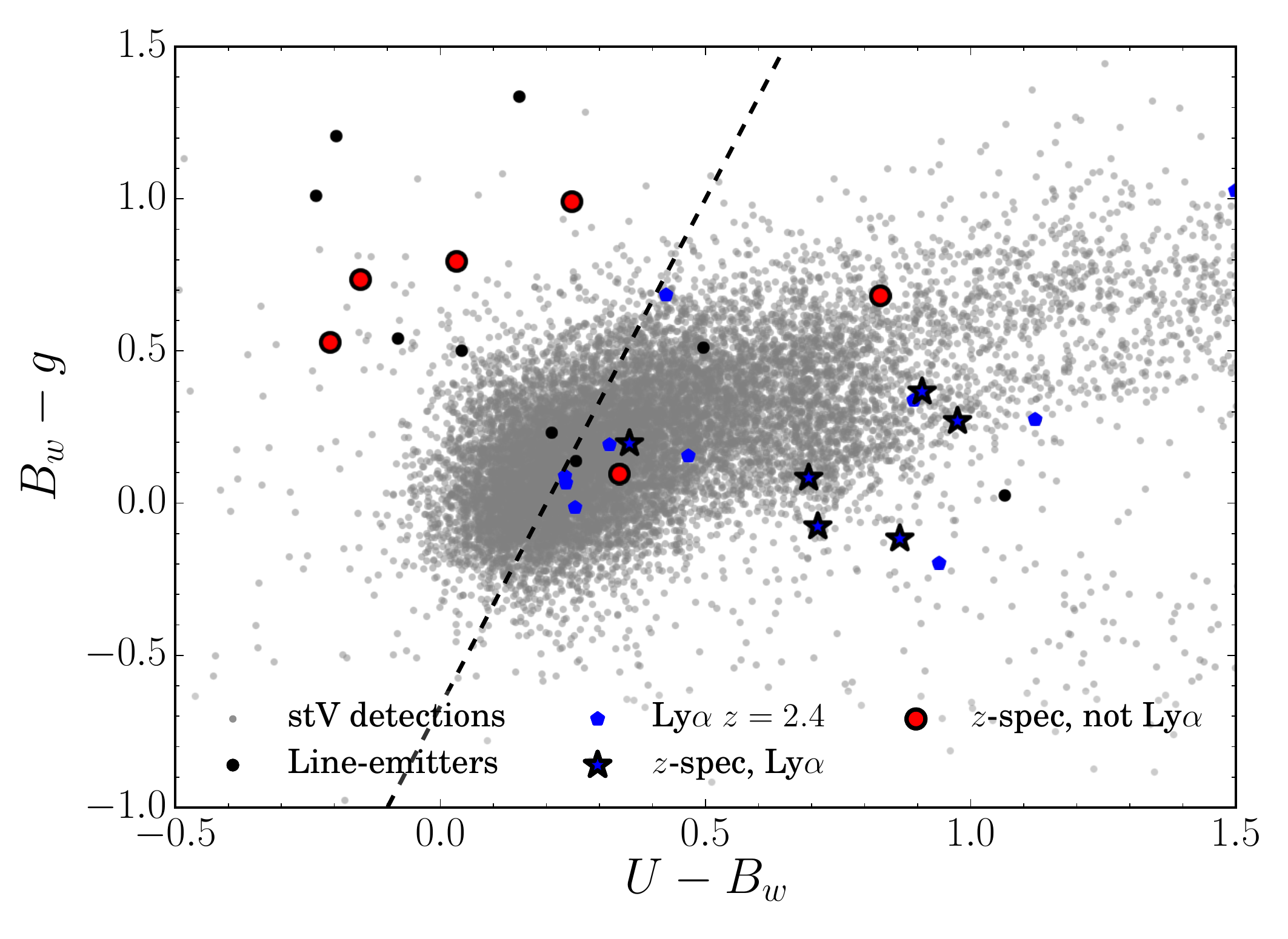} \\
	\end{tabular}
    \caption{Colour selection criteria used to select LAEs at $z=2.2$ and $z=2.4$. These are based on the BzK criterion from \citet{Daddi2004}. The $B_w$ band magnitude is adjusted for the contribution to flux in the $U$ band. We highlight the sources with spectroscopic redshifts in larger symbols.}
    \label{fig:colsel_LAEz2}
\end{figure*}

\begin{figure*}
\begin{tabular}{cc}

	\includegraphics[width=8.6cm]{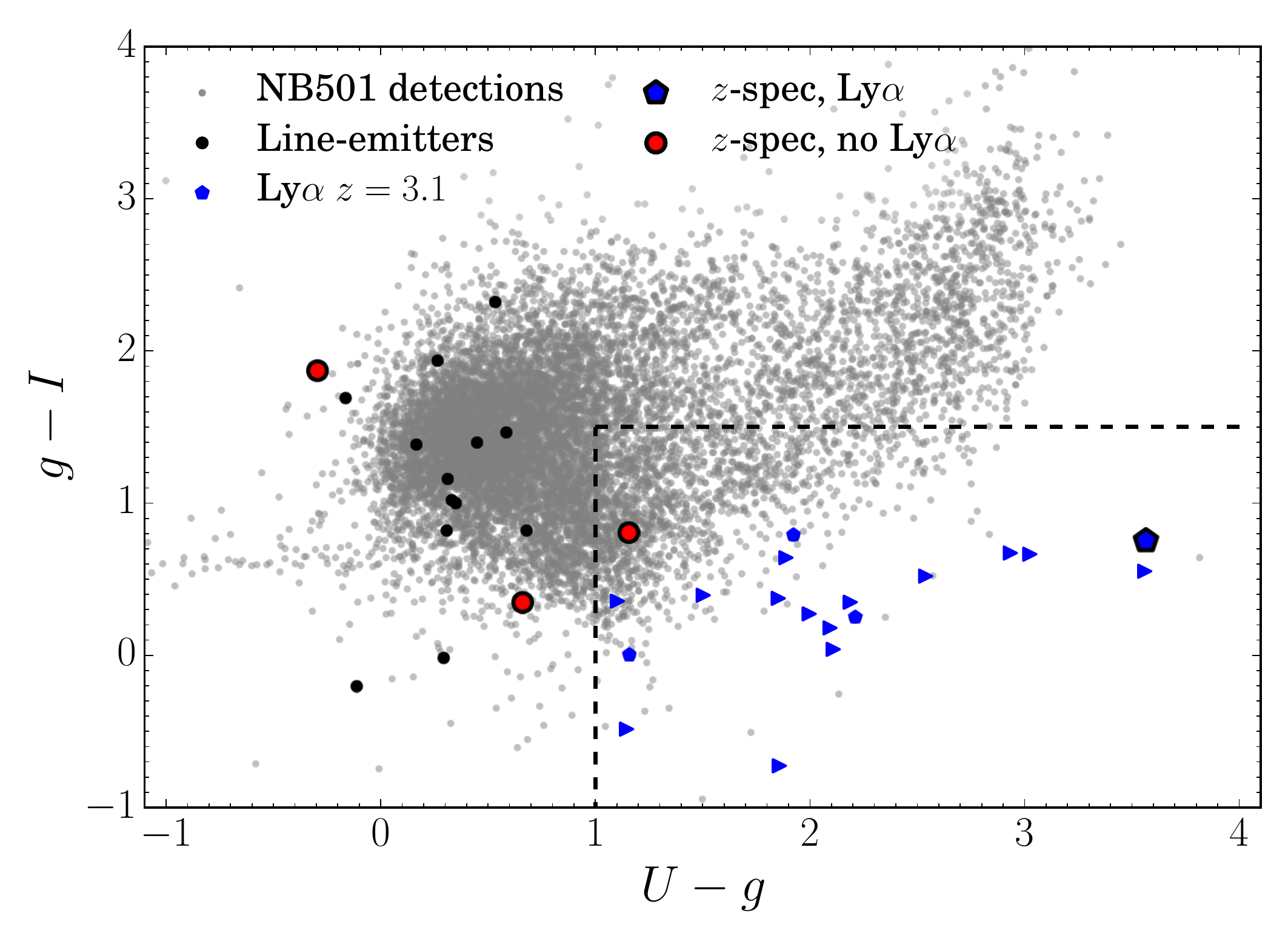} &
		\includegraphics[width=8.6cm]{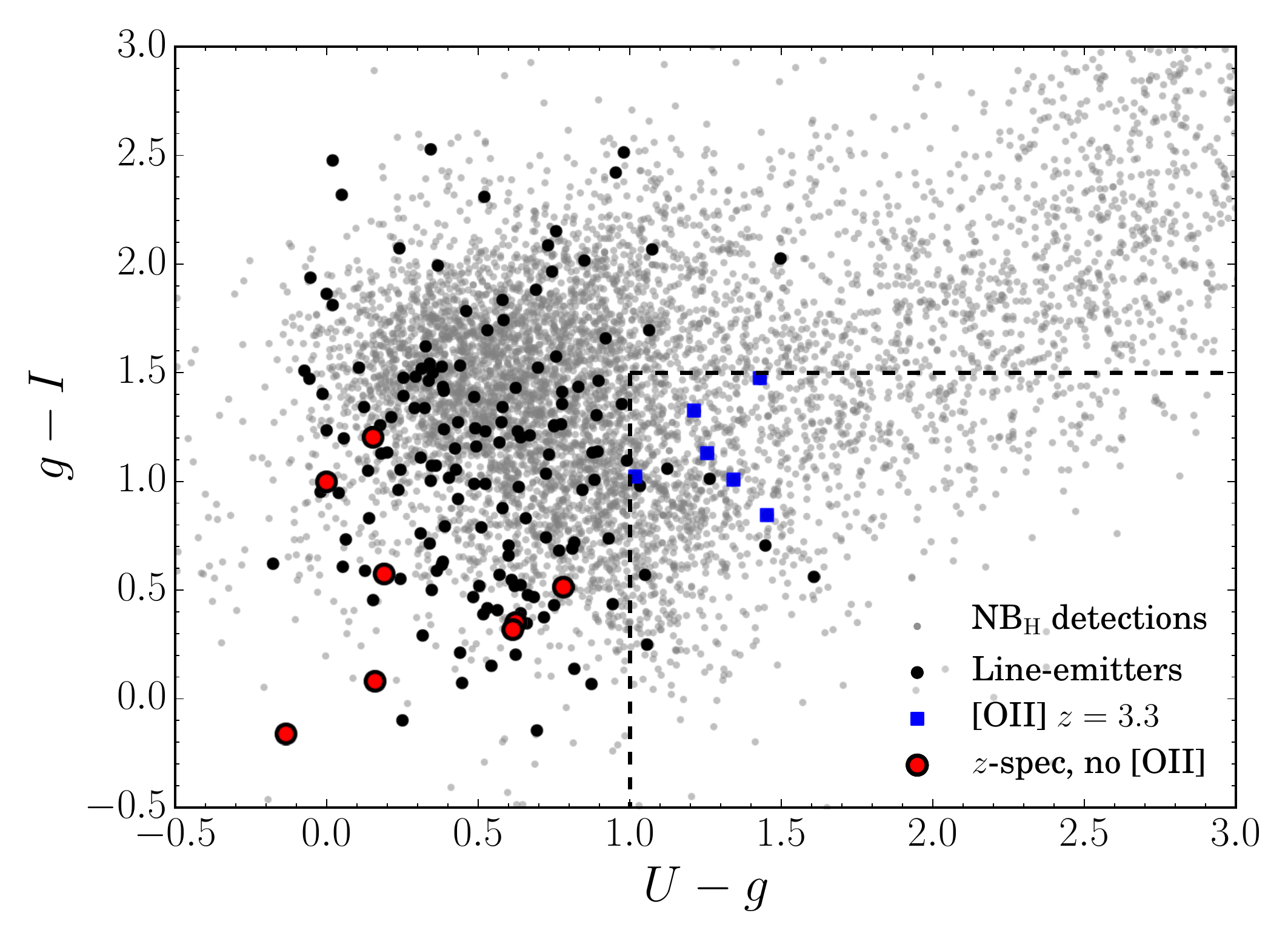} \\
\end{tabular}
    \caption{Colour selection criterion used to select LAEs at $z=3.1$ (from NB501) and [O{\sc iii}]/H$\beta$ at $z=3.2$ (from NB$_{\rm H}$), based on the $U$ drop-out criterion for Lyman-break galaxies at $z\approx3$ from \citet{Hildebrandt2009}. The $U-g$ colour identifies the Lyman-break, while the $g-I$ criterion removes any sources for which the Balmer break is mimicked by a strong Balmer break (particularly important in the case a galaxy is very dusty).}
    \label{fig:colsel_LAEz3}
\end{figure*}

\begin{figure*}
\begin{tabular}{cc}
	\includegraphics[width=8.6cm]{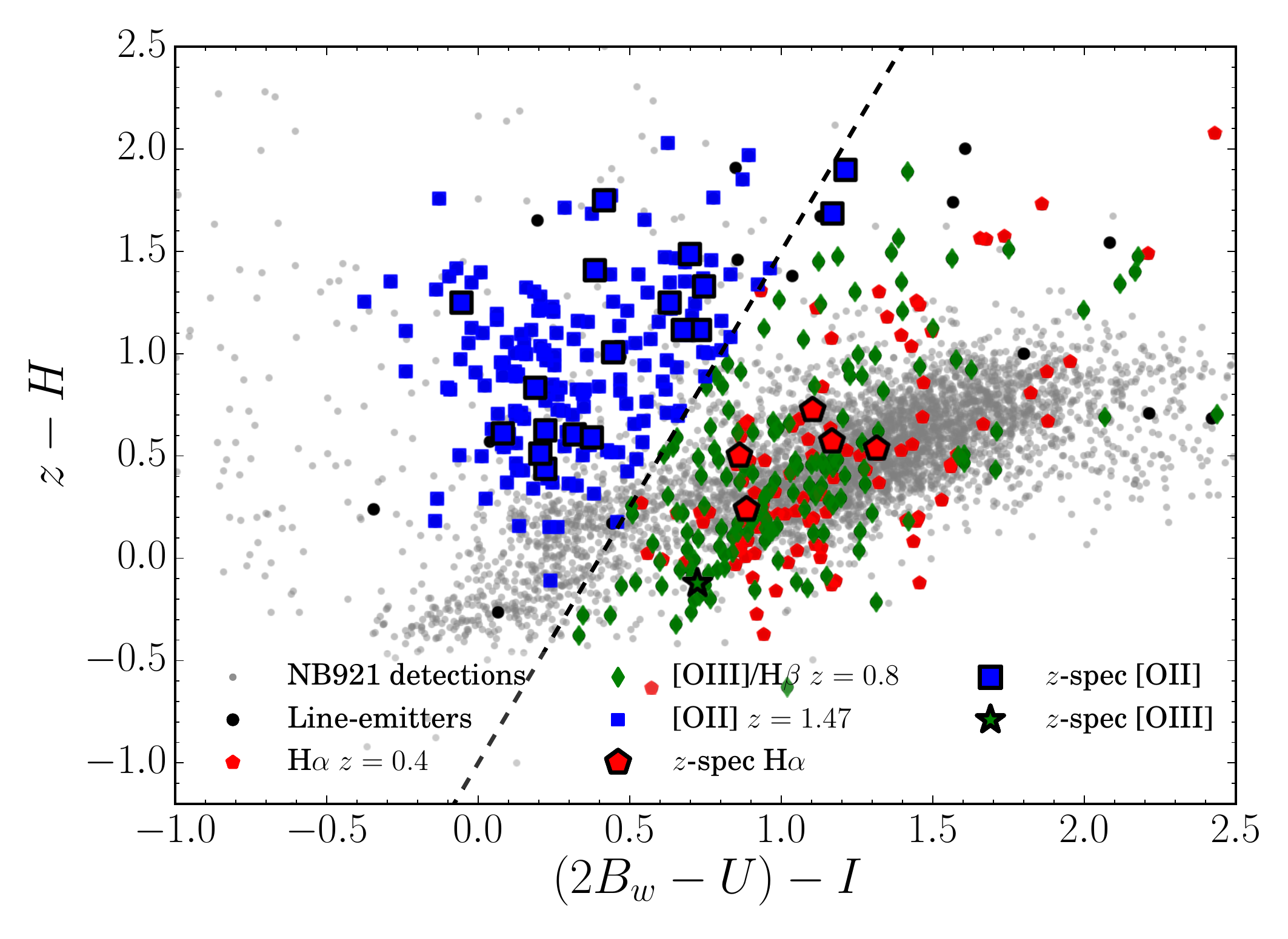} &
	\includegraphics[width=8.6cm]{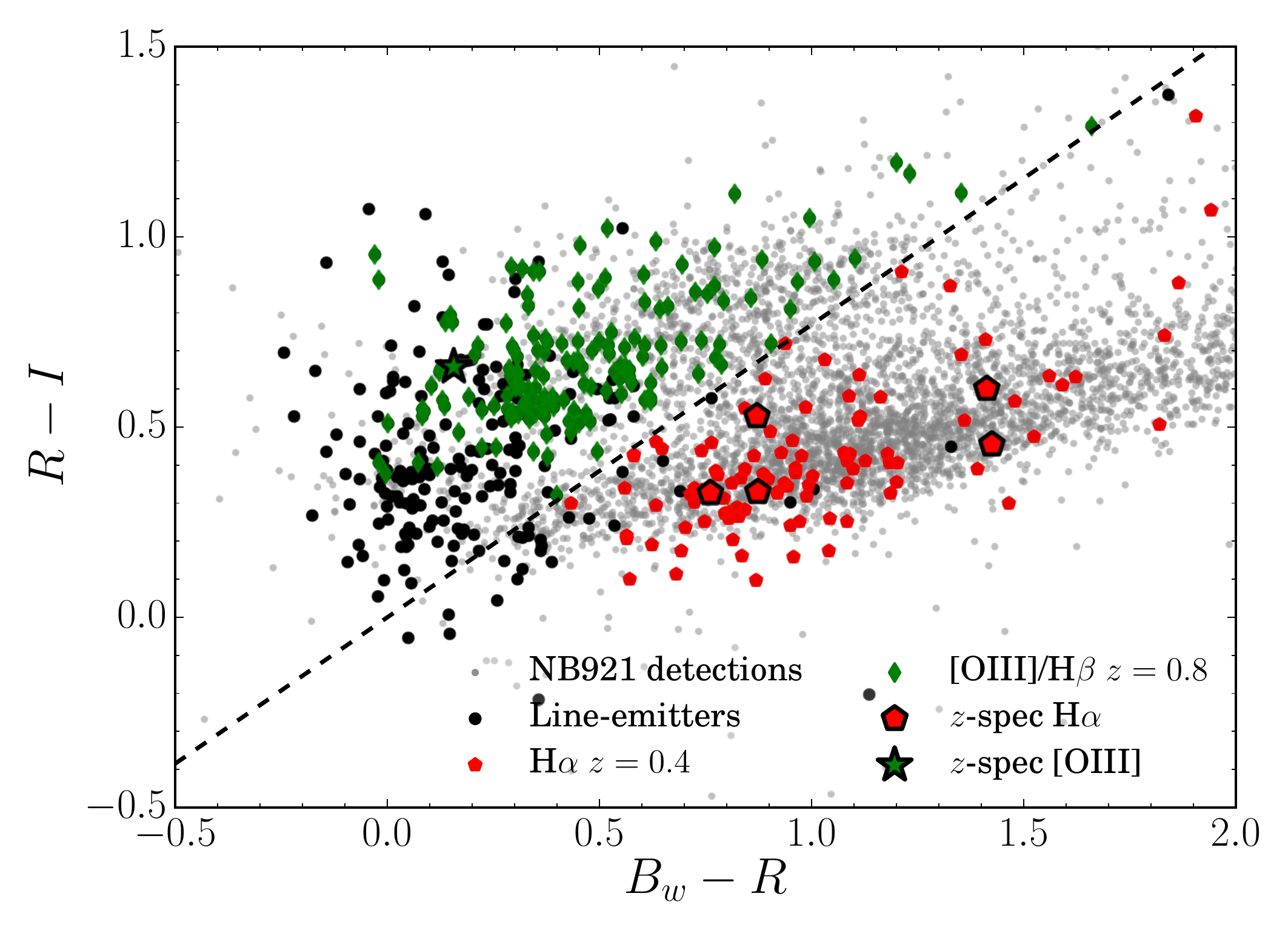} \\
	\end{tabular}
    \caption{Colour selection criteria used to distinguish H$\alpha$, [O{\sc iii}]/H$\beta$ and [O{\sc ii}] emitters among the line-emitters identified with NB921. The criterion in the left panel identifies Balmer breaks at specific redshift intervals. Because this criterion can not distinguish between $z=0.4$ and $z=0.8$, we use the additional $B_wRI$ colours to distinguish between a Balmer break between $B_w$ and $R$ (around 600 nm, or $z\approx0.5$) and between $R$ and $I$ (around 700 nm, or $z\approx0.75$).}
    \label{fig:colsel_NB921}
\end{figure*}

\begin{figure*}
\begin{tabular}{cc}
	\includegraphics[width=8.6cm]{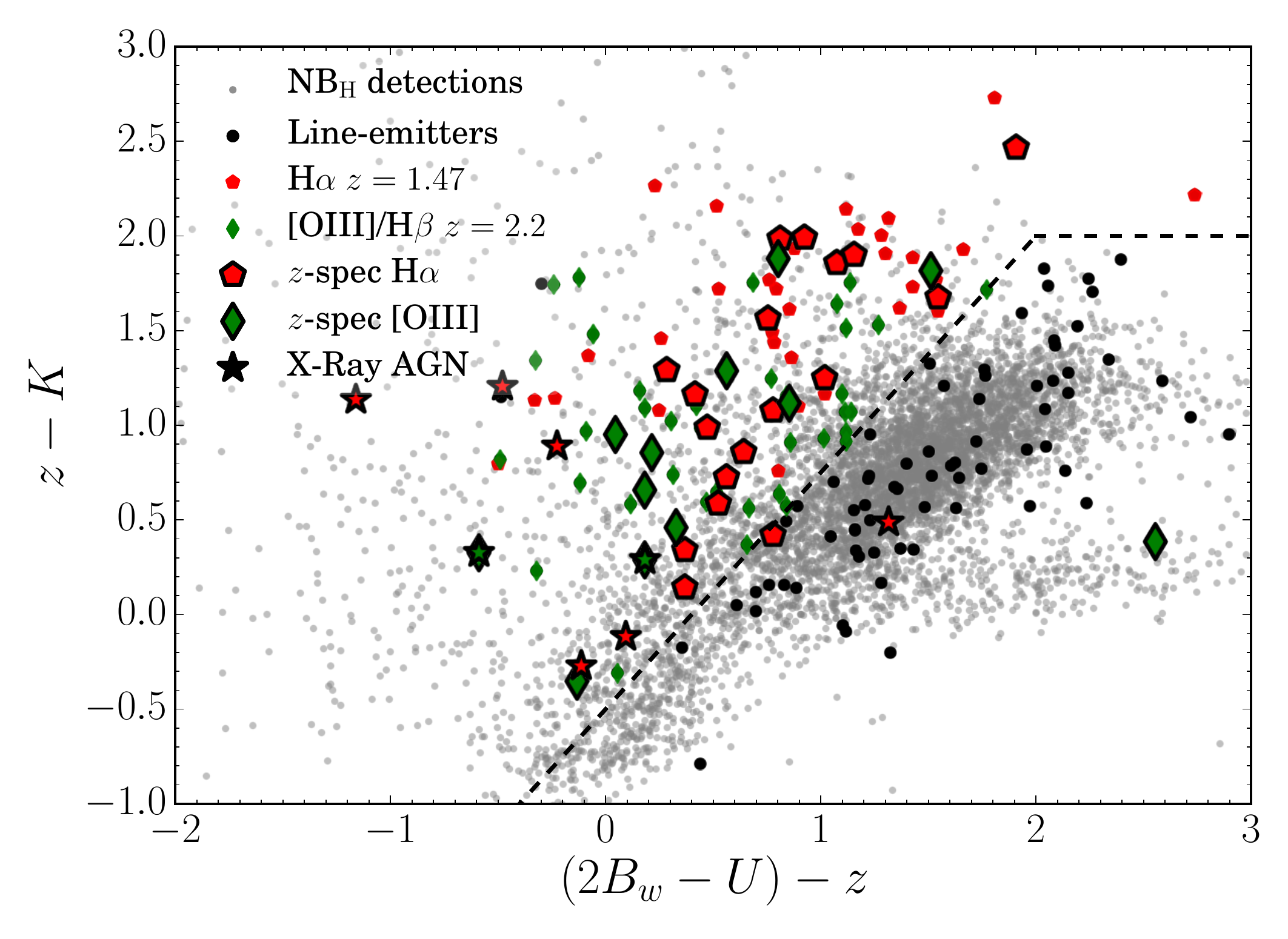} &
	\includegraphics[width=8.6cm]{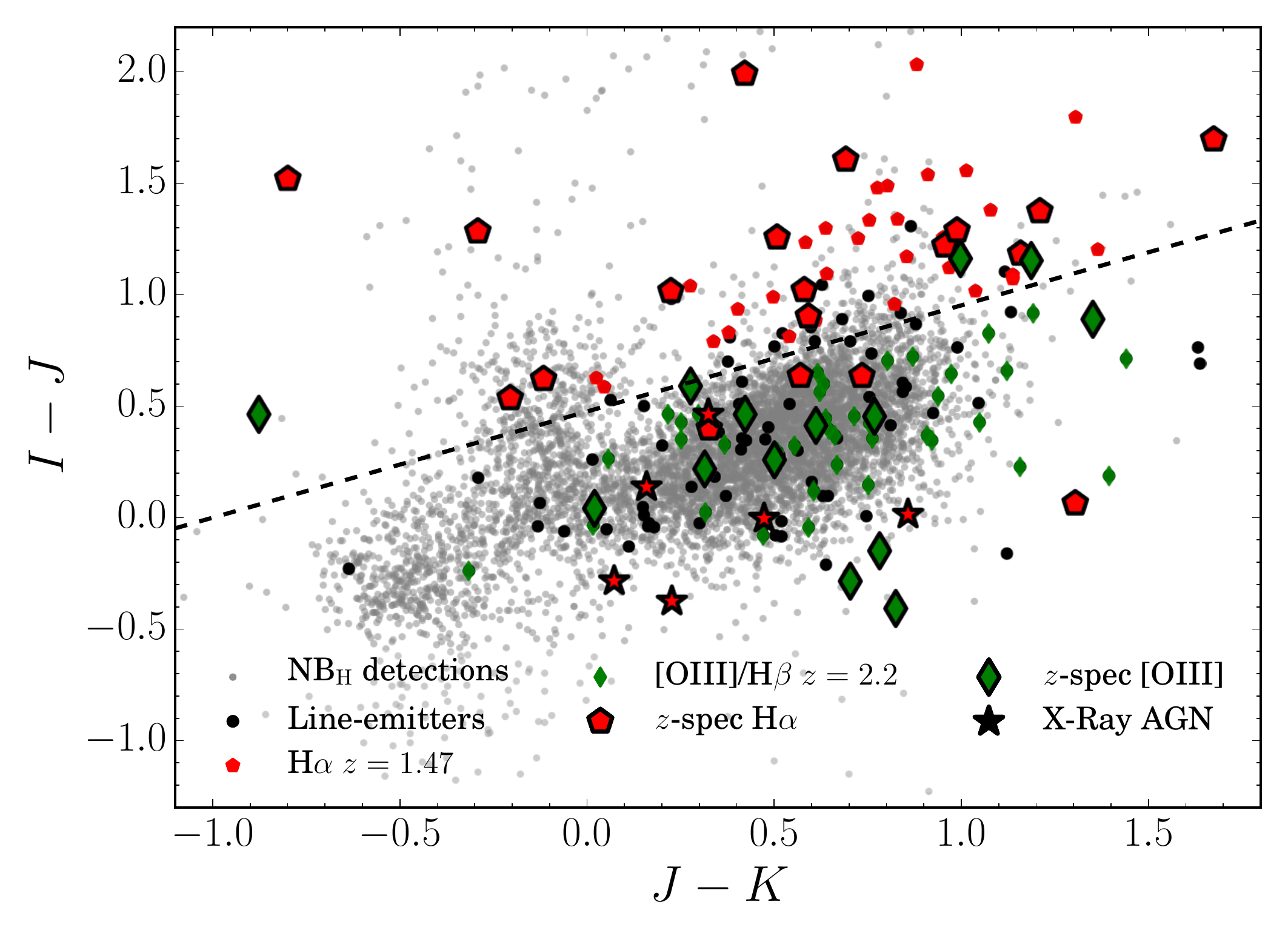} \\
	\end{tabular}
    \caption{Colour criteria used to classify line-emitters identified with the NB$_{\rm H}$ filter. The modified $B_w z K$ criterion is used to distinguish $z>1$ line-emitters from $z<1$ line-emitters. We then distinguish between $z=1.47$ and $z=2.2$ based on the positions of Balmer breaks. At $z\approx1.5$, the Balmer break lies around an observed 1000 nm, hence between $I-J$ (note that we use $I$ rather than $z$ because $I$ is deeper and broader, see Table $\ref{tab:numbers}$). At $z\approx2.2$ the Balmer break lies between the $J$ and $K$ bands. Note that several X-ray detected H$\alpha$ emitters at $z=1.47$ mimic the colours of $z=2.2$ [O{\sc iii}]/H$\beta$ emitters. These sources typically have high line-fluxes and their spectroscopic follow-up completeness is high. [O{\sc ii}] emitters at $z=3.3$ are selected based on the position of their Lyman-break, similar to LAEs at $z=3.1$ in Fig. $\ref{fig:colsel_LAEz3}$.}
    \label{fig:colsel_NBH}
\end{figure*}

\begin{figure*}
\begin{tabular}{cc}
	\includegraphics[width=8.6cm]{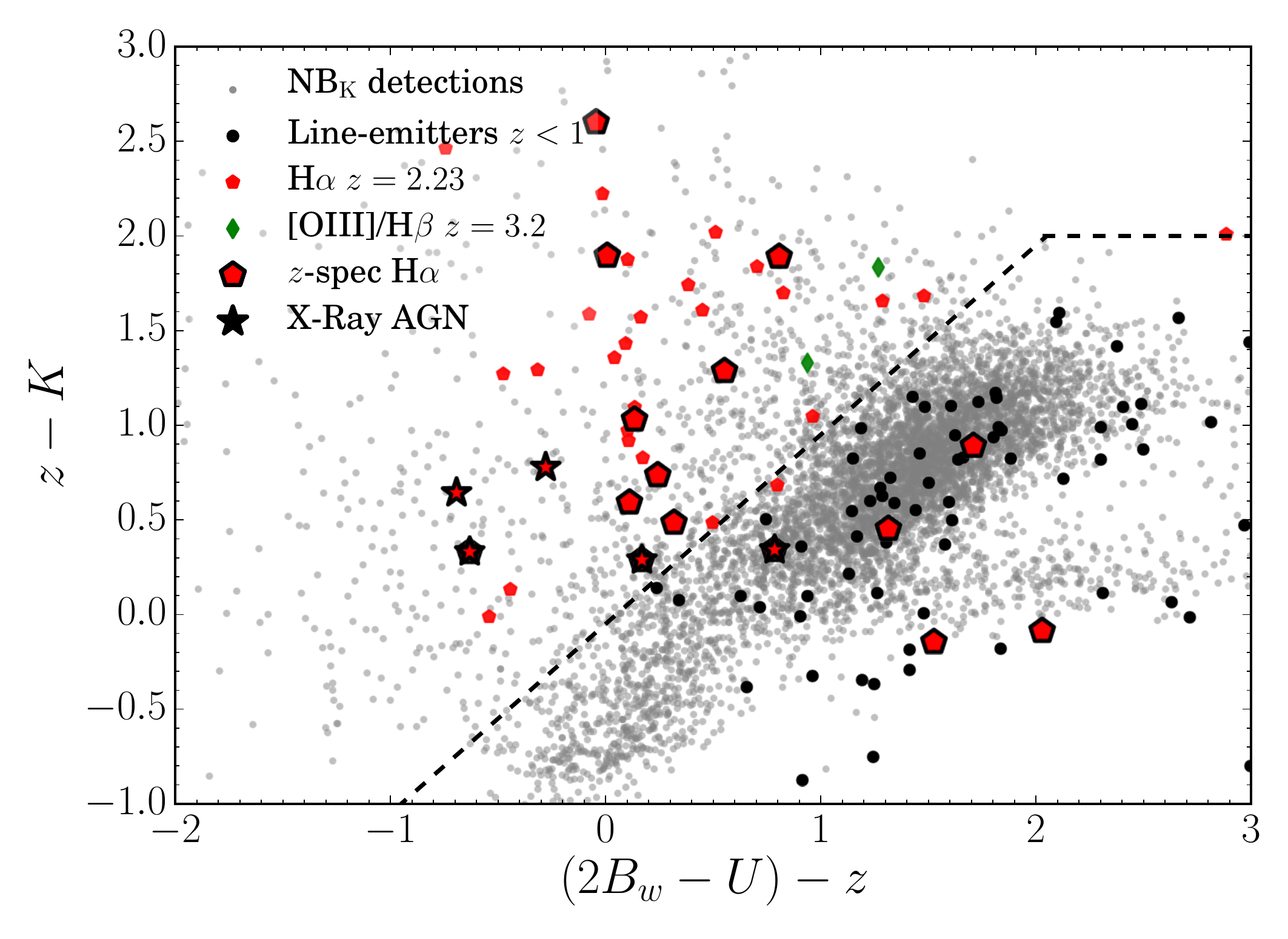} &
	\includegraphics[width=8.6cm]{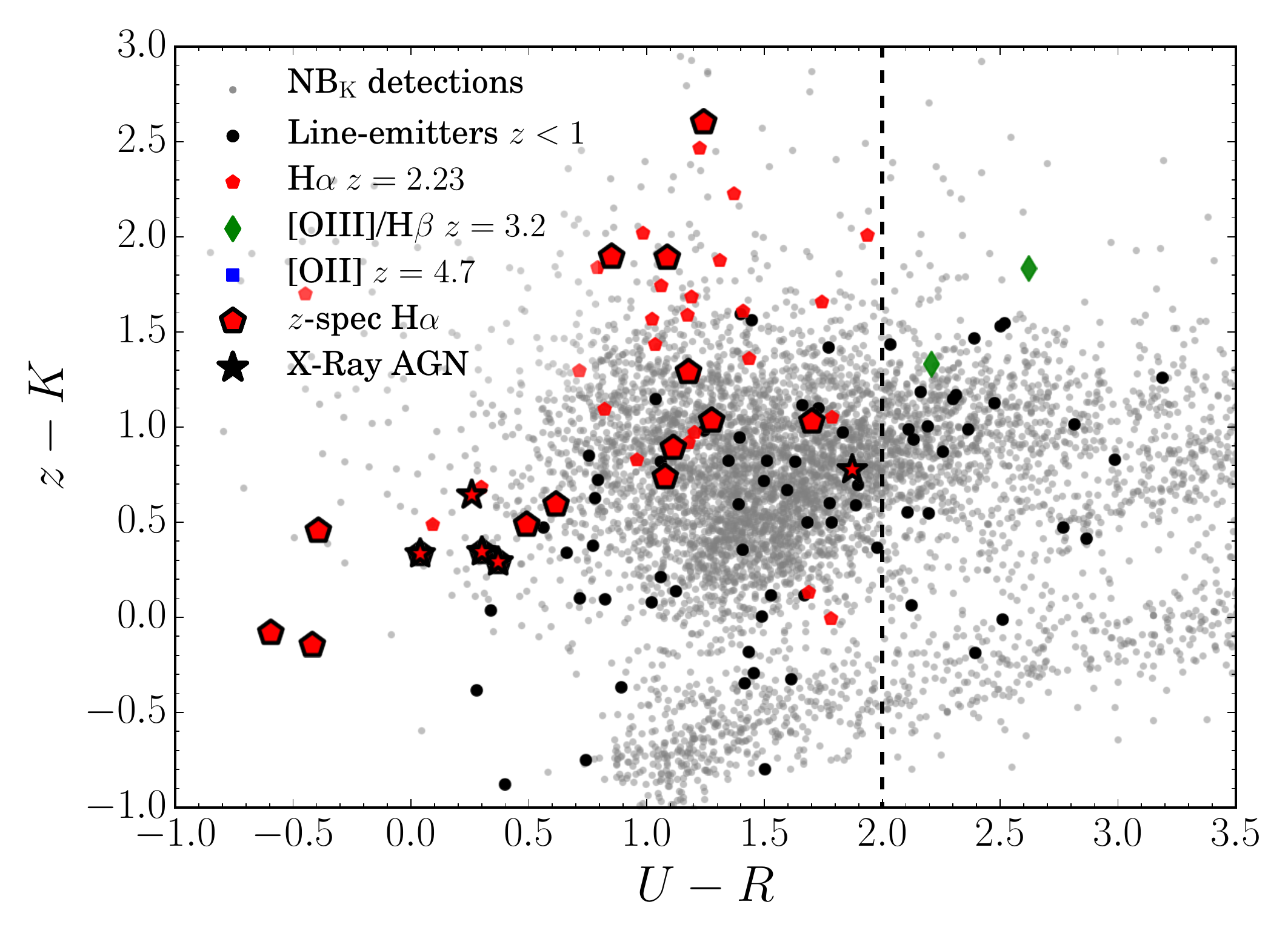} \\
	\end{tabular}
    \caption{Colour criteria used to classify line-emitters identified with the NB$_{\rm K}$ filter. Similarly to the line-emitters in the NB$_{\rm H}$ filter, we use the modified $B_w z K$ criterion to selected line-emitters at $z>1$ (left panel). We then distinguish those between H$\alpha$ at $z=2.23$ and  [O{\sc iii}]/H$\beta$ emitters at $z=3.2$ based on the positions of their Lyman-breaks.}
    \label{fig:colsel_NBK}
\end{figure*}

In this section we illustrate the colour-colour selection criteria outlined in Table $\ref{tab:criteria}$. In all figures (Fig. $\ref{fig:colsel_LAEz2}$, $\ref{fig:colsel_LAEz3}$, $\ref{fig:colsel_NB921}$, $\ref{fig:colsel_NBH}$ and $\ref{fig:colsel_NBK}$), line-emitters with secure redshift-identifications from spectroscopy or dual-NB detections are shown with larger symbols. X-ray detected AGN are shown with a star symbol.

\section{Catalogue of dual-emitters}
\begin{table*}
\caption{List of sources that are observed as dual-emitters (line-emitters in at least two narrow-bands). We present the coordinates, spectroscopic and dual-NB redshifts, $I$ band magnitudes and the list of emission-lines that are detected. X-Ray and LOFAR detected sources are marked. We note that because the coverage in NB392, stV and NB921 is not homogeneous, the number of Ly$\alpha$ detections at $z=2.23$ and [O{\sc ii}] at $z=1.47$ is likely underestimated. An electronic version of this table is available online.}
\begin{tabular}{|l|l|l|r|r|l|l|l|}
\hline
  \multicolumn{1}{|c|}{ID} &
  \multicolumn{1}{c|}{R.A.} &
  \multicolumn{1}{c|}{Dec.} &
  \multicolumn{1}{c|}{$z_{\rm dual-NB}$} &
  \multicolumn{1}{c|}{$z_{\rm spec}$} &
  \multicolumn{1}{c|}{$I$} &
  \multicolumn{1}{c|}{Note} \\
\hline
B-HiZELS\_1 & 14:33:19.29 & +33:34:31.53 & 2.23 & 2.23 & 18.9 & Ly$\alpha$+[O{\sc iii}] + H$\alpha$.\\
  B-HiZELS\_2 & 14:32:58.85 & +33:25:49.33 & 2.23  &   & 19.6 & H$\alpha$ + Ly$\alpha$ + N{\sc v}, X-Ray detected.\\
  B-HiZELS\_3 & 14:31:41.51 & +33:49:11.27 & 2.23 & 2.26 & 20.4 & [O{\sc iii}] + H$\alpha$ + C{\sc iv} +Ly$\alpha$, X-Ray detected.\\
  B-HiZELS\_4 & 14:29:40.05 & +33:33:32.13 & 2.23 & 2.27 & 20.6 & [O{\sc iii}] + H$\alpha$ + Mg{\sc ii}, X-Ray detected.\\
  B-HiZELS\_5 & 14:32:01.52 & +33:16:59.86 & 2.23 &   & 20.9 & [O{\sc iii}] + H$\alpha$\\
  B-HiZELS\_6 & 14:30:38.29 & +33:20:17.85 & 2.23 &  & 21.3 & [O{\sc iii}]+H$\alpha$, X-Ray detected.\\
  B-HiZELS\_7 & 14:29:30.87 & +34:05:44.85 & 2.23 &  & 21.7 & [O{\sc iii}]+H$\alpha$ \\
  B-HiZELS\_8 & 14:32:13.88 & +33:25:57.48 & 2.23 &   & 21.9 & [O{\sc iii}] + H$\alpha$ + Ly$\alpha$, LOFAR detected.\\
  B-HiZELS\_9 & 14:31:06.64 & +33:46:19.18 & 2.23 &   & 22.1 & [O{\sc iii}] + H$\alpha$ \\
  B-HiZELS\_10 & 14:33:07.48 & +33:52:42.48 & 1.47 &   & 22.1 & [O{\sc ii}] + H$\alpha$ \\
  B-HiZELS\_11 & 14:32:37.05 & +33:33:56.18 & 1.47 &   & 22.1 & [O{\sc ii}] + H$\alpha$ \\
  B-HiZELS\_12 & 14:30:51.00 & +33:43:54.56 & 2.23 &  & 22.3 & [O{\sc iii}]+H$\alpha$ \\
  B-HiZELS\_13 & 14:33:14.40 & +33:46:53.17 & 1.47 &   & 22.4 & [O{\sc ii}] + H$\alpha$ \\
  B-HiZELS\_14 & 14:32:12.54 & +33:22:26.15 & 1.47 &   & 22.4 & [O{\sc ii}] + H$\alpha$ \\
  B-HiZELS\_15& 14:32:32.59 & +33:59:03.32 & 2.23 & 2.24 & 22.5 & [O{\sc iii}] + H$\alpha$ + C{\sc iv}, X-Ray \& LOFAR detected.\\
  B-HiZELS\_16 & 14:30:40.31 & +34:03:20.64 & 2.23 &   & 22.8 & Ly$\alpha$, C{\sc iv} and H$\alpha$.\\
  B-HiZELS\_17 & 14:31:33.54 & +34:02:48.52 & 2.23 &  & 22.8 & [O{\sc iii}] + H$\alpha$ \\
  B-HiZELS\_18 & 14:32:52.90 & +33:39:43.29 & 1.47 &   & 22.8 & [O{\sc ii}] + H$\alpha$ \\
  B-HiZELS\_19 & 14:30:16.12 & +33:17:09.56 & 1.47 &   & 23.0 & [O{\sc ii}] + H$\alpha$ \\
  B-HiZELS\_20 & 14:30:28.28 & +33:38:16.75 & 1.47 &   & 23.0 & [O{\sc ii}] + H$\alpha$ \\
  B-HiZELS\_21 & 14:30:19.38 & +33:37:10.18 & 1.47 &   & 23.1 & [O{\sc ii}] + H$\alpha$ \\
  B-HiZELS\_22 & 14:33:15.16 & +33:50:09.62 & 1.47 &   & 23.1 & [O{\sc ii}] + H$\alpha$ \\
  B-HiZELS\_23 & 14:32:37.05 & +33:33:06.64 & 1.47 &   & 23.1 & [O{\sc ii}] + H$\alpha$, LOFAR detected.\\
  B-HiZELS\_24 & 14:30:12.20 & +33:51:57.33 & 2.23 &   & 23.3 & [O{\sc iii}] + H$\alpha$ \\
  B-HiZELS\_25 & 14:30:26.29 & +33:28:51.17 & 2.23 &   & 23.3 & [O{\sc iii}] + H$\alpha$ \\
  B-HiZELS\_26 & 14:30:39.53 & +33:57:09.60 & 1.47 &   & 23.3 & [O{\sc ii}] + H$\alpha$ \\
  B-HiZELS\_27 &14:33:19.61 & +33:34:36.66 & 2.23 &   & 23.4 & [O{\sc iii}] + H$\alpha$ \\
  B-HiZELS\_28 & 14:31:50.68 & +33:18:44.15 & 1.47 &   & 23.4 & [O{\sc ii}] + H$\alpha$ \\
  B-HiZELS\_29 & 14:30:28.56 & +33:33:29.07 & 2.23 &   & 23.5 & Ly$\alpha$ and H$\alpha$\\
  B-HiZELS\_30 & 14:30:54.03 & +33:33:01.26 & 1.47 &   & 23.5 & [O{\sc ii}] + H$\alpha$ \\
  B-HiZELS\_31 & 14:31:19.33 & +33:26:14.90 & 1.47 &   & 23.5 & [O{\sc ii}] + H$\alpha$ \\
  B-HiZELS\_32 & 14:32:48.04 & +33:57:18.93 & 2.23 &  & 23.8 & [O{\sc iii}] + H$\alpha$ \\
  B-HiZELS\_33 & 14:30:45.55 & +33:23:50.12 & 1.47 &  & 23.8 & [O{\sc ii}] + H$\alpha$ \\
  B-HiZELS\_34 & 14:30:29.53 & +33:20:49.91 & 1.47 &  & 23.8 & [O{\sc ii}] + H$\alpha$ \\
  B-HiZELS\_35 & 14:30:18.58 & +34:03:24.88 & 2.23 &  & 23.9 & [O{\sc iii}] + Ly$\alpha$ \\
  B-HiZELS\_36 & 14:31:11.70 & +33:41:28.69 & 1.47 &  & 24.0 & [O{\sc ii}] + H$\alpha$ \\
  B-HiZELS\_37 & 14:33:21.85 & +33:54:50.65 & 1.47 &  & 24.2 & [O{\sc ii}] + H$\alpha$ \\
  B-HiZELS\_38 & 14:33:22.55 & +33:48:04.57 & 1.47 &  & 24.3 & [O{\sc ii}] + H$\alpha$ \\
  B-HiZELS\_39 & 14:29:44.27 & +33:50:43.84 & 1.47 &  & 24.3 & [O{\sc ii}] + H$\alpha$ \\
  B-HiZELS\_40 & 14:31:57.69 & +33:16:37.87 & 2.23 &  & 24.9 & [O{\sc ii}]+H$\alpha$ \\
  B-HiZELS\_41 & 14:32:41.50 & +33:26:18.45 & 2.23 &  & 25.5 & [O{\sc iii}] + H$\alpha$\\
  B-HiZELS\_42 & 14:30:39.05 & +33:51:51.16 & 2.23 &  & 25.8 & [O{\sc iii}]+H$\alpha$ \\
\hline\end{tabular}
\label{tab:dualemitters}
\end{table*}

\section{Example catalogs of line-emitters}\label{appendix:B}
\begin{table*}
\caption{First five entries in the catalog of NB392 line-emitters. Coordinates are in J2000. Line-flux is in $10^{-16}$ erg s$^{-1}$ cm$^{-2}$. EW$_{\rm obs}$ is in {\AA}, where -99 entries mark sources without secure continuum measurement, and have EW$_{\rm obs}>550$ {\AA}. Entries with 99 are undetected in $I$ band. Flag\_Class: 0: Unclassed, 1: LAE at $z=2.2$. A full version of this table is available online.  }
\begin{tabular}{|l|r|r|r|r|r|r|}
\hline
  \multicolumn{1}{|c|}{ID} &
  \multicolumn{1}{c|}{R.A.} &
  \multicolumn{1}{c|}{Dec.} &
  \multicolumn{1}{c|}{Line-flux} &
     \multicolumn{1}{c|}{EW$_{\rm obs}$} &
  \multicolumn{1}{c|}{$I$} &
  \multicolumn{1}{c|}{Flag\_Class} \\
\hline
  B-HiZELS\_NB392\_1 & 217.945 & 33.310 & 1.3 & 136 & 22.2  & 0\\
  B-HiZELS\_NB392\_2 & 218.017 & 33.316 &1.9 & 119 &  20.1 & 0\\
  B-HiZELS\_NB392\_3 & 217.779 & 33.340 & 1.2 & 158 & 23.6 & 1\\
  B-HiZELS\_NB392\_4 & 218.056 & 33.366 &  2.1 & -99 & 99 & 0\\
  B-HiZELS\_NB392\_5 & 217.675 & 33.386  & 4.2 & 56 &21.9 & 0\\

\hline\end{tabular}
\end{table*}

\begin{table*}
\caption{First five entries in the catalog of stV line-emitters. Coordinates are in J2000. Line-flux is in $10^{-16}$ erg s$^{-1}$ cm$^{-2}$. EW$_{\rm obs}$ is in {\AA}, where -99 entries mark sources without secure continuum measurement, and have EW$_{\rm obs}>2500$ {\AA}. Entries with 99 are undetected in $I$ band. Flag\_Class: 0: Unclassed, 1: LAE at $z=2.4$. A full version of this table is available online.}
\begin{tabular}{|l|r|r|r|r|r|r|}
\hline
  \multicolumn{1}{|c|}{ID} &
  \multicolumn{1}{c|}{R.A.} &
  \multicolumn{1}{c|}{Dec.} &
  \multicolumn{1}{c|}{Line-flux} &
    \multicolumn{1}{c|}{EW$_{\rm obs}$} &
  \multicolumn{1}{c|}{$I$} &
  \multicolumn{1}{c|}{Flag\_Class} \\

\hline
   B-HiZELS\_stV\_1 & 218.239 & 33.305 & 64.3 & 97 & 17.2 & 0\\
  B-HiZELS\_stV\_2 & 218.021 & 33.314 & 13.9 & 1131 & 22.6 & 0\\
  B-HiZELS\_stV\_3 & 217.909 & 33.351 & 13.5 & 590 & 22.5 & 0\\
  B-HiZELS\_stV\_4 & 218.053 & 33.354 & 6.4 & 207 & 22.3 & 0\\
  B-HiZELS\_stV\_5 & 218.086 & 33.353 & 25.7 & 1079 & 21.7 & 1\\ 

\hline\end{tabular}
\end{table*}

\begin{table*}
\caption{First five entries in the catalog of NB501 line-emitters. Coordinates are in J2000. Line-flux is in $10^{-16}$ erg s$^{-1}$ cm$^{-2}$. EW$_{\rm obs}$ is in {\AA}, where -99 entries mark sources without secure continuum measurement, and have EW$_{\rm obs}>550$ {\AA}. Entries with 99 are undetected in $I$ band. Flag\_Class: 0: Unclassed, 1: LAE at $z=3.1$. A full version of this table is available online.}
\begin{tabular}{|l|r|r|r|r|r|r|}
\hline
  \multicolumn{1}{|c|}{ID} &
  \multicolumn{1}{c|}{R.A.} &
  \multicolumn{1}{c|}{Dec.} &
  \multicolumn{1}{c|}{Line-flux} &
   \multicolumn{1}{c|}{EW$_{\rm obs}$} &
     \multicolumn{1}{c|}{$I$} &
  \multicolumn{1}{c|}{Flag\_Class} \\

\hline
  B-HiZELS\_NB501\_1 & 217.635 & 33.366  & 14.9 & 198 & 19.9 & 0\\
  B-HiZELS\_NB501\_2 & 217.923 & 33.820  & 13.7 & 64 & 20.4 & 0\\
  B-HiZELS\_NB501\_3 & 217.619 & 33.558  & 7.8 & 391 & 23.3& 0\\
  B-HiZELS\_NB501\_4 & 218.127 & 33.666  & 7.2 & -99 & 22.9 & 1\\
  B-HiZELS\_NB501\_5 & 217.668 & 34.056 & 5.4 & 138 & 22.6  & 0\\

\hline\end{tabular}
\end{table*}

\begin{table*}
\caption{First five entries in the catalog of NB921 line-emitters. Coordinates are in J2000. Line-flux is in $10^{-16}$ erg s$^{-1}$ cm$^{-2}$. EW$_{\rm obs}$ is in {\AA}, where -99 entries mark sources without secure continuum measurement, and have EW$_{\rm obs}>1100$ {\AA}. Entries with 99 are undetected in $I$ band. Flag\_Class: 0: Unclassed, 1: H$\alpha$ at $z=0.4$, 2: H$\beta$/[O{\sc iii}] at $z=0.8$, 3: [O{\sc ii}] at $z=1.47$. A full version of this table is available online. }
\begin{tabular}{|l|r|r|r|r|r|r|}
\hline
  \multicolumn{1}{|c|}{ID} &
  \multicolumn{1}{c|}{R.A.} &
  \multicolumn{1}{c|}{Dec.} &
  \multicolumn{1}{c|}{Line-flux} &
  \multicolumn{1}{c|}{EW$_{\rm obs}$} &
  \multicolumn{1}{c|}{$I$} &
  \multicolumn{1}{c|}{Flag\_Class} \\
\\
\hline
  B-HiZELS\_NB921\_1 & 217.417 & 33.559 & 3.9 & 40 & 20.5 &0 \\
  B-HiZELS\_NB921\_2 & 217.990 & 33.277 & 0.3 & 142 & 24.6 & 0 \\
  B-HiZELS\_NB921\_3 & 217.831 & 33.437 & 0.3 & 59 & 23.7 & 3 \\
  B-HiZELS\_NB921\_4 & 218.220 & 33.662 & 1.3 & 101 & 22.8 & 3 \\
  B-HiZELS\_NB921\_5 & 217.799 & 33.691 & 0.3 & 70 & 23.9 & 3 \\

\hline\end{tabular}
\end{table*}

\begin{table*}
\caption{First five entries in the catalog of NB$_{\rm H}$ line-emitters. Coordinates are in J2000. Line-flux is in $10^{-16}$ erg s$^{-1}$ cm$^{-2}$. EW$_{\rm obs}$ is in {\AA}, where -99 entries mark sources without secure continuum measurement, and have EW$_{\rm obs}>1200$ {\AA}. Entries with 99 are undetected in $I$ band. Flag\_Class: 0: Unclassed, 1: H$\alpha$ at $z=1.47$, 2: H$\beta$/[O{\sc iii}] at $z=2.2$, 3: [O{\sc ii}] at $z=3.3$. A full version of this table is available online. }
\begin{tabular}{|l|r|r|r|r|r|r|}
\hline
  \multicolumn{1}{|c|}{ID} &
  \multicolumn{1}{c|}{R.A.} &
  \multicolumn{1}{c|}{Dec.} &
  \multicolumn{1}{c|}{Line-flux} &
  \multicolumn{1}{c|}{EW$_{\rm obs}$} &
  \multicolumn{1}{c|}{$I$} &
  \multicolumn{1}{c|}{Flag\_Class} \\
\\
\hline
  B-HiZELS\_NBH\_1 & 218.345 & 33.265 & 1.7 & 188 & 99 & 0\\
  B-HiZELS\_NBH\_2 & 218.329 & 33.299 & 2.1 & 212 & 99 & 0\\
  B-HiZELS\_NBH\_3 & 217.990 & 33.312 & 1.8 & 474 & 25.5 & 0\\
  B-HiZELS\_NBH\_4 & 218.165 & 33.317 & 1.8 & 139 & 22.1 & 0\\
  B-HiZELS\_NBH\_5 & 217.670 & 33.344 & 1.6 & 180 & 22.8 & 0\\

\hline\end{tabular}
\end{table*}

\begin{table*}
\caption{First five entries in the catalog of NB$_{\rm K}$ line-emitters. Coordinates are in J2000. Line-flux is in $10^{-16}$ erg s$^{-1}$ cm$^{-2}$. EW$_{\rm obs}$ is in {\AA}, where -99 entries mark sources without secure continuum measurement, and have EW$_{\rm obs}>1250$ {\AA}. Entries with 99 are undetected in $I$ band. Flag\_Class: 0: Unclassed, 1: H$\alpha$ at $z=2.23$, 2: H$\beta$/[O{\sc iii}] at $z=3.2$, 3: [O{\sc ii}] at $z=4.7$, 4: $z<1$. A full version of this table is available online. }
\begin{tabular}{|l|r|r|r|r|r|r|}
\hline
  \multicolumn{1}{|c|}{ID} &
  \multicolumn{1}{c|}{R.A.} &
  \multicolumn{1}{c|}{Dec.} &
  \multicolumn{1}{c|}{Line-flux} &
  \multicolumn{1}{c|}{EW$_{\rm obs}$} &
  \multicolumn{1}{c|}{$I$} &
    \multicolumn{1}{c|}{Flag\_Class} \\
\hline
  B-HiZELS\_NBK\_1 & 218.080 & 33.261 & 0.8 & 121 & 22.3 & 4 \\
  B-HiZELS\_NBK\_2 & 218.345 & 33.265 & 1.1 &-99 & 99 & 0 \\
  B-HiZELS\_NBK\_3 & 217.878 & 33.276 & 0.7 & 410 & 25.4 & 0 \\
  B-HiZELS\_NBK\_4 & 217.991 & 33.277 & 1.0 & 494 & 25.1 & 1 \\
  B-HiZELS\_NBK\_5 & 217.996 & 33.280 & 1.0 & -99 & 24.8 & 1 \\
\hline\end{tabular}
\end{table*}


\bsp	
\label{lastpage}
\end{document}